\documentclass[prx,reprint,groupedaddress,superscriptaddress]{revtex4-2}
\usepackage[resetlabels]{multibib}
\usepackage{lipsum}
\usepackage{graphicx}
\usepackage{dcolumn}
\usepackage{bm}
\usepackage{amsmath,amssymb,amstext,mathtools}
\usepackage{physics}
\usepackage{siunitx}
\usepackage{csquotes}
\usepackage[svgnames]{xcolor}
\usepackage[colorlinks=true,
            linkcolor=red,
            urlcolor=blue,
            citecolor=DarkGreen]{hyperref}
\usepackage[capitalize]{cleveref}
\usepackage{natbib}
\usepackage{dsfont}

\renewcommand{\selectlanguage}[1]{}

\begin{document}
\title{Automated Discovery of Coupled Mode Setups}

\author{Jonas~Landgraf}\email{Jonas.Landgraf@mpl.mpg.de}
\affiliation{Max Planck Institute for the Science of Light, Staudtstr.~2, 91058 Erlangen, Germany}
\affiliation{Physics Department, University of Erlangen-Nuremberg, Staudstr.~5, 91058 Erlangen, Germany}
\author{Vittorio~Peano}
\affiliation{Max Planck Institute for the Science of Light, Staudtstr.~2, 91058 Erlangen, Germany}
\author{Florian~Marquardt}
\affiliation{Max Planck Institute for the Science of Light, Staudtstr.~2, 91058 Erlangen, Germany}
\affiliation{Physics Department, University of Erlangen-Nuremberg, Staudstr.~5, 91058 Erlangen, Germany}

\date{\today}

\begin{abstract}
    In optics and photonics, a small number of building blocks, like resonators, waveguides, arbitrary couplings, and parametric interactions, allow the design of a broad variety of devices and functionalities, distinguished by their scattering properties. These include transducers, amplifiers, and nonreciprocal devices, like isolators or circulators. Usually, the design of such a system is handcrafted by an experienced scientist in a time-consuming process where it remains uncertain whether the simplest possibility has indeed been found. In our work, we develop the discovery algorithm \textsc{AutoScatter} that automates this challenge. By optimizing the continuous and discrete system properties our automated search identifies the minimal resources required to realize the requested scattering behavior. In the spirit of artificial scientific discovery, it produces a complete list of interpretable solutions and leads to generalizable insights, as we illustrate in several examples. This opens the door towards automated discovery of scattering setups for photonics, microwaves and optomechanics, with possible future extensions to periodic structures, sensing, and electronic devices. 
\end{abstract}

\maketitle

\date{\today}

Waves represent one of the most basic physical phenomena. This explains the importance of applications enabled by wave transport in suitably designed structures, ranging from integrated photonics to microwave circuits. 
A large class of on-chip devices, from  isolators and circulators for non-reciprocal transport \cite{kamal_noiseless_2011,sliwa2015reconfigurable, peterson_demonstration_2017,herrmann2022mirror} to hybrid frequency converters \cite{andrews_bidirectional_2014,hill_coherent_2012} and new classes of amplifiers \cite{macklin2015near,fang_generalized_2017},
can be engineered by combining the right building blocks in a well-chosen scattering topology. So far, each new design in this domain has been proposed based on human ingenuity and experience. 
It would be desirable to speed up the pace of this laborious exploration process in the high-dimensional space of possible setups and obtain a complete overview of all conceptually distinct options for any given desired functionality.
Here, we present the automated discovery algorithm \textsc{AutoScatter} \cite{github_autoscattering} that achieves these goals and also helps to uncover new  insights.



\begin{figure*}
    \centering
    \includegraphics[width=\linewidth]{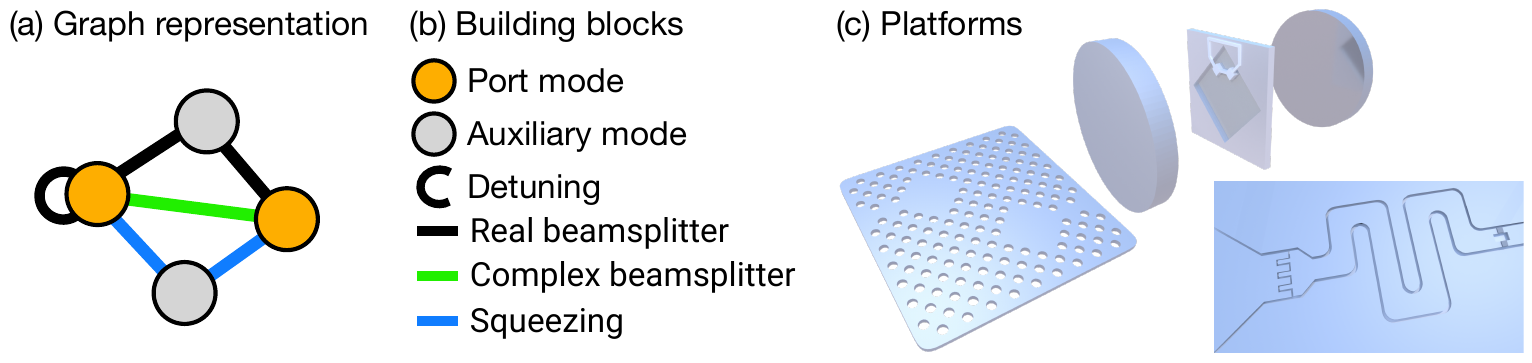}
    \caption{\textbf{Graph representation of coupled-mode systems.} (a) Physical modes correspond to graph nodes, couplings to edges. Port modes (orange) are connected to waveguides which relay input and output signals. Auxiliary modes (gray) have an intrinsic loss channel whose output signal is not monitored. The edges' weights are dimensionless coupling rates which are rescaled over the loss rates.
    (b) Discrete building blocks available to the discovery algorithm. (c) Illustration of some of the many hardware platforms which are efficiently described by coupled-mode theory. These include photonic circuits, e.g., based on photonic crystals (left), optomechanical setups (top), and microwave circuit setups (right).}
    \label{fig:model}
\end{figure*}

Tackling challenges of this kind is the domain of artificial scientific discovery \cite{wang2023scientific}. This rapidly evolving field has the ambitious aim to automate all aspects of the scientific process using tools of machine learning and artificial intelligence, with an emphasis on interpretability and discovery of new conceptual insights. Pioneering  examples are the identification of organic molecules from mass spectrometry by the computer program \textsc{Dendral} \cite{lindsay1993dendral} and the generation and experimental testing of hypotheses in biochemistry by a robot scientist \cite{king2004functional}. Important challenges in this field include automatic extraction of symbolic expressions \cite{schmidt2009distilling,udrescu2020ai}, discovery of collective coordinates \cite{iten2020discovering,sarra2021renormalized}, automated planning and execution of experiments \cite{lennon2019efficiently,moon2020machine,duris2020bayesian}, and generating new experimental setups. 
The latter task has been successfully demonstrated for the preparation of entangled states in quantum optics \cite{krenn2016automated,krenn2020computer}, for the optimization of superconducting circuits \cite{menke_automated_2021}, gravitational wave detectors \cite{krenn2023digital}, and signal processing using optical components \cite{maclellan2024inverse}. 


Automated discovery relies crucially on casting the complex search space into a unifying language.  For the domain envisaged here, we identified coupled-mode theory \cite{yariv1973coupled,haus1991coupled} as providing a  sweet spot on the continuum between more  microscopic hardware-specific descriptions (such as engineered refractive-index distributions \cite{jensen2011topology,piggott2015inverse,molesky2018inverse,ma2021deep}) and  descriptions based on macroscopic building blocks on the level of entire functional elements (like amplifiers). The latter require a larger number of building blocks and need more human design choices; for example \cite{maclellan2024inverse} focuses on signal processing setups that can be decomposed into networks of specific technical components. 
In contrast, coupled-mode designs are transferable between different platforms and can always be translated back into several different concrete physical structures, ranging from photonic crystals to optomechanical setups and microwave circuits. Of equal importance is the formalization of the goal. Here, we chose a description in terms of the scattering matrix, specifying the desired functionality via the externally visible characteristics.

Our discovery algorithm \textsc{AutoScatter} is based on an efficient search process that exploits the recursive modular structure of the problem to prune the exponentially large space of possibilities and provides a complete list of setups fulfilling specified constraints. It automatically ends up with interpretable and generalizable results both by using discrete building blocks and by suggesting only dimensionless parameters of universal, implementation-independent meaning. The final outcome is a menu of irreducible setups that allows to select optimality criteria even afterwards. Moreover, we introduce a fruitful general extension of the toolbox available in artificial discovery, observing that conceptual insights in design are often based on thinking about idealized structures realizing asymptotic limits of parameters. We describe a modification of our algorithm that exploits this meta-principle. 

We illustrate the benefits of \textsc{AutoScatter} for a variety of target functionalities, including couplers, circulators and directional amplifiers, where it produces new discoveries and generalizable insights.

\section{Coupled-mode theory}
\label{sec:coupled_mode_theory}
We consider a network of coupled bosonic modes, which can be implemented as  an optical, electric, mechanical, or hybrid setup, comprising different types of modes. This device is probed with multiple input signals and its linear response is encoded in its scattering matrix $S$. The element $S_{ji}$ is the transmission coefficient  of a signal  from the input port $i$ to the output port $j$. In our work, we aim to automatically discover the simplest setups that fulfill a certain target response, which is encoded in a target scattering matrix $S^\mathrm{target}$. 

The  precise characterization of a specific physical implementation of an ideal target device requires a carefully crafted system-specific model. In contrast, the ideal behavior is typically captured by a simple and general high-level description.  Like a human scientist would do, we choose our artificial scientist to adopt such high-level description. This approach allows us to  discover ideal and transferable solutions. We discuss in \cref{sec:implementation} how to taylor our discovered solutions to specific hardware platforms.

At the highest level of description, any multi-mode circuit can be represented as a graph with colored edges \cite{ranzani_graph-based_2015} (see \cref{fig:model}(a,b)), regardless of the underlying hardware platform. Nodes correspond to modes and edges represent resonant interactions between the modes. 
``Port modes'' (orange) are connected to waveguides transmitting the input and output signals of interest. The carrier frequencies $\omega_{L,j}$ of the input fields should be close to the respective mode resonances $\omega_j$ and can be different for different port modes, e.g., an optical and a mechanical mode will have very different frequencies. Here, we focus on the response of each port mode at the respective carrier frequency, which is described by a single scattering matrix encoding the ideal behaviour of the device. \textsc{AutoScatter} could be extended to also consider aspects like optimizing the bandwidth \cite{naaman_synthesis_2022} but this would lead to less transferable solutions, see \cref{sec:implementation}, and is therefore  beyond the scope of this work.

The port modes are often insufficient to implement a given scattering matrix. Therefore, we permit to add a variable number of auxiliary modes (gray). The auxiliary modes are also damped but we do not monitor the corresponding loss channels. Therefore, these channels are not included in the target scattering matrix.




We use colored edges to distinguish between different types of interactions. The edges' weights are appropriately rescaled dimensionless coupling constants. We choose the rescaling such that the weights fully determine the scattering matrix and, thus, also incorporate the essential information about the relative strength of the different coherent and dissipative processes. This rescaling is an essential step in our method and merits a separate discussion below. Before that we discuss the coherent interactions between the modes of the multimode device.




The response of state-of-the-art multimode devices can be precisely characterized by linearizing the full nonlinear dissipative dynamics about a steady state. In this setting, the interactions between $N$ high-quality-factor modes are well described by a time-independent bosonic quadratic Hamiltonian of the form
\begin{equation}
\label{eq:second_quantized_Hamiltonian}
\frac{1}{2}\sum_{ij=1}^N \left( g_{ij} \hat{a}_i^\dagger \hat{a}_j+\nu_{ij}\hat{a}_i^\dagger \hat{a}_j^\dagger \right) + \mathrm{H.c.} 
\end{equation}
Here, $\mathrm{H.c.}$ denotes the Hermitian conjugate, and $\hat{a}_j$ and $\hat{a}^\dagger_j$ are the ladder operators for mode $j$. Any off-diagonal non-zero coupling constant, $\nu_{ij}\neq 0$ or $g_{ij}\neq 0$, corresponds to a graph edge. 
Importantly, there are two distinct types of interactions: Squeezing interactions with coupling constants $\nu_{ij}$ (corresponding to blue edges in our graph representation) create pairs of excitations in the modes $i$ and $j$, while beamsplitter interactions with coupling constants $g_{ij}$ exchange excitations between the modes $i$ and $j$. In our optimisation, we also distinguish between complex-valued beamsplitter couplings (green) with $g_{ij}\!\in\!\mathbb{C}$ which involve a complex phase depending on the direction of the exchange and real-valued beamsplitter couplings (black) with $g_{ij}\!\in\!\mathbb{R}$ which are direction independent. It is useful to distinguish between the two different types of beamsplitter interactions in view of identifying couplings that can be implemented passively in many common experimental scenarios, see \cref{sec:implementation} for more details.

An important aspect of the Hamiltonian in \cref{eq:second_quantized_Hamiltonian} is that it is defined in a rotating frame. By applying a so-called ``Rotating-Wave-Approximation'' (RWA), we drop the fast oscillating terms describing non-resonant interactions (see \cref{app:rot_wave} for more details). In this rotating frame, each mode $j$ rotates at its corresponding carrier frequency $\omega_{L,j}$ (for an auxiliary mode $j$, $\omega_{L,j}$ is the vibration frequency induced by a signal entering the device at any port $l$ with the respective carrier frequency $\omega_{L,l}$). Thus, $-g_{jj}$ is to be interpreted  as the detuning $\Delta_j$ of the carrier frequency $\omega_{L,j}$ from the mode resonance $\omega_j$, $\Delta_j=\omega_{L,j}-\omega_j$. In our graph representation, any non-zero detuning is visualized as self-loop, see \cref{fig:model}(a).
Importantly, the RWA Hamiltonian already captures the ideal behavior of the devices even though it does not directly depend on the resonance frequencies $\omega_j$. This  is an appealing feature that makes any insights gained using this approximation highly  transferable across various platforms. For this reason, it  is  widely adopted for hand-crafted design of multi-mode circuits \cite{sliwa2015reconfigurable,peterson_demonstration_2017,herrmann2022mirror,andrews_bidirectional_2014,hill_coherent_2012,fang_generalized_2017,ranzani_graph-based_2015,naaman_synthesis_2022,lecocq_nonreciprocal_2017,habraken_continuous_2012,koch_time-reversal-symmetry_2010,metelmann_nonreciprocal_2015,malz_quantum-limited_2018,abdo2013directional,liu_fully_2023,clerk2010introduction,bernier2017nonreciprocal,kwende2023josephson,aspelmeyer2014cavity,Hafezi_Optomechanically_2012}.  A more refined description including non-resonant interactions is often used to quantify how a specific implementation deviates from the ideal target behavior. 
This deviation is typically small for high-quality factor modes ubiquitous in modern quantum optics platforms. Moreover, it is device-specific, thus,  incorporating it in our approach would lead to less transferable results.

As anticipated above, a key novelty of our graph representation is our rescaling of the weights. Previous works adopting graph representations of multimode circuits \cite{ranzani_graph-based_2015,naaman_synthesis_2022} had introduced weights by expressing the coupling constants and other frequency parameters in units of a single base frequency.  This is also the most common approach in other numerical and analytical studies of the scattering matrix. In this approach, the number of free parameters/graph weights is reduced by one without losing  generality. 
Instead, we achieve a much larger reduction of the free parameters by incorporating in the graph edges $N$ decay rates (one for each mode). We first express the  Hamiltonian in \cref{eq:second_quantized_Hamiltonian} in matrix form, i.e., in first quantization as a so-called Bogoliubov-de-Gennes (BdG) Hamiltonian. Then, we rescale the BdG Hamiltonian in the following way (see \cref{Appendix:Bog_de_Gennes}):
\begin{equation}\label{eq:rescaled_Bog_de_Gennes}
	 H=
\frac{1}{\sqrt{\kappa}}
 \begin{pmatrix}
 g&\nu\\
 \nu^{*}&g^{*}\\
 \end{pmatrix}
\frac{1}{\sqrt{\kappa}}.
\end{equation}
Here, $\kappa$ is the diagonal matrix $\kappa={\rm diag}(\kappa_1,\ldots,\kappa_N,\kappa_1,\ldots,\kappa_N)$, and $\kappa_i$ is the decay rate of mode $i$ (out-coupling rate for port modes, intrinsic loss rate for auxiliary modes). Our graph weights are then the entries of this dimensionless BdG Hamiltonian. Importantly, this approach does not involve any loss of generality since in our high-level description based on the RWA, the scattering matrix is fully determined by the dimensionless BdG Hamiltonian $H$ (see \cref{app:scattering_matrix}). 

The  importance of the rescaling in \cref{eq:rescaled_Bog_de_Gennes} is two-fold. On one hand, the weights $H_{ij}$ acquire a universal  implementation-independent meaning as they are now directly related to the dimensionless physical parameters that determine the scattering behavior, i.e.~the dimensionless detuning $(\omega_{L,i}-\omega_i)/\kappa_i=-H_{ii}$,
the cooperativities $C_{ij}=4|H_{ij}|^2$, 
and the gauge invariant phases  accumulated on  closed loops (synthetic field fluxes),  $\Phi_{i,j,l,\ldots,k}=\arg (H_{ij}H_{jl}\ldots H_{ki})$. On the other hand,  it implies that the algorithm discovers  classes of solutions that  all obey the desired target scattering behavior. 
When translating back to a concrete physical setup, this leaves the freedom to choose arbitrarily the decay rates $\kappa_i$ and, thus, fix the couplings accordingly (although for amplifiers some additional stability constraints might apply, see \cref{app:stability}).


For concreteness, we focus below on the important special case of phase-preserving devices. In this setting, the equations of motion for $(\hat{a}_1,\ldots, \hat{a}_N,\hat{a}_1^\dag,\ldots, \hat{a}_N^\dag)$ break down into two decoupled sets of equations which are related to each other via the particle-hole symmetry. As a result, the scattering matrix is an $N\times N$ matrix. For phase-preserving amplifiers, we enforce the block structure of the equations of motion by following the common engineering principle of dividing the modes (graph nodes) into two disjoint sets \cite{liu_fully_2023,sliwa2015reconfigurable,ranzani_graph-based_2015,lecocq_nonreciprocal_2017,malz_quantum-limited_2018}. The modes within the same set are then only coupled via beamsplitter couplings, while the modes in different sets are only coupled via squeezing interactions.  In this way, the annihilation operators for the modes in one set are only coupled to the creation operators of the modes in the other set (see \cref{app:scattering_matrix} for more details). 
This constraint restricts the set of graphs considered in our optimisation.
For example,  a triangle with just one blue edge will be excluded. We do not indicate explicitly which mode belongs to the same group because this can be easily deduced from the colors of the edges (see \cref{app:scattering_matrix}).

\begin{figure*}
    \centering
    \includegraphics[width=\linewidth]{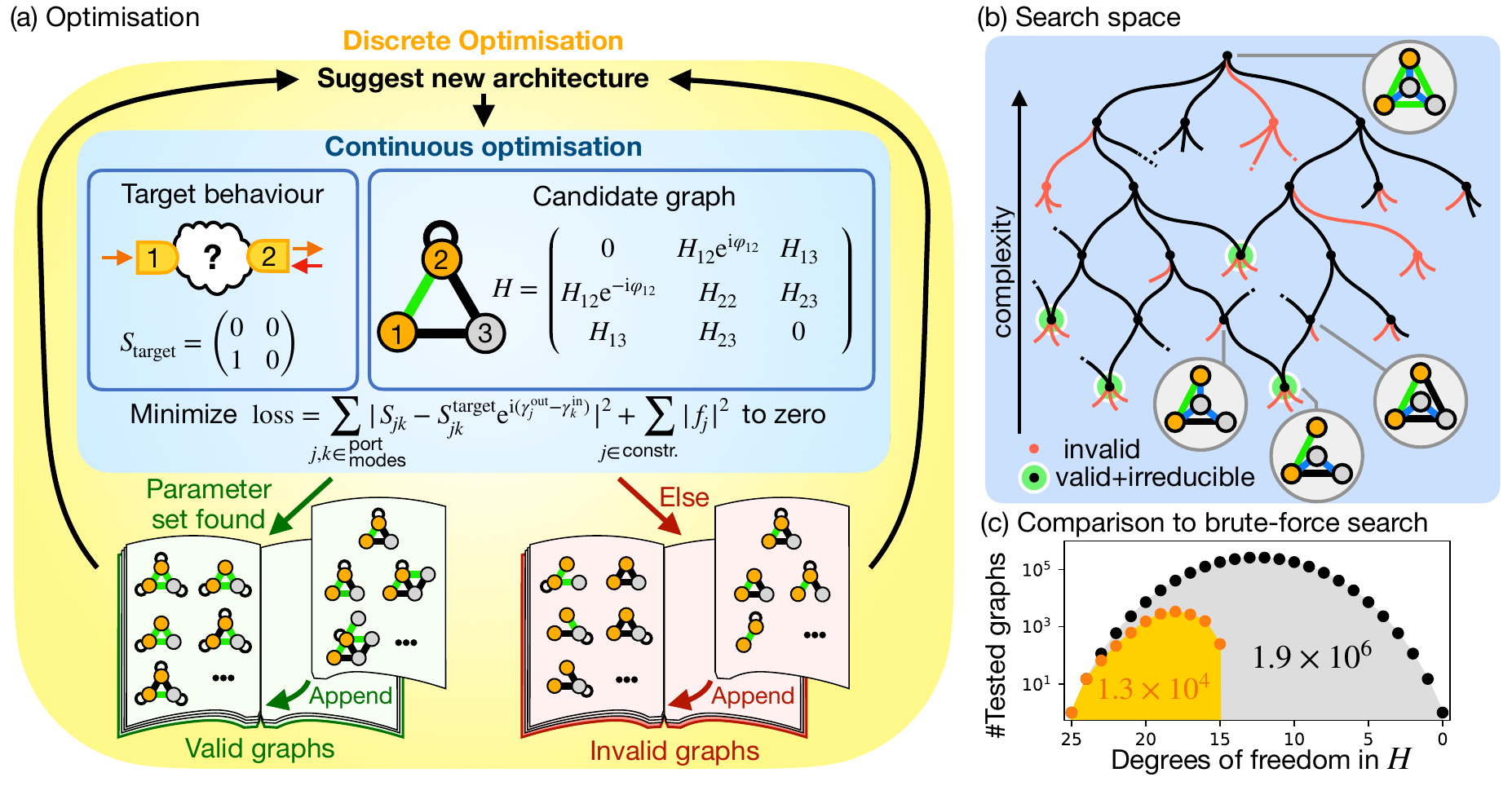}
    \caption{
    \textbf{Automated Discovery of Modular Scattering Setups}. (a) Optimisation scheme behind {\sc AutoScatter}. The discrete optimization sorts graphs into the libraries of valid (green book) and invalid graphs (red book). In every round, a new graph is suggested and, depending on the outcome of the continuous optimisation, sorted with all its extensions/subgraphs into the library of valid/invalid graphs. The continuous optimization minimizes the loss function to realize a target scattering behaviour (left block), here an isolator. The candidate graph (right block) defines the number of auxiliary modes and an allowed set of non-zero couplings rates and phases in the dimensionless Hamiltonian $H$. (b) Search space (schematic). The search starts from the smallest fully connected graph realising the target behavior (topmost graph). From there on, our method successively prunes edges, resulting in multiple irreducible solutions. (c) Comparison of the number of tested graphs for the directional coupler (see \cref{fig:results}(b)), which uses five modes. The total number of possible graphs (black) for a certain number of free variables in $H$ is compared to the number of graphs which had actually to be tested by our exhaustive search. Our scheme starts with maximally connected graphs (left end) and successively goes to more restricted graphs (right end). By testing only around 13,000 graphs we are able to characterize all 1.9 million graphs in this example.
    }
    \label{fig:method}
\end{figure*}

\section{Optimisation scheme}
\label{sec:optimisation}
To identify setups that implement the target scattering matrix, {\sc AutoScatter} has to explore the discrete search space of all possible graphs and, given a graph, find an appropriate set of values for the underlying continuous parameters, i.e., the non-zero cooperativities and coupling phases. To this end, {\sc AutoScatter} performs a two-step procedure: A discrete optimization routine suggests new graphs and an embedded continuous optimization algorithm  looks for appropriate values of the parameters (see \cref{fig:method}(a)). If the continuous optimization is successful,  we label a graph as valid, i.e., this graph fulfills the desired scattering behavior.

First, we discuss the discrete optimization, see \cref{fig:method}(a). We aim to identify the simplest valid graphs as they represent the simplest experimental implementations of the target response. Often there are a multitude of valid graphs that differ in the number of auxiliary modes and the number  and  complexity of the required couplings. 
In a first step, {\sc AutoScatter} tests only fully connected graphs for validity starting from the graph with zero auxiliary modes. It increments the number of auxiliary modes until it finds the minimum number of auxiliary modes required to achieve the target behavior. This is the smallest fully connected graph that satisfies the target behavior. From then on, {\sc AutoScatter} prunes this fully connected graph and identifies ``irreducible"  graphs that can not be further simplified by setting any coupling rate or phase to zero.  To identify all such graphs, we perform a complete enumeration.

Due to the large discrete search space, a brute-force approach testing all graphs is infeasible in many cases of interest. We solve this problem by leveraging the knowledge gained from previously tested graphs, drastically reducing the required computational resources. We note that an alternative approach could be developed based purely on continuous optimization, introducing terms in a loss function that reward sparsity (zero couplings) as in \cite{krenn_conceptual_2021,ruiz-gonzalez_digital_2022,arlt2022digital} – however, this would not enable us to obtain a complete set of solutions.

We take advantage of the fact that removing an edge from a graph is equivalent to setting a coupling rate to zero. Thus, if a graph is found to be invalid, that carries over to all graphs, which can be generated by removing some edges and/or setting coupling phases to zero. Likewise, all extensions of a valid graph are valid (see \cref{fig:method}(b)). This observation allows us to efficiently construct exhaustive libraries of valid and invalid graphs by testing only a small fraction, see \cref{fig:method}(c). In \cref{sec:search_space_reduction}, we give an in-depth analysis of the search space reduction provided by {\sc AutoScatter}. See also \cref{app:discrete_optimization} for further details on the discrete optimization.

We now turn to the continuous optimization. The scattering matrix $S$ is fully determined by the dimensionless Hamiltonian $H$ defined in \cref{eq:rescaled_Bog_de_Gennes}, which contains the free coupling parameters. We look for setups implementing the desired scattering behavior by minimizing the squared deviation loss function ${\cal L}$ with respect to those couplings:
\begin{equation}\label{eq:loss}
    {\cal L}=\sum_{j,k\in\substack{\mathrm{port}\\\mathrm{modes}}}\abs{S_{jk}-S_{jk}^\mathrm{target}\mathrm{e}^{\mathrm{i}(\gamma^{\rm out}_j-\gamma^{\rm in}_k)}}^2 +\sum_{j\in \mathrm{constr.}} |f_j|^2. 
\end{equation} 
We note that the target scattering matrix $S^\mathrm{target}$ can contain further free parameters that are not specified by the desired scattering properties. For example, we might impose that two scattering amplitudes are equal, but a suitable value needs to be discovered. In addition, we introduce the free gauge parameters $\gamma^{\rm in/out}_i$. This reflects the freedom of choosing independently for each port a reference quadrature and a reference position where the fields are measured. For more details, including how to fix the gauge in specific experimental settings, see \cref{app:gauge_freedom}. 
Finally, we allow for extra constraints $f_j=0$ that can be used, e.g., to enforce minimum added noise as detailed in \cref{App:noise_analysis}.
We  minimize the loss ${\cal L}$ as a function of the free parameters in  $H$, $S_\mathrm{target}$, and the phases $\gamma^{\rm in/out}_{i}$, see \cref{app:details_continuous_optimization} for more details. A valid solution is obtained whenever the loss reaches zero.

\begin{figure}
    \centering
    \includegraphics[width=0.6\linewidth]{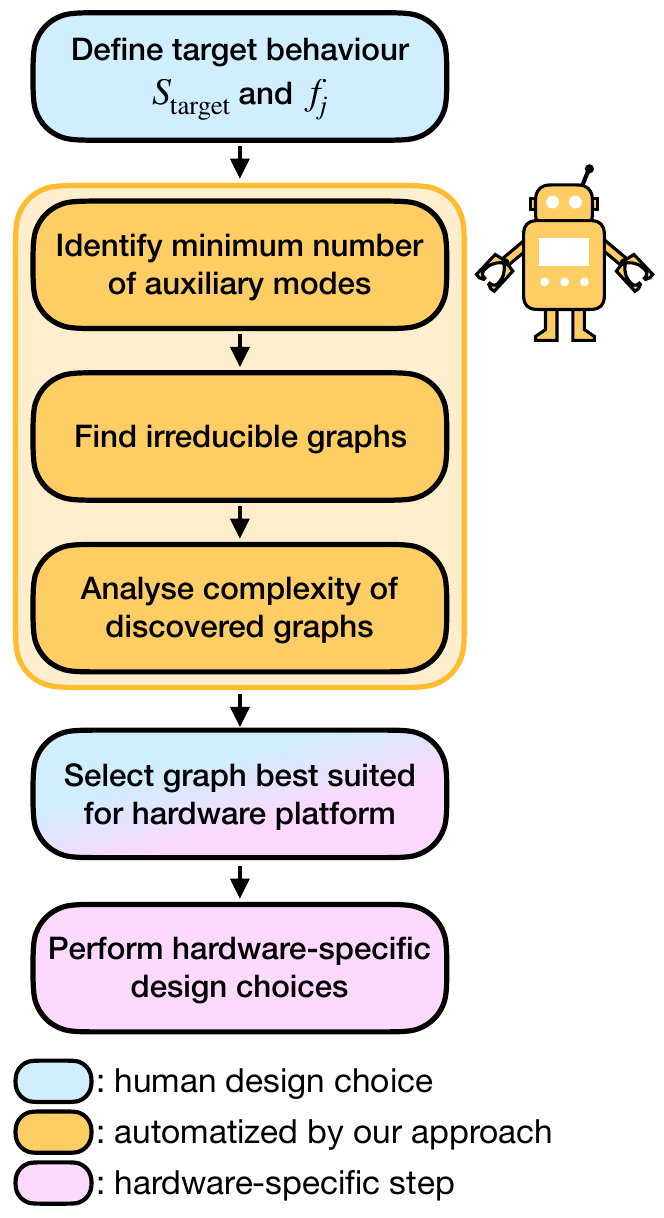}
    \caption{Flowchart showing the full design pipeline of coupled-mode systems. The steps automatized by {\sc AutoScatter} are highlighted in orange. The remaining steps, not covered by our method, are hardware-specific, see \cref{sec:implementation}.}
    \label{fig:design_flow}
\end{figure}

The optimization described within this section is the main part of {\sc AutoScatter}. In addition, {\sc AutoScatter} provides an automated analysis of the complexity of the discovered solutions, as discussed in \cref{sec:implementation}.
The broader design pipeline for a specific experimental platform including also steps not covered by {\sc AutoScatter} is shown in the flowchart in \cref{fig:design_flow}. 




In the following, we demonstrate through a series of illustrative examples how {\sc AutoScatter} can design scattering setups, some of which have been overlooked by hand-crafted methods. We also show that our approach can automate several key aspects of scientific discovery. These include the generalization of design patterns, the finding of explicit analytical expressions for continuous classes of problems, and the discovery of asymptotic solutions.


\begin{figure*}
    \centering
    \includegraphics[width=\linewidth]{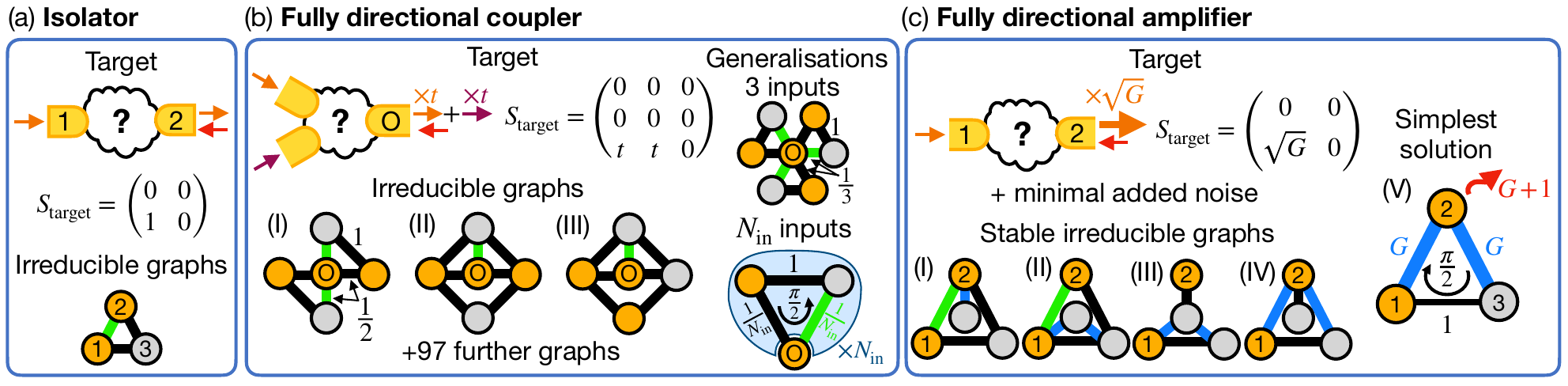}
    \caption{\textbf{Illustrative examples for automated discovery of scattering setups}. Target scattering behaviors and the corresponding irreducible graphs identified by {\sc AutoScatter}. (a) Isolator. Only signal from mode 1 to 2 is transmitted. (b) Directional coupler. Multiple input signals are summed up and transmitted to the output port. Graph (I-III) are the smallest irreducible graphs for a directional coupler with two inputs. Graph (I) has the smallest number of couplings, graph (II) and (III) have the smallest number of parametric drives. The other identified 97 further irreducible graphs (not shown) require either more resources or are gauge transformations of the shown graphs. Graph (I) can be generalized to more inputs (right panel). To realize a coupler with $N_\mathrm{in}$ inputs, one has to repeat the graph structure (bottom right, blue background) $N_\mathrm{in}$ times. Analytical terms discovered for the cooperativities and fluxes are indicated, the transmission turns out to be $t=1/\sqrt{N_\mathrm{in}}$. (c) Directional amplifier. Input from port 1 is amplified by a factor of $\sqrt{G}$, the other direction is blocked. The device has minimal back-action and the noise is quantum-limited. The graphs (I-IV) are the simplest solution without making use of intrinsic loss on the port modes. Allowing intrinsic losses, graph (V) is the simplest graph. Analytical terms discovered for the cooperativities $C_{ij}$, the  intrinsic loss rescaled by the outcoupling (red arrow), and the flux $\Phi_{132}$ are indicated. All shown graphs are stable. The unstable solutions are further discussed in the Supplemental Material.}
    \label{fig:results}
\end{figure*}

\section{A simple example: Isolator}\label{Section:Isolator}

As a first proof of concept, we apply {\sc AutoScatter} to implement an isolator, i.e., a two-port device with perfect transmission from the input to the output and zero reverse transmission, see \cref{fig:results}(a). 
Our method shows that the simplest solution has one auxiliary mode and a single complex-valued beamsplitter coupling. The solution  has the cooperativities $C_{12}=C_{23}=C_{31}=1$ and the synthetic flux $\Phi_{123}=\pi/2$. In this case, the unique irreducible solution had already been discovered with hand-crafted approaches as in \cite{koch_time-reversal-symmetry_2010,habraken_continuous_2012,ranzani_graph-based_2015,sliwa2015reconfigurable,lecocq_nonreciprocal_2017}.

Since our solutions are valid for arbitrary decay rates, they also cover the special limiting case in which the auxiliary modes are fastly decaying and act as Markovian baths, mediating dissipative interactions \cite{metelmann_nonreciprocal_2015}. In Supplemental Material C, we provide a mapping from the graph's weights to the Lindblad Master equation of a reservoir engineering scheme. Coming back to the specific isolator example, this mapping allows us to recover
the dissipative isolator discussed in \cite{metelmann_nonreciprocal_2015}. 

In addition to finding the simplest solution, any valid reducible graph provided by our method can be used as an ansatz by an experienced researcher to find the analytical expression for the corresponding manifold of solutions. The numerical data can provide further insights to refine the initial ansatz, e.g.~suggesting constraints in the graph parameters. As an example of this approach, we find the analytical expressions that describe (up to gauge transformations) all three-modes solutions of the isolator problem and generalize the solution presented in \cite{koch_time-reversal-symmetry_2010,habraken_continuous_2012,ranzani_graph-based_2015,sliwa2015reconfigurable,lecocq_nonreciprocal_2017, metelmann_nonreciprocal_2015}.
By using the empirically observed constraints that all dimensionless detunings and all cooperativities  assume the same values, we find a single analytical expression parametrized by the flux $\Phi_{123}$ , 
\begin{equation}
    \label{equ:general_circulator}
    C_{ij}=1 + 4\left(\frac{\Delta_i}{\kappa_i}\right)^2,\quad \frac{\Delta_i}{\kappa_i}=\frac{1}{2}\tan (\Phi_{123}-\pi/2),  
\end{equation}
for $j\neq i $, and $ i,j=1,2,3$.    We note for later discussion that after replacing the auxiliary mode with a third port mode,  the same solution above describes the most general three-mode circulator.

\section{Discovering generalizable design patterns}\label{Section:Fully_directional_coupler}

An important  goal of scientific discovery is to create insights that can be used to derive more general solutions.
A common example of such an insight is a design pattern discovered in solving a simple problem, but which can also be reused as a building block to solve more complex problems. 
Graphical representations naturally provide insights that can be exploited to create insights. For example, chemical graphs are used by the \textsc{Dendral} to interpret mass spectra \cite{lindsay1993dendral}. As well, the authors of \cite{krenn2016automated,krenn_conceptual_2021,ruiz-gonzalez_digital_2022,arlt2022digital} discovered a new graph-based scheme 
to create many-body entangled quantum states, which allows them to automatically discover graph solutions with a small number of particles and to generalize them to more particles.

In our setting, we demonstrate how our graph-based solutions can be generalized to realize devices  with an arbitrary number of ports. This is an example of how the fact that the weights of {\sc AutoScatter} solutions are typically unique or in a low-dimensional manifold, contributes to their interpretability. This appealing feature is achieved by focusing on irreducible graphs and by our rescaling in \cref{eq:rescaled_Bog_de_Gennes}. This is also the reason why we have been able to derive analytical expressions from most of the numerical solutions provided by {\sc AutoScatter}, see the following example and Sections V and VI. 

\subsection*{Example: Fully directional coupler}
As an example, we consider a fully directional coupler. This device combines signals from multiple input ports and transmits them with equal transmission amplitude to an output port. Here, we aim for a fully directional coupler with zero reflection and reverse-transmission amplitudes, see \cref{fig:results}(b). We leave the transmission amplitude $t$ as a free parameter. We aim to find a general solution valid for an arbitrary number of input ports $N_{\rm in}$.

First, we use {\sc AutoScatter} to find possible solutions for $N_{\rm in}=2$. It turns out that at least two auxiliary modes are required to realize a two-port device. In total, we find 100 different irreducible graphs that differ in the number and complexity of the underlying couplings. We select the graphs with the minimal number of couplings, Graph (I), and the graphs that require the smallest number of parametric pumps, Graph (II) and (III). For an explanation of how to count the number of pumps, see \cref{sec:implementation}. The remaining irreducible graphs are either gauge transformations of these graphs or require more resources in terms of couplings and parametric pumps. 

Inspecting Graph (I), we note that it can be viewed as the combination of two isolator graphs. This inspires a general ansatz for an $N_{\rm in}$-port coupler: a graph that comprises an isolator building block for each input port, see the right panel in \cref{fig:results}(b). We then use our continuous optimization routine to verify the validity of the ansatz and calculate the free parameters for the first few $N_{\rm in}$, up to $N_{\rm in}=5$ corresponding to $N=11$ nodes. For such large graphs, an exhaustive search would be daunting. 
From the solutions, we infer the weights for arbitrary $N_{\rm in}$: 
All cooperativities corresponding to the interaction between an input mode and its respective auxiliary mode take the same value,  $C_{\rm in, aux}=1$, the cooperativities corresponding to the coupling between the output mode and other modes take the value $C_{\rm out, in/aux}=1/N_{\rm in}$. The synthetic field fluxes are the same for all isolator-like loops, $\Phi_{\mathrm{in},\mathrm{out},\mathrm{aux}}=\pi/2$. Finally, the transmission turns out to be $t=1/\sqrt{N_\mathrm{in}}$, implying that no noise from the auxiliary modes is injected into the output port.

\section{Solving a continuous class of problems}
\label{Sec:Directional_amplifier}
Many interesting physical problems are parameterized by continuous parameters, e.g., the temperature or the value of an external field. Human researchers typically attempt to tackle these problems by deriving analytical solutions that are naturally interpretable. This becomes even more interesting when applied to optimization problems where the objective is continuous, e.g., when optimizing a variational quantum algorithm \cite{cerezo2021variational} to realize a continuous set of quantum gates. 

In our multi-mode circuit setting, one is often interested in a class of problems parameterized by one or more continuous parameters in the target scattering matrix. This type of challenge can be solved using our algorithm by first identifying a valid graph using a fixed set of values of the parameters. Then, we use the continuous optimization to create a dataset of solutions for varying parameter values (but fixed graph structure). Finally, we attempt to derive closed analytical formulas for the cooperativities and fluxes by symbolic regression. This scheme can also be used in similar optimization problems with continuous objectives as discussed above.

\subsection*{Example: Directional quantum-limited amplifier}
To illustrate this, consider several recent works \cite{metelmann_nonreciprocal_2015,ranzani_graph-based_2015,lecocq_nonreciprocal_2017,malz_quantum-limited_2018,sliwa2015reconfigurable,abdo2013directional,liu_fully_2023} that have proposed schemes to realize on-chip directional quantum-limited amplifiers using  multi-mode circuits to replace bulky state-of-the-art ferrite-based devices. The target device has arbitrary gain $G$ and is fully directional, i.~e., has zero back-reflection, and zero reverse-transmission. Moreover,  both at the input and the output ports the added noise should reach the so-called quantum limit, i.e., the fundamental limit set by the laws of quantum mechanics \cite{clerk2010introduction}. These requirements can be translated into constraints $f_j$ (see \cref{eq:loss}) on the scattering matrix $S$ that depend on the temperatures of the input and output ports (see \cref{App:noise_analysis} for more details). 
For both input and output ports at zero-temperature {\sc AutoScatter} recovers the solution used in \cite{ranzani_graph-based_2015, sliwa2015reconfigurable,metelmann_nonreciprocal_2015,lecocq_nonreciprocal_2017} (not shown). In this solution, the added noise in the output port is injected into the multi-mode system from the same port. For this reason, the device will not be quantum limited if the output port is hot.


An interesting open problem in this context is to find the simplest design for a fully directional amplifier that is still quantum-limited in the presence of a hot output port \cite{liu_fully_2023}, see \cref{fig:results}(c). Liu et.~al.~\cite{liu_fully_2023} addressed this problem with a handcrafted approach,  discovering two small architectures comprising two auxiliary modes and six couplings. However, applying {\sc AutoScatter} we discovered several architectures, see graphs (I-IV) in \cref{fig:results}(c), which require less couplings and can also be implemented with less parametric pumps (see \cref{sec:implementation}).

In our description in \cref{sec:coupled_mode_theory} we assumed for simplicity that the port modes' only loss channel is the out-coupling to some waveguide. In general, port modes can be subjected to further losses, e.g., an intrinsic loss channel, which is not monitored. Using the loss rates of these additional loss channels as degrees of freedom in the continuous optimisation (see \cref{app:scattering_matrix} for more details), {\sc AutoScatter} finds graph (V) in \cref{fig:results}(c). This solution comprises a single auxiliary mode, three couplings, and has an under-coupled output port, i.e., most of the signal is  dissipated into a (zero-temperature) loss channel. While this setup is not power-efficient, it fulfills all the imposed goals and represents an instructive counter-intuitive solution, since one could naively expect that the loss of a large part of the signal  would be incompatible with quantum-limited amplification. 

To discover the graphs in \cref{fig:results}(c), we run the optimization with a constant value for the gain factor $G$. In a second step, we use our continuous optimization to create a dataset of solutions for different values of $G$, while keeping the graph structure fixed. Using simple symbolic regression, we were able to derive closed analytical formulas for the cooperativities, and fluxes as well as the full scattering matrix for all discovered irreducible graphs, see Supplemental Material D and \cref{fig:results}(c). 


\section{Discovering asymptotic solutions}

In many physical systems, an ideal target behavior can be reached only in the special limiting case of one or more system parameters that tend to infinity or zero, e.g., a strong coupling limit. Even though this ideal limit is strictly unattainable in practice, adopting it promotes simplification and conceptual understanding. How can we introduce such fruitful asymptotic analysis in the context of artificial scientific discovery? One key complication is that often multiple parameters have to assume ideal limits simultaneously and that the way these limits are approached determines the desired target behavior.

\begin{figure}
    \centering
    \includegraphics[width=\linewidth]{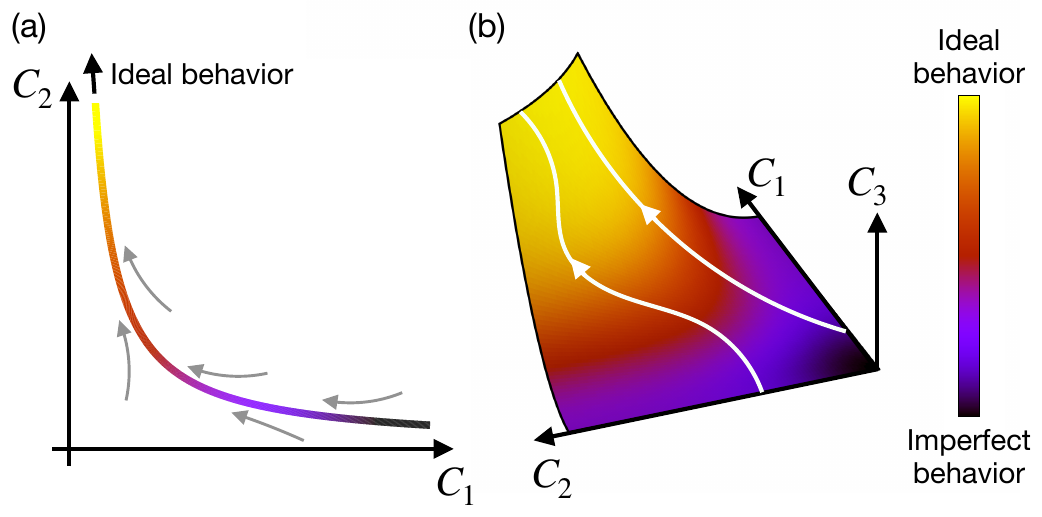}
    \caption{
    \textbf{Asymptotic solutions in automated discovery.} (a) Possible asymptotic behavior in a two-dimensional parameter space. The ideal behavior is achieved when $C_1$ tends to zero and $C_2$ to infinity while reaching asymptotically a curve. The color represents the loss on this curve, as indicated by the color bar.
    The gray arrows indicate possible paths in parameter space that a gradient-based optimization could take. In the driven lossy qubit example, the parameter $C_1$ corresponds to the gate time $T$, and $C_2$ to the drive amplitude $\lambda$.
    (b) Asymptotic behavior in higher-dimensional parameter spaces. In this case, the three parameters $C_1$, $C_2$, and $C_3$ should diverge while approaching a specific plane to achieve the ideal behavior. The loss on this plane is shown together with two, out of infinitely many, possible paths of approach to the ideal behavior. 
    }
    \label{fig:asymptotic}
\end{figure}

We first discuss a simple scenario featuring only two parameters. Consider the task of applying an ideal gate to a dissipative qubit by driving it with a pulse of appropriate amplitude $\lambda$ and duration $T$. During a gradient-based optimization run,  the gate time $T$  approaches the asymptotic limit $T\to 0$ to reduce dissipative effects. At the same time, the amplitude $\lambda$  goes to infinity to compensate for the shorter pulse. Finally, the asymptotic behavior, e.g., $\lambda\approx \pi/ T$ for a $\pi$ pulse, can be extrapolated from the data,  see \cref{fig:asymptotic}(a). In this approach, one can deduce both the scaling law $\lim_{T\to 0}\lambda T=\mathrm{const.}$ and the numerical value of the constant from the data. In quantum circuits consisting of multiple gates, one now has multiple amplitudes and gate times. Instead of optimizing those gate parameters directly, one can reformulate the optimization task assuming instantaneous pulses parametrized by the pulse area $\lambda T$. In this way, one builds the prior knowledge of the scaling law into the model and obtains the respective values for $\lambda T=\mathrm{const.}$ directly without requiring extrapolation. This instantaneous pulse concept can be reused as a building block and is naturally used to solve efficiently complex optimization tasks involving more than one qubit and many pulses of different types in a variational quantum circuit \cite{cerezo2021variational}.

\label{sec:OM_circulator}

Now let us consider a more general setting with three or more dimensionless parameters that reach an unattainable limit, see \cref{fig:asymptotic}(b). As for the qubit problem, we would like to reparametrize the problem in a way that allows us to take the asymptotic limit and subsequently perform our optimization without requiring extrapolation. For the reparametrization, we need the scaling laws of all parameters in terms of one of the parameters reaching the unattainable limit.  However, these scaling laws are a priori unknown. 
Moreover, in a less constrained problem, the parameters could approach the unattainable limit on trajectories with different asymptotic limits making it difficult to extrapolate the potentially infinite number of asymptotic solutions or even the scaling laws of the parameters, see \cref{fig:asymptotic}(b).

To address these general challenges, we propose an approach based on asymptotic building blocks. This concept can be applied to a broad range of systems that can be represented as weighted graphs. 
Just like the instantaneous pulse building block for the previously mentioned example of variational circuits,  our building blocks build into the graph model an assumption regarding the scaling laws of the parameters. Here, a building block is a node or group of nodes that can be added to the graph and is defined by the scaling laws for the corresponding weights.  
This approach allows us to combine different building blocks and graph topologies to test a variety of asymptotic ansatzes automatically.

To be more specific, we come back to the discovery of scattering setups. In our experience, an asymptotic solution is often required in the presence of experimental constraints that prevent the port modes from interacting directly \cite{hill_coherent_2012,andrews_bidirectional_2014,bernier2017nonreciprocal,malz_quantum-limited_2018}. This scenario is widespread for hybrid devices including transducers \cite{hill_coherent_2012,andrews_bidirectional_2014}, isolators \cite{bernier2017nonreciprocal}, and amplifiers \cite{malz_quantum-limited_2018}. One or more auxiliary bus modes (for example, in optomechanics mechanical modes) are needed to mediate the interactions between port modes. To achieve the ideal behavior,  the cooperativities $ C_{a_jb_k}$ of the couplings between the bus modes $\hat{b}_k$ and the port modes $\hat{a}_j$  and possibly also the dimensionless detunings of the bus modes $\sqrt{C_{b_kb_k}}$ must diverge. On the other hand, the ratios of cooperativities $C_{a_jb_k}/ C_{a_lb_k}$ or $ C_{b_kb_k}/ C^\alpha_{a_lb_k}$ (with suitable exponents $\alpha$, depending on the situation) tend to a constant.

\begin{figure*}
    \centering
    \includegraphics[width=0.8\linewidth]{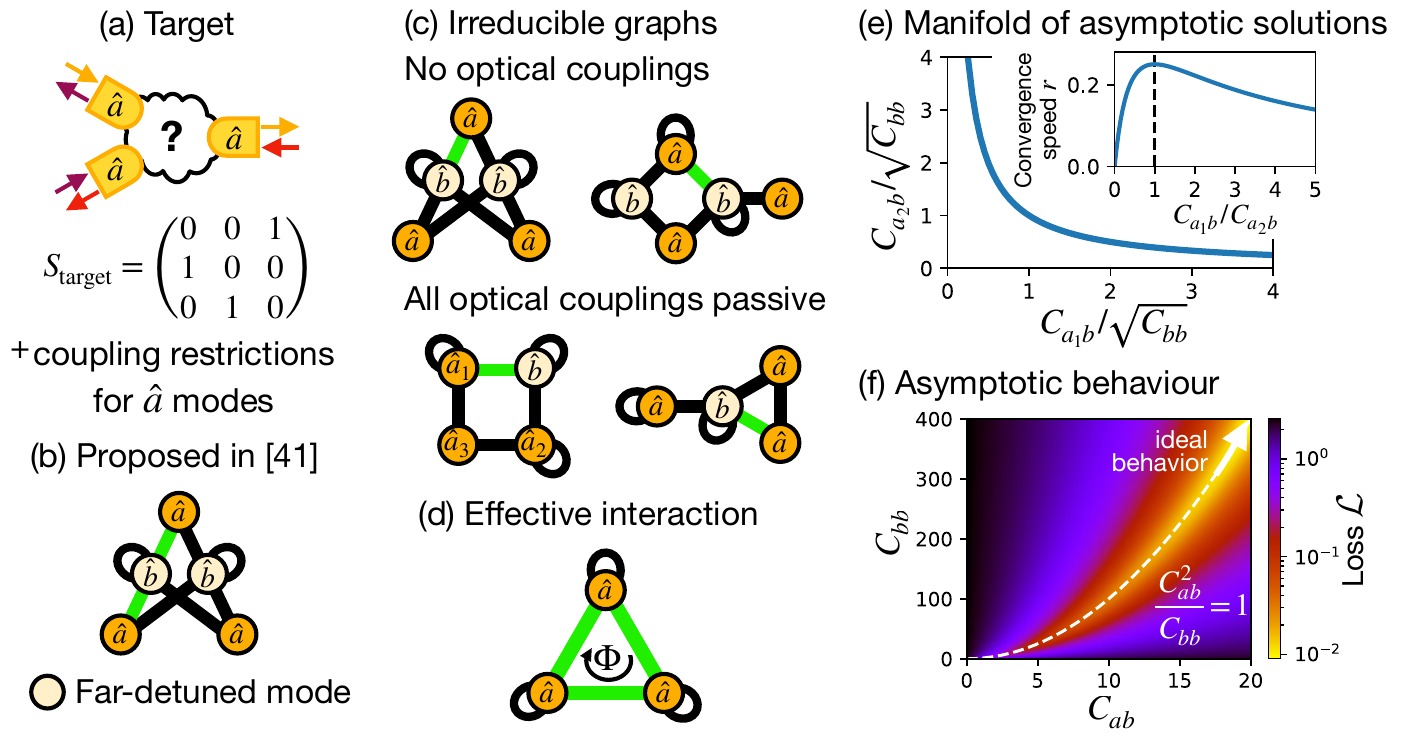}
    \caption{\textbf{Optomechanical circulator}. (a) Target operation: Realize a circulator between optical modes ($\hat{a}$) by mediating the interaction using mechanical modes ($\hat{b}$). 
    (b) Architecture proposed in \cite{bernier2017nonreciprocal} when no direct couplings are allowed between the optical modes. Instead, the interaction is mediated by far-detuned mechanical modes.
    (c) Irreducible graphs identified by {\sc AutoScatter} when no direct couplings are allowed between the optical modes (top) or when the couplings between two optical modes must be passive (bottom). The shown graphs have the smallest possible number of couplings. There exist more irreducible graphs which require more couplings (not shown).
    (d) Effective interaction between the optical modes after integrating out the far-detuned modes in any of our discovered graphs. This graph resembles a generalized form of the well known circulator scheme discussed in \cite{sliwa2015reconfigurable,lecocq_nonreciprocal_2017,habraken_continuous_2012,koch_time-reversal-symmetry_2010,ranzani_graph-based_2015}.
    Panels (e) and (f) refer to the bottom left graph in (c). (e) Manifold of asymptotic solutions. Every point on the curve $(C_{a_1b}/\sqrt{C_{bb}})(C_{a_2b}/\sqrt{C_{bb}}) = 1$ represents an asymptotic solution which approaches the ideal behavior for $C_{bb}\to\infty$. The inset shows the convergence speed $r$ as a function of the ratio $C_{a_1b}/C_{a_2b}$ of the asymptotic parameters $C_{a_1b}/\sqrt{C_{bb}}$, and $C_{a_2b}/\sqrt{C_{bb}}$.
    (f) Cross-section of the square deviation loss ${\cal L}$ between the scattering matrix $S$ and the target matrix $S_{\rm target}$. 
    We fix $C_{ab}\equiv C_{a_1b}=C_{a_2b}$.
     ${\cal L}\to 0$ in the asymptotic limit  $C_{ab}, C_{bb}\to\infty$ with $C_{ab}^2/C_{bb}=1$ (dashed white line).
    }
    \label{fig:optomechanical_circulator}
\end{figure*}

We can gain intuition on these asymptotic solutions and incorporate the respective scaling law as a building block in {\sc AutoScatter} by integrating out one of the modes ($\hat{b}$ or $\hat{a}$ depending on the situation). In this way, we map the original graph to a smaller effective graph whose parameters depend only on the appropriate ratios of cooperativities in the asymptotic limit. This allows us to take explicitly the asymptotic limit and later obtain the asymptotic ratios using the continuous optimization of {\sc AutoScatter}. For example, consider a set of far-detuned auxiliary bus modes $b_k$ as in \cite{bernier2017nonreciprocal}. The bus modes are only virtually occupied and, thus, mediate a purely coherent interaction between two or more port modes $a_j$. In this case, we can integrate them out to obtain an effective graph for the port modes $a_j$. In the new graph, the effective cooperatives $C^{\rm eff}_{jl}=C_{a_j b_k}C_{b_k a_l}/C_{b_kb_k}$ depend only on the asymptotic parameters $C_{a_j b_k}/\sqrt{C_{b_k b_k}}$, see Supplemental Material E. 

Currently, our approach is limited to using certain asymptotic building blocks, which have to be predefined by a human expert. The same expert-aided approach could be adopted in other artificial scientific discovery problems, e.g., featuring a varying number of masses or charges of very different sizes. An important goal for future research is to automatically identify such limits and automatically define such new concepts.

\subsection*{Example: Optomechanical circulator}

We apply this extended optimization scheme to discover the simplest setup implementing an optomechanical circulator in the experimental scenario considered in \cite{bernier2017nonreciprocal}. Three microwave port modes are coupled via the optomechanical interactions mediated by a varying number of mechanical modes. No direct coupling is available (see \cref{fig:optomechanical_circulator}(a)). Since each optomechanical coupling requires a different laser drive, the key measure of experimental complexity in this case is the overall number of couplings.  The hand-crafted architecture proposed in \cite{bernier2017nonreciprocal} requires two far-detuned mechanical modes and six couplings (see \cref{fig:optomechanical_circulator}(b)). Our optimization scheme finds a simpler setup, which requires the same number of far-detuned modes but only five couplings (see \cref{fig:optomechanical_circulator}(c)). If we apply the softer constraint where passive couplings are allowed between the optical modes, our scheme identifies even simpler graphs, only involving one far-detuned mode (see \cref{fig:optomechanical_circulator}(c)).

As we explained above, {\sc Autoscatter} translates the asymptotic graphs with far-detuned modes into an effective graph featuring only port modes. Our effective graphs improve the interpretability of the solutions. For example,  all valid asymptotic graphs in our optomechanical circulator  are translated into  the same effective  three-mode circulator graph shown in  \cref{fig:optomechanical_circulator}(d), see Supplemental Material E for details.  As we discussed in  \cref{Section:Isolator}, this graph encodes the most general three-modes solution for an ideal circulator/isolator, see \cref{equ:general_circulator}. We can leverage this solution to discover the manifold of asymptotic parameters corresponding to ideal  solutions for our discovered asymptotic graphs. As an example, consider the lower left graph in \cref{fig:optomechanical_circulator}(c). Here, all passive couplings between the optical modes would be allowed. However, they can not realize the required synthetic field flux (see \cref{sec:implementation}). 
This role is fulfilled by the mechanical mode, which mediates an effective complex-valued beamsplitter coupling between $\hat{a}_1$ and $\hat{a}_2$. We empirically find that this architecture has only solutions where the effective detuning is zero. Therefore, the effective cooperativity $C_{a_1 a_2}^\mathrm{eff}$ should fulfill $C_{a_1{a_2}}^\mathrm{eff} = 1$, see Eq.~(\ref{equ:general_circulator}). Taking into account that $C_{a_1 a_2}^\mathrm{eff}=C_{a_1b}C_{a_2b}/C_{bb}$, this constraint identifies a manifold of ideal solutions for the asymptotic parameters  $C_{a_1b}/\sqrt{C_{bb}}$ and $C_{a_2b}/\sqrt{C_{bb}}$, 
as visualized in \cref{fig:optomechanical_circulator}(e).    

In realistic experimental scenarios, the coupling strength and loss rates have upper and lower bounds limiting how close one can approach the asymptotic limits, see \cref{fig:optomechanical_circulator}(f). For example, for the lower left graph in \cref{fig:optomechanical_circulator}(c), the loss function $\cal{L}$ scales like ${\cal L}\sim r / C_{bb}$ in the asymptotic limit. Here, $r$ is a constant that can be interpreted as a convergence speed.
Even though in the asymptotic limit, all solutions are ideal, they differ in the convergence speed. 
In our example, when we send the dimensionless detuning $C_{bb}$ to infinity, the convergence speed $r$ depends on the ratio of the cooperativities, see inset in \cref{fig:optomechanical_circulator}(e). For practical applications, similar considerations can be used to identify a preferred path to approach the asymptotic limit.

\section{Analysis of the search space reduction}
\label{sec:search_space_reduction}

\begin{figure}
    \centering
    \includegraphics[width=\linewidth]{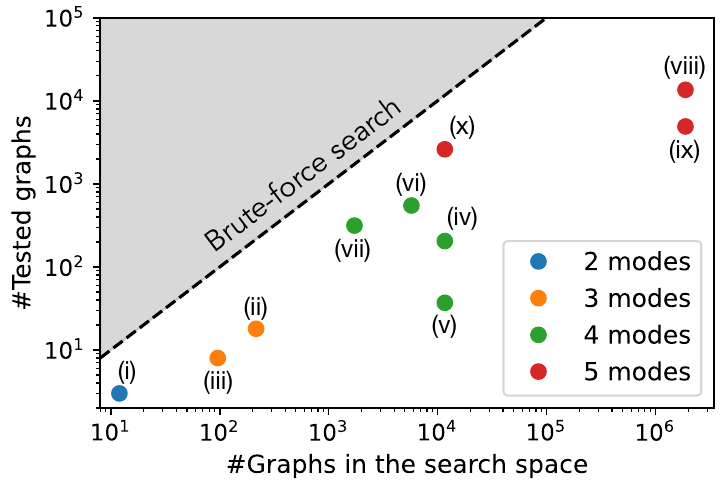}
    \caption{\textbf{Search space reduction.} We compare the number of graphs tested by \textsc{AutoScatter} to the total number of graphs within the search space. A brute-force search would have to test every graph (dashed black line). Each data point corresponds to a different target functionality/device: (i) Gyrator, (ii) Isolator$^\ast$, (iii) Amplifier with intrinsic port losses$^\ast$, (iv) Three-port gyrator, (v) Wilkinson power divider, (vi) Amplifier without intrinsic port losses$^\ast$, (vii) Circulator with only passive couplings between optical modes$^\ast$, (viii) Directional coupler$^\ast$, (ix) Resistive power divider, and (x) Circulator without couplings between optical modes$^\ast$. All examples marked with an asterisk are discussed in the \cref{Section:Isolator,Section:Fully_directional_coupler,Sec:Directional_amplifier,sec:OM_circulator}. See Supplemental Material G for a summary of all further examples.}
    \label{fig:search_space_reduction}
\end{figure}


The main achievement of our work is the automated discovery of interpretable and transferable coupled mode setups. In addition, we have introduced an efficient discrete optimization algorithm that allows us to reduce the search space compared to a brute-force search. An interesting open question is how large the reduction in the search space will be.  
Considering that our algorithm can be directly applied to solve a large class of problems whose potential solutions can be encoded in graphs with continuous weights,  this question is of interest beyond our setting of coupled mode setups.

In \cref{fig:search_space_reduction}, we show the number of graphs tested by {\sc AutoScatter} in comparison to the total number of graphs within the search space. Each data point corresponds to a different target device,  as identified by the target scattering matrix. 


As we can see, the performance gain in comparison to a brute-force approach was highly significant in some of the most complex problems. If we quantify the performance gain as the ratio of the number of graphs in the search space over the number of tested graphs, it varied in a broad range from $\sim 5$ (see, e.g., (vii)) to $\sim 400$ in our experiments (see (ix)). As one might intuitively expect, the largest gains are obtained for complex devices featuring four or five modes. This is because the brute-force search space grows exponentially with the squared number of modes in the device. It is not possible to predict in general how the performance gain evolves with mode number, since that depends on the target device.
In Supplemental Material F, we provide more details on our analysis and discuss how the search space structures of the individual target devices affect the respective performance gain. Furthermore, we provide in Supplemental Material G a summary of all target devices included in \cref{fig:search_space_reduction}. With the help of the substantial performance gains displayed here, and for the mode numbers relevant for the variety of practically important target devices, we found our approach to be able to provide the complete enumeration of solutions that is one of its key advantages. It should be noted that eventually, for even further increased mode numbers, one would have to resort to a combination of our search-space reduction with heuristic approaches.

\section{Implementation in specific experimental platforms}
\label{sec:implementation}




In this section, we review the main steps to design a scattering device in a specific experimental platform. This section has two main motivations. The first motivation is to demonstrate how the ideal solutions discovered using {\sc AutoScatter} can be implemented in a wide range of experimental platforms including purely optical platforms,  microwave circuits, optomechanics and other hybrid platforms, covering a wide range of frequencies from \SI{}{\mega \hertz} to hundreds of \SI{}{\tera \hertz}. 
In this framework, we show how to exploit the flexibility of our solutions to optimise secondary objectives, such as the device's bandwidth or the number of active couplings.
The second motivation to view our method in the broader context of the full experimental design pipeline is to highlight how it stands out as the only design step that is dependent on human ingenuity and leads to interpretable and transferable findings. 



Typically, the first concrete choice in view of designing a scattering device is the type of port modes. This choice will depend on the functionality of the target device, e.g., for a microwave to optics converter, one needs a microwave and an optical mode with resonance frequency close to the respective target carrier frequencies. 

Pair of modes of the same type and with similar resonance frequency can naturally be coupled via a passive beamsplitter interaction \cite{kwende2023josephson,fang_generalized_2017},  
e.g., the evanescent coupling between two defect modes in a photonic crystal \cite{fang_generalized_2017}. 
By itself, even a large difference between the resonance frequencies of two modes does not pose a fundamental constraint to their available linearized couplings because signals can be up and down converted by pumping a weak nonlinearity with a coherent drive. The specific coupling scheme depends on the experimental platform. In a purely optical setting, one can exploit the optical nonlinearities in the material, i.e., the Kerr or the $\chi^{(2)}$ nonlinearity. Mechanical modes and optical modes can be coupled via radiation pressure. 
In superconducting circuits, the coupling of microwave modes of different frequencies is based on the  Josephson nonlinearity, driving the circuit with an AC current or a time-dependent magnetic flux piercing a Superconducting Quantum Interference Device (SQUID) \cite{SQUID_handbook}.
While active couplings are widely available, some constraints to the quadratic Hamiltonian  \cref{eq:second_quantized_Hamiltonian} and, thus, our graphs connectivity might be present in specific experimental settings. For example,  coupling a circuit microwave mode to an optical cavity mode typically requires a mechanical auxiliary mode \cite{andrews_bidirectional_2014}.

The next and most important step in designing a new scattering experiment is to find a suitable RWA Hamiltonian that enforces the desired scattering behavior, taking into account possible hardware-specific constraints on the couplings. 
This step typically requires human ingenuity and is the central focus of many experimental and theoretical works \cite{peterson_demonstration_2017,kwende2023josephson,hill_coherent_2012,fang_generalized_2017,ranzani_graph-based_2015,naaman_synthesis_2022,metelmann_nonreciprocal_2015,malz_quantum-limited_2018,liu_fully_2023,bernier2017nonreciprocal}.
Our method fully automates this task finding all possible solutions and representing them using our graph language (see \cref{fig:model}(a)). 

At the level of the RWA, the solutions are highly transferable between hardware platforms as one needs only to commit to a certain number of modes, coupling types, and the underlying cooperativites and synthetic field fluxes, as we have shown in \cref{app:scattering_matrix}. Since they do not depend on the resonance frequency of the modes, they leave open the choice of whether to implement certain couplings as active or passive couplings. For example, the isolator graph shown in \cref{fig:results}(a) can be implemented fully actively, requiring three parametric pumps, as in \cite{sliwa2015reconfigurable}. Alternatively, it can be implemented using two parametric pumps and a passive coupling as in \cite{kwende2023josephson}. 

If one is willing to employ active couplings for every link, any desired graph can certainly be implemented.
This approach, combining several pump tones at different frequencies, is common in different platforms \cite{malz_quantum-limited_2018,bernier2017nonreciprocal,peterson_demonstration_2017,andrews_bidirectional_2014,abdo2013directional,lecocq_nonreciprocal_2017,kamal_noiseless_2011,sliwa2015reconfigurable,herrmann2022mirror,macklin2015near}. 
However, implementing active couplings that rely on parametric driving may not be the most economical design in many scenarios. For example, in the optical regime, it is more challenging to combine many pump tones at different frequencies in this higher frequency range. For this reason, it is valuable to identify those setups that require the minimum amount of parametric pumps.

\begin{figure}
    \centering
    \includegraphics[width=\linewidth]{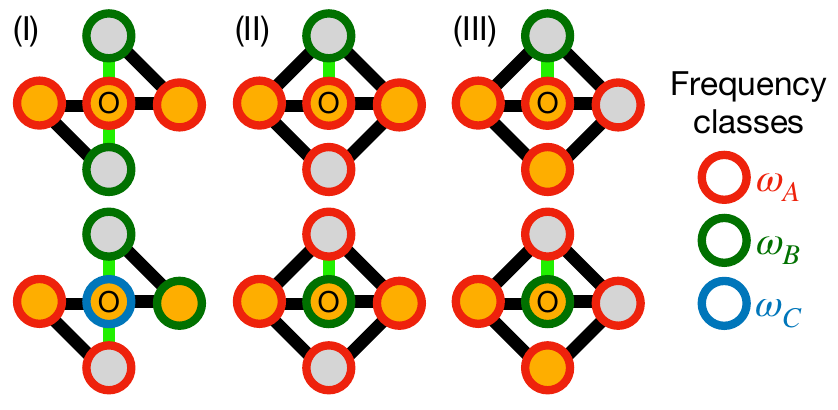}
    \caption{
    \textbf{Minimization of the number of active couplings.}
    The graphs are annotated with the carrier frequency classes that minimize the number of active couplings for the directional coupler graphs from \cref{fig:results}(b).
    Edges that connect two modes belonging to the same frequency class are coupled passively. The remaining couplings are active.
    A sorting is correct when it fulfills the constraint that complex-valued beamsplitter couplings (green) and squeezing (blue) interactions have to connect modes of different frequency classes. We find all possible sortings by using a brute-force search, see \cref{app:graph_coloring} for more details. The shown sortings minimize the number of active couplings. Graph (I) has six more such solutions (not shown).}
    \label{fig:frequency_labeling}
\end{figure}

In many experimental platforms, one needs at least one active coupling to engineer a synthetic gauge flux \cite{malz_quantum-limited_2018,bernier2017nonreciprocal,peterson_demonstration_2017,abdo2013directional,lecocq_nonreciprocal_2017,kamal_noiseless_2011,sliwa2015reconfigurable,herrmann2022mirror}
,  $\Phi_{i,j,l,\ldots,k}=\arg (g_{ij}g_{jl}\ldots g_{ki})\neq 0$. A notable exception are setups featuring passive coupling between chiral modes \cite{hafezi_robust_2011,ruesink2016nonreciprocity}, see Supplemental Material B.  Here, we exclude such exceptions and assume that all loops formed exclusively by passive couplings have zero flux. In this scenario, we can always choose a gauge in which all passive couplings are real-valued. 
However, only beamsplitter interactions coupling modes with the same carrier frequency can be realised passively, and this could be forbidden by other constraints in the graph. In particular, all complex-valued beamsplitter couplings and squeezing couplings are active and, thus, connect modes of different frequencies.

To determine the minimum number of active couplings, one has to solve a sorting problem, in which one sorts modes into frequency classes. If two modes are in the same class, they have the same carrier frequency, otherwise, they do not. 
The goal is to minimize the number of edges connecting modes of different classes under the aforementioned constraints. 
We solve this problem with a brute-force search, see \cref{app:graph_coloring} for more details. This search yields the minimum number of pumps as well as all possible combinations of passively coupled modes that allow to reach the minimum. As an example, we analyze the number of active pumps for 
the fully directional coupler of \cref{fig:results}(b). Our analysis shows that Graph (I) requires at least four parametric pumps, while graph (II) and (III) only require three pumps. 
\cref{fig:frequency_labeling} displays two different optimal solutions for each graph. Graphs (II) and (III) have only these two solutions, while graph (I) has eight in total. 

\begin{figure}
    \centering
    \includegraphics[width=\linewidth]{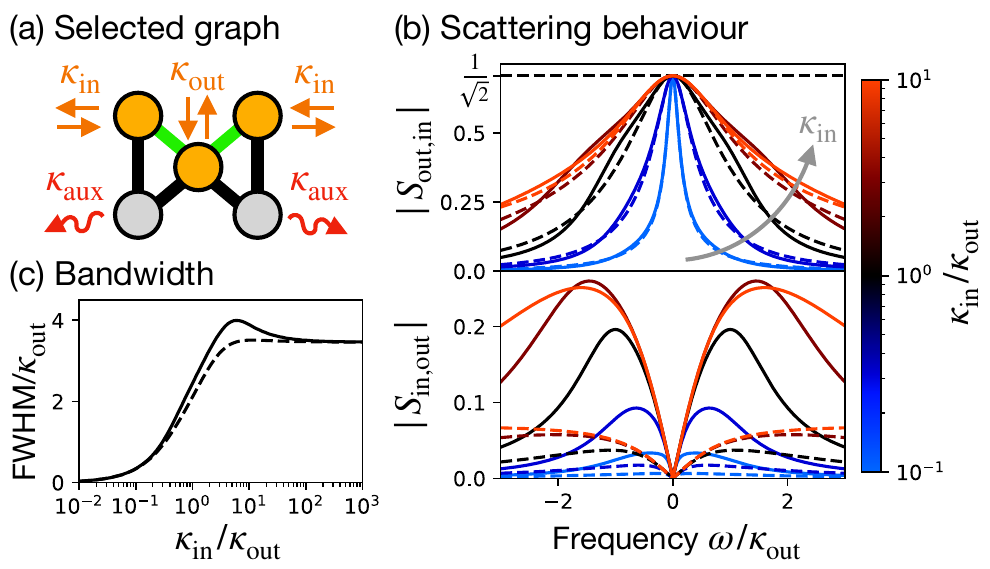}
    \caption{\textbf{Frequency-dependent scattering response.} We 
    show the scattering response of a directional coupler (Graph (1) in \cref{fig:results}(b)) for different values  of the loss rates.
    (a) Graph representation  labeled with  the out-couplings (orange arrows) and intrinsic losses (red arrows). 
    (b) Transmission from an input to the output port (top) and reverse transmission (bottom) as a function of the frequency (counted off from the carrier frequency) for different outcoupling ratios  $\kappa_\mathrm{in}/\kappa_\mathrm{out}$ (see colorbar), and $\kappa_\mathrm{aux}/\kappa_\mathrm{out}=2$ (solid lines) and $10$ (dashed lines). 
    The ideal transmission is  $\abs{S_\mathrm{out,in}}=1/\sqrt{2}$ (dashed black line) while the ideal reverse transmission $S_\mathrm{in,out}$ is  zero. 
    Our approach optimised the system, such that the coupler always displays the ideal behaviour at the carrier frequency (for $\omega=0$).  
    (c) Bandwidth  as function of $\kappa_\mathrm{in}$ for $\kappa_\mathrm{aux}/\kappa_\mathrm{out}$ as in (b). We define the bandwidth as the full width at half maximum (FWHM) of the transmission $\abs{S_\mathrm{out,in}}$.
    }
    \label{fig:frequency_dependency}
\end{figure}

Even after choosing a graph and deciding which couplings are active and passive, our solutions still leave the freedom to choose the modes' decay rates.
We recall that the weights $H_{ij}$ of our graphs are proportional to square roots of cooperativities,  $H_{ij}=g_{ij}/\sqrt{\kappa_i \kappa_j}$ or $H_{ij}=\nu_{ij}/\sqrt{\kappa_i \kappa_j}$ cf. \cref{eq:rescaled_Bog_de_Gennes}.
Independent of the concrete values for the modes' decay rates $\kappa_i$, the scattering amplitudes at the carrier frequencies do not change as long as the coupling rates $g_{ij}$ and $\nu_{ij}$ are rescaled accordingly. At the same time, the outcouplings and other loss rates can typically be tuned to a certain degree in experiments. This freedom can be used to change the frequency-dependence of the scattering process, as illustrated 
in \cref{fig:frequency_dependency} for one of our directional couplers. By increasing the outcoupling rate of the input ports, one can increase the bandwidth of the transmission from the input to the output. With large losses on the auxiliary modes, one can suppress the reverse transmission over a broad frequency range.

Beyond the design choices discussed above, any hardware platform possesses its own hardware-specific challenges.
The device should be engineered to host the desired high-quality-factor modes with the target passive quadratic couplings and strong enough high-order nonlinear interactions. This task requires domain-specific simulations of, e.g., the Maxwell equations or the elasticity equations. We note in passing that this task typically requires human ingenuity and, for this reason,  has been the focus of intense efforts aimed at automatizing it, see e.g.~\cite{chan2012optimized,asano2018optimization,asano2019iterative}. 
However, its platform-specific character makes any finding less transferable and generalizable. For this reason, it is outside of the scope of our work.

As a final step, the dynamics of the multimode system can be modelled using a set of coupled Langevin equations featuring also the nonlinear interactions, with bare couplings calculated from the domain-specific simulations. One can straightforwardly use this more general description to find the driving protocol to implement the target active couplings as detailed below.
The  pump frequencies $\omega_{ij}^P$ are chosen to up and down-convert the signals to the target carrier frequencies $\omega_{L,j}$, 
e.g.~for a nonlinearity based on three-wave mixing
$\omega_{ij}^P=\abs{\omega_{L,i} \pm \omega_{L,j}}$ ($+$ for squeezing, $-$ for beamsplitter couplings). Then, the steady-state amplitudes of the modes (including also pump modes that do not directly appear in the RWA description) are calculated from a set of nonlinear equations derived from the  Langevin equations. 
By linearizing the Langevin equations around this steady state, one can express the quadratic coupling $g_{ij}$ and $\nu_{ij}$ in terms of the bare nonlinear couplings and the steady amplitudes and calculate target values for the latter quantities. In turn, the steady amplitudes can be adjusted by varying the powers and phases of the pumps. The appropriate values of the powers and phases can be estimated from the nonlinear equations for the steady state and can also be fine-tuned directly in the experiments.  For most scattering devices, it is possible to achieve the ideal behaviour without taking any asymptotic limit. In this scenario, the cooperativities are typically of order $1$. Asymptotic solutions require large cooperativities. Such cooperativities can be reached in a variety of different hardware platforms featuring optical, microwave, and mechanical modes.

\section{Conclusion}


Our automated discovery algorithm {\sc AutoScatter} finds ideal high-level designs for real-world scattering devices like transducers, amplifiers,
isolators, or circulators. It provides interpretable and transferable solutions thanks to its graph representation, which encodes in the same set of interpretable weights a whole class of setups, unified by their scattering behavior but with different physical parameters.  The flexibility in translating each solution back  into physical parameters, allows the user to implement them in a range of state-of-the-art experimental platforms and transfer them to different dissipative regimes,  covering both Hamiltonian and reservoir engineering schemes.  {\sc AutoScatter} can discover types of solutions that are normally difficult to obtain in an automated approach, such as asymptotic solutions.
Through a series of examples, we have shown  that the  solutions 
generated by {\sc AutoScatter}  lead to new insights, in the form of counterintuitive design principles, generalizable design patterns, and analytical solutions to continuous classes of problems. 


{\sc AutoScatter} is amenable to numerous future extensions, such as searching for optimized implementations in specific hardware platforms, neural-network-guided heuristics for larger devices where complete enumeration is infeasible, or targeting specific quantum noise characteristics.
More generally, our results now open the door towards automated discovery in additional related domains of importance, like photonic  transport through periodic structures composed of small modular blocks or scattering-based sensing experiments, as well as transfer to areas like electronic scattering transport. Nonlinear setups or even devices with functionalities based on time-dependent control require significant new developments but still could be treated with the same graph-based efficient search algorithm. Overall, the results shown here validate the fruitfulness of the main ideas guiding the field of artificial scientific discovery.

\section*{Reproducibility}

We open-source all code used to produce the results of this manuscript in the GitHub
repository \textsc{AutoScatter} \cite{github_autoscattering}.

\begin{acknowledgments}

We acknowledge fruitful discussions with Mario Krenn, Clara Wanjura, and Flore Kunst.

All authors contributed to the ideas, their implementation, and the writing of the manuscript. The implementation of the discovery algorithm and the numerics were performed by J.~L.

\end{acknowledgments}

\appendix

\section{Details of the discrete optimisation}
\label{app:discrete_optimization}

During the optimisation we keep two libraries, one for valid, and one for invalid graphs. At the beginning of the discrete optimization, our method tests only fully connected graphs for validity using the continuous optimisation starting from the graph with zero auxiliary modes. It increments the number of auxiliary modes until it finds the minimum number of auxiliary modes required to achieve the target behavior. This is the smallest fully connected graph that satisfies the target behavior. From there, we investigate which couplings or coupling phases can be removed. This search space corresponds to a directed acyclic graph (see visualization in \cref{fig:method}(b)), where the previously identified fully connected graph is the root node. To search through this space, we perform a breadth-first search. This means that our search always explores all nodes in the search space at a the current depth before proceeding on to the next level starting from the root node.

In each iteration, we first check if the selected graph (the next node in the search space) is already in one of the two libraries. If so, the graph is skipped. Otherwise, it is tested by the continuous optimization. 
If no solution is found, the tested graph and all its subgraphs are added to the library of invalid graphs. Otherwise, the graph and all its extensions are added to the library of valid graphs. If some of the discovered parameters turn out to be zero, the corresponding simpler graph and all its extensions are added instead. The optimization is finished when all possible graphs have been sorted into one of the two libraries.


\section{Details of the continuous optimisation}
\label{app:details_continuous_optimization}
For the minimization of the loss function in \cref{eq:loss}, we use the Broyden–Fletcher–Goldfarb–Shanno algorithm \cite{BFGS_reference}, an iterative gradient-descent method for solving non-linear optimization problems. A valid solution is obtained whenever the loss reaches zero. The algorithm fails to find a solution if the loss gets stuck in a non-zero minimum.  Since the optimization landscape supports local minima, there is a risk of false-negative errors. To reduce this risk, we repeat the continuous optimization $N_{\rm rep}$ times. The hyperparameter $N_{\rm rep}$ is chosen empirically with a default value $N_{\rm rep}=10$.

During the discrete optimisation we assume that there always exist a valid graph. However, in our work we identified two occasions, where the loss function cannot be minimized to zero for any graph:
(i) The desired target scattering behaviour violates fundamental physics. For example, if you ask the noise to adopt a value below the quantum limit, no valid graph can be found. (ii) The desired scattering behaviour can only be realised in a limiting case not covered by our optimisation. If you ask our optimisation scheme to search for the circulator behaviour discussed in \cref{fig:optomechanical_circulator} without allowing for far-detuned modes, our continuous optimization will label any graph as invalid.

Please note, that we have assumed in the main text that each port mode is coupled in reflection to a single waveguide. In the Supplemental Material we discuss how some alternative coupling schemes can be translated to our setups and how one can extend our approach to further coupling schemes, covering also setups including running-waves modes that are not invariant under time-reversal, as in \cite{Hafezi_Optomechanically_2012}. 

\section{Bogoliubov-de Gennes Hamiltonian}\label{Appendix:Bog_de_Gennes}
We  consider a hybrid system comprising $N$ modes of different types and the most general quadratic time-independent Hamiltonian
\begin{equation}\label{eq:gen_Ham}
    \hat{H} = \frac{1}{2}\sum_{j,k=1}^N \left(g_{jk} \hat{a}_j^\dag \hat{a}_k + \nu_{jk} \hat{a}_j^\dag \hat{a}_k^\dag \right)+ \mathrm{H.c.}
\end{equation}
Here, $\hat{a}_j$ and $\hat{a}_j^\dagger$ are the ladder operators of mode $j$, $g$ is a Hermitian  matrix, and $\nu$ can be chosen to be symmetric. For a more compact notation, it is convenient to group all the modes in a $2N$ dimensional vector,  $\hat{\boldsymbol{\xi}}=\{\hat{a}_1,\ldots,\hat{a}_{N},\hat{a}_{1}^{\dagger},\ldots,\hat{a}_{N}^{\dagger}\}$, and to introduce the first-quantized Bogoliubov de Gennes Hamiltonian
\begin{equation}\label{eq:BdG_dimensional}
	\hat{H}=\frac{1}{2}\hat{\boldsymbol{\xi}}^\dagger H_{\rm BdG}\hat{\boldsymbol{\xi}},\quad H_{\rm BdG}=
 \begin{pmatrix}
 g&\nu\\
 \nu^{*}&g^{*}\\
 \end{pmatrix}.
\end{equation}

\section{Rotating Frame}
\label{app:rot_wave}
The Hamiltonian in \cref{eq:gen_Ham} is defined in a frame in which the phase space of each mode $j$ rotates at the corresponding carrier frequency $\omega_{L,j}$. Thus,  $-g_{jj}$ is to be interpreted  as the detuning  $\Delta_j$  of the carrier frequency $\omega_{L,j}$ from the mode resonance $\omega_j$, $\Delta_j=\omega_{L,j}-\omega_j$. For multi-mode circuits, such time-independent Hamiltonian is typically referred to as the rotating-wave Hamiltonian as it is obtained after  a Rotating-Wave Approximation (RWA) that drops the fast-oscillating terms, which result from non-resonant interactions. The corrections to the RWA approximation are  small if all resonant frequencies $\omega_j$ are much larger than the coupling rates $g_{ij}$ and $\nu_{ij}$ as well as the decay rates $\kappa_j$. 

We note that if all couplings are active the carrier frequencies $\omega_{L,j}$ for the different modes can be chosen independently of each other, adjusting the frequencies $\omega_{ij}^P$ of the pump mediating the interaction such that $\omega_{ij}^P=\abs{\omega_{L,i} \pm \omega_{L,j}}$ ($+$ for squeezing, $-$ for beamsplitter couplings). Therefore, in experimental scenarios in which the mode resonances $\omega_j$ are not easily tunable, one can choose the carrier frequencies to implement the detunings $\Delta_j$ required in the scheme. On the other hand, two modes whose coupling is implemented passively (without a parametric pump) have the same carrier frequency and, thus, one of the resonance frequencies together with the common carrier frequency have to be tuned to implement the target detunings. 

Besides these constraints, the mode resonances $\omega_j$ are irrelevant in the high-level description based on the RWA. This  is an appealing feature that makes any insights gained using our method highly  transferable across a variety of platforms.  A more refined description including non-resonant interactions is typically used to quantify how a specific implementation deviates from the ideal target behavior. This deviation is device-specific and, thus, goes beyond the scope of our work.

\section{Langevin equations}
We use input-output theory to describe the dynamics of the open multi-mode circuit. 
As described in the main text, we distinguish between port modes, which are used as input and output ports for the signals, and auxiliary ports, which are not used as input ports and whose  output is not monitored. Accordingly, all the losses in an auxiliary mode can be incorporated into a single loss channel. On the other hand, whenever we want to incorporate intrinsic losses in the port modes we have to introduce two loss channels for these modes, an out-coupling and an intrinsic loss channel.  Based on these considerations, we arrive at the Langevin equations
\begin{align}
    \label{equ:equation_of_motion}
    \begin{split}
        \dot{\hat{a}}_j(t) = &-\mathrm{i} \sum_k g_{jk} \hat{a}_k(t) - \mathrm{i} \sum_k \nu_{jk} \hat{a}_k^\dag (t) \\
        &- \frac{\kappa_j+\Gamma_j}{2} \hat{a}_j(t) - \sqrt{\kappa}_j \hat{a}_j^\mathrm{in} (t) - \sqrt{\Gamma}_j \hat{a}_j^\mathrm{noise} (t)
    \end{split}
\end{align}
Here, $\hat{a}_j^\mathrm{in}$ is the input field for the main decay channel (for the port modes the out-coupling channel), with decay rate $\kappa_j$. Likewise, $\hat{a}_j^\mathrm{noise}$ is the noise entering port mode $j$ from its intrinsic loss channel, which has decay rate $\Gamma_j$.
The Langevin equations can be rewritten in a compact form as
\begin{align}
    \label{equ:compact_equation_of_motion}
    \begin{split}
        \dot{\hat{\boldsymbol{\xi}}}(t) = &\left(-i\sigma_z H_{\rm BdG}  - \frac{\kappa+\Gamma}{2}\right)\hat{\boldsymbol{\xi}}(t) \\
        &- \sqrt{\kappa} \hat{\boldsymbol{\xi}}^\mathrm{in} (t) - \sqrt{\Gamma} \hat{\boldsymbol{\xi}}^\mathrm{noise} (t) 
    \end{split}
\end{align}
Here, we have grouped all input and noise fields in  the  vectors, 
$\hat{\boldsymbol{\xi}}^{\rm in}=\{\hat{a}_1^{\rm in},\ldots,\hat{a}_{N}^{\rm in},\hat{a}_{1}^{{\rm in}\dagger},\ldots,\hat{a}_{N}^{{\rm in}\dagger}\}$
and $\hat{\boldsymbol{\xi}}^{\rm noise}=\{\hat{a}_1^{\rm noise},\ldots,\hat{a}_{N}^{\rm noise},\hat{a}_{1}^{{\rm noise}\dagger},\ldots,\hat{a}_{N}^{{\rm noise}\dagger}\}$, respectively. Moreover, we have introduced the diagonal matrices $\kappa={\rm diag}(\kappa_1,\ldots,\kappa_N,\kappa_1,\ldots,\kappa_N)$, $\Gamma={\rm diag}(\Gamma_1,\ldots,\Gamma_N,\Gamma_1,\ldots,\Gamma_N)$ and 
\begin{align}
	\sigma_z=
 \begin{pmatrix}
 \mathds{1}_N&0\\
 0& -\mathds{1}_N\\
 \end{pmatrix}.
\end{align}
We note that the Langevin equations have the embedded particle-hole symmetry $\sigma_x{\cal K}$ where ${\cal K}$ denotes the complex conjugation and 
\begin{align}
	\sigma_x=
 \begin{pmatrix}
 0&\mathds{1}_N\\
 \mathds{1}_N&0\\
 \end{pmatrix}.
\end{align}

This simply reflect that the last $N$ equations are the adjoint of the first $N$ equations.

\section{Scattering matrix}
\label{app:scattering_matrix}
To fully characterize the linear response of a multi-mode circuit, one has to take into account  that it depends on the frequency of the input field via the time-derivative in the left-hand side of \cref{equ:compact_equation_of_motion}. However, the ideal scattering behavior is only realized in a high-level description that focuses on signals that have a very smooth envelope and, thus,  are spectrally well localized (in each port about the respective carrier frequency $\omega^s_j$).  Taking into account that in the rotating frame the frequency is counted off from the respective carrier frequency $\omega^s_j$, the response of interest is the ``zero-frequency" response obtained by setting  $\dot{\hat{\boldsymbol{\xi}}}(t)=0$  in \cref{equ:compact_equation_of_motion}. After calculating $\hat{\boldsymbol{\xi}}$ in this way and  plugging it into the input-output relations 
\begin{equation}
    \label{equ:boundary_condition}
    \hat{\boldsymbol{\xi}}^\mathrm{out}=\hat{\boldsymbol{\xi}}^\mathrm{in} + \sqrt{\kappa}\hat{\boldsymbol{\xi}},
\end{equation}
we find
\begin{align}
    \label{equ:full_scattering}
    \hat{\boldsymbol{\xi}}^\mathrm{out} = S \hat{\boldsymbol{\xi}}^\mathrm{in} + \mathcal{N} \hat{\boldsymbol{\xi}}^\mathrm{noise}
\end{align}
with 
\begin{subequations}
	\begin{align}\label{eq:S_matrix_of_h}
		S &= \mathds{1}_{2N} +  \left( -\mathrm{i}\sigma_z H-\frac{\gamma}{2} -\frac{\mathds{1}_{2N}}{2}\right)^{-1} \\
  \label{eq:N_matrix_of_h}
		\mathcal{N} &= (S-\mathds{1}_{2N})\sqrt{\gamma}.
	\end{align}
\end{subequations}
Here, $S$ is the scattering matrix and the matrix $\mathcal{N}$ describes the linear response to the fluctuations entering the circuit from the intrinsic loss channels. In view of discovering  classes of solutions displaying the same scattering behavior, we found that it is possible to eliminate the explicit dependence of $S$ and $\mathcal{N}$ on the decay rates $\kappa_i$ by appropriately rescaling the Bogoliubov de Gennes Hamiltonian and the intrinsic losses by introducing the rescaled  Bogoliubov Hamiltonian $H$ as:
\begin{equation}\label{eq:Bog_rescaled_sup}
 H =\frac{1}{\sqrt{\kappa}}  H_{\rm BdG}\frac{1}{\sqrt{\kappa}} ,\quad \gamma=\Gamma \kappa^{-1}.
\end{equation}
We note that $H$ is parametrized by $N$ rescaled detunings $H_{ii}=-\Delta_i/\kappa_i$ with $1\leq i\leq N$, $N(N-1)/2$ dimensionless couplings  $g_{ij}/\sqrt{\kappa_j\kappa_i}$ with $1\leq i<j\leq N$, and $N(N+1)/2$ squeezing amplitudes $\nu_{ij}/\sqrt{\kappa_j\kappa_i}$ with $1\leq i\leq j\leq N$. These parameters are learning parameters in the continuous optimization. As pointed out in the main text, the functionality of the device depends only on the dimensionless detunings $H_{ii}$,
the cooperativities $C_{ij}=4|H_{ij}|^2$, and the gauge-invariant geometrical phases $\Phi_{i,j,l,\ldots,k}=\arg (H_{ij}H_{jl}\ldots H_{ki})$  accumulated in  closed loops (synthetic field fluxes), see \cref{eq:S_matrix_of_h}. 
If we allow for intrinsic loss channels in the port modes, the behaviour is furthermore influenced by the dimensionless intrinsic loss rates $\gamma$, see \cref{eq:S_matrix_of_h,eq:N_matrix_of_h}.
 
We note that in the most general case, the $S$ matrix is a $2N\times 2N$ matrix, cf \cref{eq:BdG_dimensional,eq:S_matrix_of_h,eq:Bog_rescaled_sup}.  This dimensionality reflects the fact that the intensity of the transmitted field can depend on the  phase of the input field.


To simplify the discussion in the main text we have focused on phase-preserving devices. For such devices, the  scattering matrix as defined in \cref{eq:S_matrix_of_h} has a block-diagonal form with  two $N\times N$ blocks mapped one into the other by the particle-hole symmetry \cite{liu_fully_2023}. This allows one to define the scattering matrix $S$ as one of the two blocks obtained from \cref{eq:S_matrix_of_h} and, thus, to arrive at  an  $N\times N$ matrix, consistent with \cref{eq:loss}. In this cases, our program does not calculate the``large" $2N\times 2N$ matrix but rather  only one block.   

For purely routing device (without squeezing interaction, $\nu=0$), the diagonal blocks correspond to the annihilation and creation operators, respectively. Thus, the ``smaller" scattering matrix can be taken to be the block $S_{1:N,1:N}$ of the ``larger matrix" per \cref{equ:full_scattering}. This is also the standard definition of the scattering matrix for excitation conserving systems.

For phase-preserving amplifiers, we enforce the block structure of the ``large" scattering matrix by  constraining the matrix $H$ in the following way: We divide the modes into two subsets $M_1$ and $M_2$ with $N_1$ and $N_2=N-N_1$ modes, respectively.  Then, we couple modes in the same set exclusively with  beamsplitter interactions, and modes in different sets via two-mode squeezing interactions.  Consequently, the set of equations of motion for $\hat{\boldsymbol{\xi}}=\{\hat{a}_1,\ldots,\hat{a}_{N},\hat{a}_{1}^{\dagger},\ldots,\hat{a}_{N}^{\dagger}\}$ (see \cref{equ:equation_of_motion}) breaks down into two decoupled sets of equation for $\hat{\tilde{\xi}}$ and $\hat{\tilde{\xi}}^\dag$ which are related with each other via particle-hole symmetry. $\hat{\tilde{\xi}}$ contains all $\hat{a}_j$ with $j\in M_1$ and all $\hat{a}_j^\dag$ with $j\in M_2$ and decouples from $\hat{\tilde{\xi}}^\dag$ which contains the respective conjugate operators. Both decoupled sets correspond to the same graph. Our optimisation automatically tests all possible assignments of the modes into the two subsets $M_1$ and $M_2$.

Given a graph, one can read out which modes are in which subset in the following way:
Modes coupled via a beamsplitter interaction are within the same subset, modes coupled via a squeezing interaction are in different subsets. By applying this rule step by step for each coupling, one has characterised all modes quickly.

The resulting scattering matrix has a block structure because the output fields $a^{\rm out}_{i\in M_1}$ and $a^{{\rm out}\dagger}_{i\in M_2}$ depend only on the input fields $a^{\rm in}_{i\in M_1}$ and $a^{{\rm in}\dagger}_{i\in M_2}$, and, likewise, for the  fields $a^{\rm out}_{i\in M_2}$, $a^{{\rm out}\dagger}_{i\in M_1}$,  $a^{\rm in}_{i\in M_2}$, and $a^{{\rm in}\dagger}_{i\in M_1}$. This allows us to take as scattering matrix the block enconding the response of the output fields $a^{\rm out}_{i\in M_1}$ and $a^{{\rm out}\dagger}_{i\in M_2}$ to the the input fields $a^{\rm in}_{i\in M_1}$ and $a^{{\rm in}\dagger}_{i\in M_2}$. In practice, we calculate the ``small"  $N\times N$ scattering matrix  by replacing all the  $2N\times 2N$ matrices ($\mathds{1}_{2N}$, $\sigma_z$, $H$, and $\gamma$) in  \cref{eq:S_matrix_of_h} with the relevant   $N\times N$  block, e.g.~we replace $\sigma_z$ with
\begin{equation}
	\begin{pmatrix}
 \mathds{1}_{N_1}&0\\
 0& -\mathds{1}_{N_2}\\
 \end{pmatrix}.
\end{equation}

\section{Stability}
\label{app:stability}
In the presence of  squeezing interactions, the  Langevin equations in \cref{equ:equation_of_motion} can describe unstable motions drifting away from a saddle point. We can rewrite \cref{equ:compact_equation_of_motion} in the form:
\begin{equation}
    \dot{\hat{\boldsymbol{\xi}}}(t) = D\hat{\boldsymbol{\xi}}(t) - \sqrt{\kappa} \hat{\boldsymbol{\xi}}^\mathrm{in} (t) - \sqrt{\Gamma} \hat{\boldsymbol{\xi}}^\mathrm{noise} (t) 
\end{equation}
with $D=-i\sigma_z H_{\rm BdG}  - (\kappa+\Gamma)/2$ as the dynamical matrix of the coupled system. The motion is stable as long as the dynamical matrix $D$ has only eigenvalues with a negative real part. 

Since the solutions obtained using our method are not necessarily stable, we exclude  unstable solutions after checking their stability. We note that the eigenvalues of $D$ as well as the signs of their real parts can depend on the actual choice of $\kappa_j$. This can lead to constraints on these parameters.

\section{Noise analysis}\label{App:noise_analysis}
The field operators for the input and noise channels $\hat{a}_j^\mathrm{in}$ and $\hat{a}_j^\mathrm{noise}$ obey the correlators:
\begin{subequations}
	\begin{align}
		\langle \hat{a}^{\dag,s}_j(t) \hat{a}^s_j(t^\prime) \rangle &= n_j^s \delta (t-t^\prime)\\
		\langle \hat{a}^s_j(t) \hat{a}^{\dag,s}_j(t^\prime)\rangle &= (n_j^s+1) \delta (t-t^\prime) 
	\end{align}
\end{subequations}
with $s$ standing either for {\it in} or {\it noise}. $n_j^\mathrm{in}$ and $n_j^\mathrm{noise}$ are the occupation numbers of the baths connected to the respective mode.

For the phase-preserving amplifier discussed in \cref{fig:results}(b) the number of added noise photons on the input (mode 1) and output channel (mode 2) equals:
\begin{subequations}
    \begin{align}
        N_\mathrm{input}^\mathrm{add}=\frac{1}{2} \sum_{k} \left( (n_k^\mathrm{in}+1) \abs{S_{1k}}^2 + (n_k^\mathrm{noise}+1) \abs{\mathcal{N}_{1k}}^2 \right) \\
        N_\mathrm{output}^\mathrm{add}=\frac{1}{2} \sum_{k\neq 1} \left( (n_k^\mathrm{in}+1) \abs{S_{2k}}^2 + (n_k^\mathrm{noise}+1) \abs{\mathcal{N}_{2k}}^2 \right)
    \end{align}
\end{subequations}
The quantum limit for a phase-preserving amplifiers is $N_\mathrm{input}^\mathrm{add}=1/2$ and $N_\mathrm{output}^\mathrm{add}=(G-1)/2$ \cite{clerk2010introduction}. 
If the two modes are in different subsets $M_1$ and $M_2$ (see \cref{app:scattering_matrix}), the added noise on the output is increased by one photon, meaning that the minimum number of added photons equals $N_\mathrm{output}^\mathrm{add}=(G+1)/2$ in this case.

We assume that all the baths are at zero temperature, except the bath connected to the output channel. So, $n_j^\mathrm{in}=0$ if $j\neq 2$, $n_j^\mathrm{noise}=0$ for all $j$, and $n_2^\mathrm{in} > 0$.

During optimization, we enforce that the noise on the input and output channel equals their respective quantum limits by adding the two additional terms $|f_1|^2=(N_\mathrm{input}^\mathrm{add}-1/2)^2$ and $|f_2|^2=(N_\mathrm{output}^\mathrm{add}-(G\pm 1)/2)^2$ to the cost function Eq.~(\ref{eq:loss}).
Therefore, all identified architectures automatically fulfill these equality constraints. The sign in constraint $f_2$ is chosen according to the tested graph. If input and output modes are in the same subset $M_1$ or $M_2$, we choose the minus sign. Otherwise, we choose the plus sign.

\section{Minimal number of parametric interactions}
\label{app:graph_coloring}
Here, we summarize our algorithm for finding the minimum number of parametric pumps needed to realize a given graph. The key problem is to sort the graph's nodes into frequency classes. If two modes are in the same class, they have the same frequency, otherwise, they do not. Each graph edge connecting modes of two different classes requires a pump. The goal is to minimize the number of such edges under the constraint that the modes coupled by squeezing (blue) or complex beamsplitter interactions (green) must be in different classes. 

We solve this problem in the following way: Iterating through all nodes, we assign to the current node either one of the frequency classes already assigned to a previous node, or we assign a new one, which was not used before. After all nodes have been labeled, we test whether modes coupled by squeezing (blue) or complex beamsplitter interactions (green) are in different classes. 
If this is true, we count the number of pumps. Otherwise, this labeling is skipped. 
We repeat this algorithm until every possible distribution of frequency classes has been tested. 
This search yields the minimum number of pumps, and also all possible combinations of carrier frequencies that must be chosen equally to each other so that this minimum is reached. 

The task described above is in fact a so-called \enquote{graph coloring} problem. In our setting, the frequency class plays the role of the \enquote{color}. 


\section{Gauge freedom and Gauge fixing}
\label{app:gauge_freedom}
 The scattering matrix is a complex matrix that encodes the amplitudes and phases of the output fields in response to a set of input fields. Many experiments focus  on the amplitude of the fields. Nevertheless, the phases can also be observed  after fixing a gauge by committing to specific experimental settings and physical interpretations of the fields $\hat{a}_{ j}^{\rm in/out}$. However, we do not want to distinguish between phases that depend on the detailed arrangement of the experimental setup or equivalent physical interpretations. For this purpose, we allow for the free parameters $\gamma^{\rm in/out}_j$ in our cost function Eq.~(\ref{eq:loss}). The goal of this appendix is to make an explicit connection between this gauge freedom and the aforementioned  experimental settings and physical interpretations. 
 

We start discussing a gauge freedom that depends on the detailed arrangement of the experimental setup. These are the reference positions where the input and output fields are evaluated. Assume that we change the reference position for port $j$ and that the new reference position is further away from the device. An incoming signal  acquires the additional phase  $\phi_j$ while traveling from the new reference point to the old reference point. The same phase will also be acquired by an outgoing signal when traveling from the old to the new reference point.  In other words, the transfer matrix will change according to the rule $S_{jk}\to S_{jk} e^{i(\phi_j+\phi_k)}$ where $\phi_j+\phi_k$ is the overall additional phase acquired by a signal traveling  from the new reference position in port $k$ to the new reference position in port $j$.  Comparing to Eq.~(\ref{eq:loss}), we see that  $\gamma_j^{\rm out}=-\gamma_j^{\rm in}=\phi_j$.

An additional gauge freedom for the fields $\hat{a}_{ j}^{\rm in/out}$
comes from the arbitrary phase $\phi_j$ in the transformation to the rotating frame, $\hat{a}^{\rm in/out}_{ j}=\exp [i(\omega_{L,j}t+\phi_j)]\hat{a}^{\rm in/out}_{{\rm lab}, j}$ where  $\hat{a}^{\rm in/out}_{{\rm lab}, j}$ are the fields in the lab frame. The phase $\phi_j$ determines the physical interpretation of the quadratures $\hat{X}^{\rm in/out}_{ j}$ and $\hat{P}^{\rm in/out}_{j}$ (real and imaginary part of the field),
\begin{align}
\label{eq:rot_quadrature}
\hat{X}_j^{\rm in}=\hat{X}^{\rm in}_{{\rm lab}, j}\cos(\omega_{L,j}t+\phi_j)-\hat{P}^{\rm in}_{{\rm lab}, j}\sin(\omega_{L,j}t+\phi_j)
\end{align}
The quadratures of the field in the lab frame have the same physical interpretation at all ports, e.g., in electromagnetism $\hat{X}^{\rm in}_{{\rm lab}, j}$ could correspond to the amplitude of an electric field with a fixed polarization and $\hat{P}^{\rm in}_{{\rm lab}, j}$ to the amplitude of the transverse magnetic field. For incommensurate carrier frequencies  $\omega_{L,j}$, any special initial choice of phases $\phi_j$, e.g., $\phi_j=\phi$,  would quickly dephase. As a consequence, all gauges $\phi_j$ are physically equivalent. To set them on equal footing  we allow for arbitrary $\gamma_j^{\rm out}=\gamma_j^{\rm in}=\phi_j$ in \cref{eq:loss}. Combining this gauge freedom with the choice of a reference position in each port waveguide we arrive at independent values for $\gamma_j^{\rm out}$ and $\gamma_j^{\rm in}$  in \cref{eq:loss}.

In the special case of equal carrier frequency $\omega_{L,j}=\omega_L$,  the uniform gauge  $\phi_j=\phi$  has a special physical meaning. In this gauge, the physical interpretation  of the quadratures  depends on time but remains the same at every port, see \cref{eq:rot_quadrature}. 
In this experimental scenario, the target scattering matrix is usually written in the  uniform gauge and one should restrict the gauge freedom to $\gamma_j^{\rm out}=-\gamma_j^{\rm in}=\phi_j$. To highlight the importance of the right choice for the gauge freedom we consider a concrete example in the gyrator, see Supplemental Material G. This is a non-reciprocal device with scattering matrix 
\begin{equation}
    \label{equ:gyrator}
    S_\mathrm{target}=
    \begin{pmatrix}
        0 & 1 \\
        -1 & 0
    \end{pmatrix}.
\end{equation}
For a gyrator, it is assumed that the carrier frequencies and the physical meaning of the fields at the two ports are the same. With this assumption, we can interpret the scattering matrix as describing a non-reciprocal device in which a field injected into the first port exits the second port as the same field, e.g., an electric field, but with an opposite sign. On the other hand, exchanging the emitter and the receiver, a field entering from the second port exits the first port without any sign change.

On the contrary, without the assumption that the quadratures have the same physical interpretation at both ports, the sign difference between the two off-diagonal elements loses any meaning. Indeed, it can be gauged away  with the appropriate   choice of $\gamma_j^{\rm out}$ and $\gamma_j^{\rm in}$ in Eq.~(\ref{eq:loss}), e.g. $\gamma_2^{\rm out}=\pi$ and $\gamma_2^{\rm in}=\gamma_1^{\rm in/out}=0$.

As the choice of gauge freedom is platform dependent, {\sc AutoScatter} allows the user to choose between four different options for handling the gauge freedoms: (i) no gauge freedom, (ii) allow only for free reference positions of the ports ($\gamma_j^{\rm out}=-\gamma_j^{\rm in}$), (iii) allow only for the gauge freedoms of the fields $\hat{a}_j^\mathrm{in/out}$ ($\gamma_j^{\rm out}=\gamma_j^{\rm in}$), and (iv) allow for both degrees of freedom (independent $\gamma_j^{\rm out}$ and $\gamma_j^{\rm in}$).

In all examples presented in the main text,  one could use both options (ii) and (iv) to arrive at the same results. However, as we discussed using the gyrator as an example, option (ii) is sometimes implicitly assumed for certain non-reciprocal devices. For this reason, we have used this option to produce the data in \cref{fig:search_space_reduction} and Supplemental Material F and G.






\bibliography{artificial_discovery}

\begin{thebibliography}{55}%
\makeatletter
\providecommand \@ifxundefined [1]{%
 \@ifx{#1\undefined}
}%
\providecommand \@ifnum [1]{%
 \ifnum #1\expandafter \@firstoftwo
 \else \expandafter \@secondoftwo
 \fi
}%
\providecommand \@ifx [1]{%
 \ifx #1\expandafter \@firstoftwo
 \else \expandafter \@secondoftwo
 \fi
}%
\providecommand \natexlab [1]{#1}%
\providecommand \enquote  [1]{``#1''}%
\providecommand \bibnamefont  [1]{#1}%
\providecommand \bibfnamefont [1]{#1}%
\providecommand \citenamefont [1]{#1}%
\providecommand \href@noop [0]{\@secondoftwo}%
\providecommand \href [0]{\begingroup \@sanitize@url \@href}%
\providecommand \@href[1]{\@@startlink{#1}\@@href}%
\providecommand \@@href[1]{\endgroup#1\@@endlink}%
\providecommand \@sanitize@url [0]{\catcode `\\12\catcode `\$12\catcode
  `\&12\catcode `\#12\catcode `\^12\catcode `\_12\catcode `\%12\relax}%
\providecommand \@@startlink[1]{}%
\providecommand \@@endlink[0]{}%
\providecommand \url  [0]{\begingroup\@sanitize@url \@url }%
\providecommand \@url [1]{\endgroup\@href {#1}{\urlprefix }}%
\providecommand \urlprefix  [0]{URL }%
\providecommand \Eprint [0]{\href }%
\providecommand \doibase [0]{https://doi.org/}%
\providecommand \selectlanguage [0]{\@gobble}%
\providecommand \bibinfo  [0]{\@secondoftwo}%
\providecommand \bibfield  [0]{\@secondoftwo}%
\providecommand \translation [1]{[#1]}%
\providecommand \BibitemOpen [0]{}%
\providecommand \bibitemStop [0]{}%
\providecommand \bibitemNoStop [0]{.\EOS\space}%
\providecommand \EOS [0]{\spacefactor3000\relax}%
\providecommand \BibitemShut  [1]{\csname bibitem#1\endcsname}%
\let\auto@bib@innerbib\@empty
\bibitem [{\citenamefont {Kamal}\ \emph {et~al.}(2011)\citenamefont {Kamal},
  \citenamefont {Clarke},\ and\ \citenamefont
  {Devoret}}]{kamal_noiseless_2011}%
  \BibitemOpen
  \bibfield  {author} {\bibinfo {author} {\bibfnamefont {A.}~\bibnamefont
  {Kamal}}, \bibinfo {author} {\bibfnamefont {J.}~\bibnamefont {Clarke}},\ and\
  \bibinfo {author} {\bibfnamefont {M.~H.}\ \bibnamefont {Devoret}},\
  }\bibfield  {title} {\bibinfo {title} {Noiseless non-reciprocity in a
  parametric active device},\ }\href
  {https://www.nature.com/articles/nphys1893} {\bibfield  {journal} {\bibinfo
  {journal} {Nature Physics}\ }\textbf {\bibinfo {volume} {7}},\ \bibinfo
  {pages} {311} (\bibinfo {year} {2011})}\BibitemShut {NoStop}%
\bibitem [{\citenamefont {Sliwa}\ \emph {et~al.}(2015)\citenamefont {Sliwa},
  \citenamefont {Hatridge}, \citenamefont {Narla}, \citenamefont {Shankar},
  \citenamefont {Frunzio}, \citenamefont {Schoelkopf},\ and\ \citenamefont
  {Devoret}}]{sliwa2015reconfigurable}%
  \BibitemOpen
  \bibfield  {author} {\bibinfo {author} {\bibfnamefont {K.~M.}\ \bibnamefont
  {Sliwa}}, \bibinfo {author} {\bibfnamefont {M.}~\bibnamefont {Hatridge}},
  \bibinfo {author} {\bibfnamefont {A.}~\bibnamefont {Narla}}, \bibinfo
  {author} {\bibfnamefont {S.}~\bibnamefont {Shankar}}, \bibinfo {author}
  {\bibfnamefont {L.}~\bibnamefont {Frunzio}}, \bibinfo {author} {\bibfnamefont
  {R.~J.}\ \bibnamefont {Schoelkopf}},\ and\ \bibinfo {author} {\bibfnamefont
  {M.~H.}\ \bibnamefont {Devoret}},\ }\bibfield  {title} {\bibinfo {title}
  {{Reconfigurable Josephson circulator/directional amplifier}},\ }\href
  {https://journals.aps.org/prx/abstract/10.1103/PhysRevX.5.041020} {\bibfield
  {journal} {\bibinfo  {journal} {Physical Review X}\ }\textbf {\bibinfo
  {volume} {5}},\ \bibinfo {pages} {041020} (\bibinfo {year}
  {2015})}\BibitemShut {NoStop}%
\bibitem [{\citenamefont {Peterson}\ \emph {et~al.}(2017)\citenamefont
  {Peterson}, \citenamefont {Lecocq}, \citenamefont {Cicak}, \citenamefont
  {Simmonds}, \citenamefont {Aumentado},\ and\ \citenamefont
  {Teufel}}]{peterson_demonstration_2017}%
  \BibitemOpen
  \bibfield  {author} {\bibinfo {author} {\bibfnamefont {G.~A.}\ \bibnamefont
  {Peterson}}, \bibinfo {author} {\bibfnamefont {F.}~\bibnamefont {Lecocq}},
  \bibinfo {author} {\bibfnamefont {K.}~\bibnamefont {Cicak}}, \bibinfo
  {author} {\bibfnamefont {R.~W.}\ \bibnamefont {Simmonds}}, \bibinfo {author}
  {\bibfnamefont {J.}~\bibnamefont {Aumentado}},\ and\ \bibinfo {author}
  {\bibfnamefont {J.~D.}\ \bibnamefont {Teufel}},\ }\bibfield  {title}
  {\bibinfo {title} {Demonstration of {Efficient} {Nonreciprocity} in a
  {Microwave} {Optomechanical} {Circuit}},\ }\href
  {https://doi.org/10.1103/PhysRevX.7.031001} {\bibfield  {journal} {\bibinfo
  {journal} {Physical Review X}\ }\textbf {\bibinfo {volume} {7}},\ \bibinfo
  {pages} {031001} (\bibinfo {year} {2017})}\BibitemShut {NoStop}%
\bibitem [{\citenamefont {Herrmann}\ \emph {et~al.}(2022)\citenamefont
  {Herrmann}, \citenamefont {Ansari}, \citenamefont {Wang}, \citenamefont
  {Witmer}, \citenamefont {Fan},\ and\ \citenamefont
  {Safavi-Naeini}}]{herrmann2022mirror}%
  \BibitemOpen
  \bibfield  {author} {\bibinfo {author} {\bibfnamefont {J.~F.}\ \bibnamefont
  {Herrmann}}, \bibinfo {author} {\bibfnamefont {V.}~\bibnamefont {Ansari}},
  \bibinfo {author} {\bibfnamefont {J.}~\bibnamefont {Wang}}, \bibinfo {author}
  {\bibfnamefont {J.~D.}\ \bibnamefont {Witmer}}, \bibinfo {author}
  {\bibfnamefont {S.}~\bibnamefont {Fan}},\ and\ \bibinfo {author}
  {\bibfnamefont {A.~H.}\ \bibnamefont {Safavi-Naeini}},\ }\bibfield  {title}
  {\bibinfo {title} {Mirror symmetric on-chip frequency circulation of light},\
  }\href {https://doi.org/https://doi.org/10.1038/s41566-022-01026-7}
  {\bibfield  {journal} {\bibinfo  {journal} {Nature Photonics}\ }\textbf
  {\bibinfo {volume} {16}},\ \bibinfo {pages} {603} (\bibinfo {year}
  {2022})}\BibitemShut {NoStop}%
\bibitem [{\citenamefont {Andrews}\ \emph {et~al.}(2014)\citenamefont
  {Andrews}, \citenamefont {Peterson}, \citenamefont {Purdy}, \citenamefont
  {Cicak}, \citenamefont {Simmonds}, \citenamefont {Regal},\ and\ \citenamefont
  {Lehnert}}]{andrews_bidirectional_2014}%
  \BibitemOpen
  \bibfield  {author} {\bibinfo {author} {\bibfnamefont {R.~W.}\ \bibnamefont
  {Andrews}}, \bibinfo {author} {\bibfnamefont {R.~W.}\ \bibnamefont
  {Peterson}}, \bibinfo {author} {\bibfnamefont {T.~P.}\ \bibnamefont {Purdy}},
  \bibinfo {author} {\bibfnamefont {K.}~\bibnamefont {Cicak}}, \bibinfo
  {author} {\bibfnamefont {R.~W.}\ \bibnamefont {Simmonds}}, \bibinfo {author}
  {\bibfnamefont {C.~A.}\ \bibnamefont {Regal}},\ and\ \bibinfo {author}
  {\bibfnamefont {K.~W.}\ \bibnamefont {Lehnert}},\ }\bibfield  {title}
  {{\selectlanguage {en}\bibinfo {title} {Bidirectional and efficient
  conversion between microwave and optical light}},\ }\href
  {https://www.nature.com/articles/nphys2911} {\bibfield  {journal} {\bibinfo
  {journal} {Nature Physics}\ }\textbf {\bibinfo {volume} {10}},\ \bibinfo
  {pages} {321} (\bibinfo {year} {2014})}\BibitemShut {NoStop}%
\bibitem [{\citenamefont {Hill}\ \emph {et~al.}(2012)\citenamefont {Hill},
  \citenamefont {Safavi-Naeini}, \citenamefont {Chan},\ and\ \citenamefont
  {Painter}}]{hill_coherent_2012}%
  \BibitemOpen
  \bibfield  {author} {\bibinfo {author} {\bibfnamefont {J.~T.}\ \bibnamefont
  {Hill}}, \bibinfo {author} {\bibfnamefont {A.~H.}\ \bibnamefont
  {Safavi-Naeini}}, \bibinfo {author} {\bibfnamefont {J.}~\bibnamefont
  {Chan}},\ and\ \bibinfo {author} {\bibfnamefont {O.}~\bibnamefont
  {Painter}},\ }\bibfield  {title} {{\selectlanguage {en}\bibinfo {title}
  {Coherent optical wavelength conversion via cavity optomechanics}},\ }\href
  {https://doi.org/10.1038/ncomms2201} {\bibfield  {journal} {\bibinfo
  {journal} {Nature Communications}\ }\textbf {\bibinfo {volume} {3}},\
  \bibinfo {pages} {1196} (\bibinfo {year} {2012})}\BibitemShut {NoStop}%
\bibitem [{\citenamefont {Macklin}\ \emph {et~al.}(2015)\citenamefont
  {Macklin}, \citenamefont {O’Brien}, \citenamefont {Hover}, \citenamefont
  {Schwartz}, \citenamefont {Bolkhovsky}, \citenamefont {Zhang}, \citenamefont
  {Oliver},\ and\ \citenamefont {Siddiqi}}]{macklin2015near}%
  \BibitemOpen
  \bibfield  {author} {\bibinfo {author} {\bibfnamefont {C.}~\bibnamefont
  {Macklin}}, \bibinfo {author} {\bibfnamefont {K.}~\bibnamefont {O’Brien}},
  \bibinfo {author} {\bibfnamefont {D.}~\bibnamefont {Hover}}, \bibinfo
  {author} {\bibfnamefont {M.~E.}\ \bibnamefont {Schwartz}}, \bibinfo {author}
  {\bibfnamefont {V.}~\bibnamefont {Bolkhovsky}}, \bibinfo {author}
  {\bibfnamefont {X.}~\bibnamefont {Zhang}}, \bibinfo {author} {\bibfnamefont
  {W.~D.}\ \bibnamefont {Oliver}},\ and\ \bibinfo {author} {\bibfnamefont
  {I.}~\bibnamefont {Siddiqi}},\ }\bibfield  {title} {\bibinfo {title} {{A
  near--quantum-limited Josephson traveling-wave parametric amplifier}},\
  }\href {https://www.science.org/doi/10.1126/science.aaa8525} {\bibfield
  {journal} {\bibinfo  {journal} {Science}\ }\textbf {\bibinfo {volume}
  {350}},\ \bibinfo {pages} {307} (\bibinfo {year} {2015})}\BibitemShut
  {NoStop}%
\bibitem [{\citenamefont {Fang}\ \emph {et~al.}(2017)\citenamefont {Fang},
  \citenamefont {Luo}, \citenamefont {Metelmann}, \citenamefont {Matheny},
  \citenamefont {Marquardt}, \citenamefont {Clerk},\ and\ \citenamefont
  {Painter}}]{fang_generalized_2017}%
  \BibitemOpen
  \bibfield  {author} {\bibinfo {author} {\bibfnamefont {K.}~\bibnamefont
  {Fang}}, \bibinfo {author} {\bibfnamefont {J.}~\bibnamefont {Luo}}, \bibinfo
  {author} {\bibfnamefont {A.}~\bibnamefont {Metelmann}}, \bibinfo {author}
  {\bibfnamefont {M.~H.}\ \bibnamefont {Matheny}}, \bibinfo {author}
  {\bibfnamefont {F.}~\bibnamefont {Marquardt}}, \bibinfo {author}
  {\bibfnamefont {A.~A.}\ \bibnamefont {Clerk}},\ and\ \bibinfo {author}
  {\bibfnamefont {O.}~\bibnamefont {Painter}},\ }\bibfield  {title}
  {{\selectlanguage {en}\bibinfo {title} {Generalized non-reciprocity in an
  optomechanical circuit via synthetic magnetism and reservoir engineering}},\
  }\href {https://www.nature.com/articles/nphys4009} {\bibfield  {journal}
  {\bibinfo  {journal} {Nature Physics}\ }\textbf {\bibinfo {volume} {13}},\
  \bibinfo {pages} {465} (\bibinfo {year} {2017})}\BibitemShut {NoStop}%
\bibitem [{\citenamefont {Landgraf}(2024)}]{github_autoscattering}%
  \BibitemOpen
  \bibfield  {author} {\bibinfo {author} {\bibfnamefont {J.}~\bibnamefont
  {Landgraf}},\ }\href@noop {} {\bibinfo {title} {\textsc{AutoScatter}}},\
  \bibinfo {howpublished} {\url{https://github.com/jlandgr/autoscatter}}
  (\bibinfo {year} {2024})\BibitemShut {NoStop}%
\bibitem [{\citenamefont {Wang}\ \emph {et~al.}(2023)\citenamefont {Wang},
  \citenamefont {Fu}, \citenamefont {Du}, \citenamefont {Gao}, \citenamefont
  {Huang}, \citenamefont {Liu}, \citenamefont {Chandak}, \citenamefont {Liu},
  \citenamefont {Van~Katwyk}, \citenamefont {Deac} \emph
  {et~al.}}]{wang2023scientific}%
  \BibitemOpen
  \bibfield  {author} {\bibinfo {author} {\bibfnamefont {H.}~\bibnamefont
  {Wang}}, \bibinfo {author} {\bibfnamefont {T.}~\bibnamefont {Fu}}, \bibinfo
  {author} {\bibfnamefont {Y.}~\bibnamefont {Du}}, \bibinfo {author}
  {\bibfnamefont {W.}~\bibnamefont {Gao}}, \bibinfo {author} {\bibfnamefont
  {K.}~\bibnamefont {Huang}}, \bibinfo {author} {\bibfnamefont
  {Z.}~\bibnamefont {Liu}}, \bibinfo {author} {\bibfnamefont {P.}~\bibnamefont
  {Chandak}}, \bibinfo {author} {\bibfnamefont {S.}~\bibnamefont {Liu}},
  \bibinfo {author} {\bibfnamefont {P.}~\bibnamefont {Van~Katwyk}}, \bibinfo
  {author} {\bibfnamefont {A.}~\bibnamefont {Deac}}, \emph {et~al.},\
  }\bibfield  {title} {\bibinfo {title} {Scientific discovery in the age of
  artificial intelligence},\ }\href
  {https://www.nature.com/articles/s41586-023-06221-2} {\bibfield  {journal}
  {\bibinfo  {journal} {Nature}\ }\textbf {\bibinfo {volume} {620}},\ \bibinfo
  {pages} {47} (\bibinfo {year} {2023})}\BibitemShut {NoStop}%
\bibitem [{\citenamefont {Lindsay}\ \emph {et~al.}(1993)\citenamefont
  {Lindsay}, \citenamefont {Buchanan}, \citenamefont {Feigenbaum},\ and\
  \citenamefont {Lederberg}}]{lindsay1993dendral}%
  \BibitemOpen
  \bibfield  {author} {\bibinfo {author} {\bibfnamefont {R.~K.}\ \bibnamefont
  {Lindsay}}, \bibinfo {author} {\bibfnamefont {B.~G.}\ \bibnamefont
  {Buchanan}}, \bibinfo {author} {\bibfnamefont {E.~A.}\ \bibnamefont
  {Feigenbaum}},\ and\ \bibinfo {author} {\bibfnamefont {J.}~\bibnamefont
  {Lederberg}},\ }\bibfield  {title} {\bibinfo {title} {Dendral: a case study
  of the first expert system for scientific hypothesis formation},\ }\href
  {https://doi.org/https://doi.org/10.1016/0004-3702(93)90068-M} {\bibfield
  {journal} {\bibinfo  {journal} {Artificial Intelligence}\ }\textbf {\bibinfo
  {volume} {61}},\ \bibinfo {pages} {209} (\bibinfo {year} {1993})}\BibitemShut
  {NoStop}%
\bibitem [{\citenamefont {King}\ \emph {et~al.}(2004)\citenamefont {King},
  \citenamefont {Whelan}, \citenamefont {Jones}, \citenamefont {Reiser},
  \citenamefont {Bryant}, \citenamefont {Muggleton}, \citenamefont {Kell},\
  and\ \citenamefont {Oliver}}]{king2004functional}%
  \BibitemOpen
  \bibfield  {author} {\bibinfo {author} {\bibfnamefont {R.~D.}\ \bibnamefont
  {King}}, \bibinfo {author} {\bibfnamefont {K.~E.}\ \bibnamefont {Whelan}},
  \bibinfo {author} {\bibfnamefont {F.~M.}\ \bibnamefont {Jones}}, \bibinfo
  {author} {\bibfnamefont {P.~G.~K.}\ \bibnamefont {Reiser}}, \bibinfo {author}
  {\bibfnamefont {C.~H.}\ \bibnamefont {Bryant}}, \bibinfo {author}
  {\bibfnamefont {S.~H.}\ \bibnamefont {Muggleton}}, \bibinfo {author}
  {\bibfnamefont {D.~B.}\ \bibnamefont {Kell}},\ and\ \bibinfo {author}
  {\bibfnamefont {S.~G.}\ \bibnamefont {Oliver}},\ }\bibfield  {title}
  {\bibinfo {title} {Functional genomic hypothesis generation and
  experimentation by a robot scientist},\ }\href
  {https://www.nature.com/articles/nature02236} {\bibfield  {journal} {\bibinfo
   {journal} {Nature}\ }\textbf {\bibinfo {volume} {427}},\ \bibinfo {pages}
  {247} (\bibinfo {year} {2004})}\BibitemShut {NoStop}%
\bibitem [{\citenamefont {Schmidt}\ and\ \citenamefont
  {Lipson}(2009)}]{schmidt2009distilling}%
  \BibitemOpen
  \bibfield  {author} {\bibinfo {author} {\bibfnamefont {M.}~\bibnamefont
  {Schmidt}}\ and\ \bibinfo {author} {\bibfnamefont {H.}~\bibnamefont
  {Lipson}},\ }\bibfield  {title} {\bibinfo {title} {Distilling free-form
  natural laws from experimental data},\ }\href
  {https://www.science.org/doi/10.1126/science.1165893} {\bibfield  {journal}
  {\bibinfo  {journal} {{S}cience}\ }\textbf {\bibinfo {volume} {324}},\
  \bibinfo {pages} {81} (\bibinfo {year} {2009})}\BibitemShut {NoStop}%
\bibitem [{\citenamefont {Udrescu}\ and\ \citenamefont
  {Tegmark}(2020)}]{udrescu2020ai}%
  \BibitemOpen
  \bibfield  {author} {\bibinfo {author} {\bibfnamefont {S.-M.}\ \bibnamefont
  {Udrescu}}\ and\ \bibinfo {author} {\bibfnamefont {M.}~\bibnamefont
  {Tegmark}},\ }\bibfield  {title} {\bibinfo {title} {{AI Feynman: A
  physics-inspired method for symbolic regression}},\ }\href
  {https://www.science.org/doi/10.1126/sciadv.aay2631} {\bibfield  {journal}
  {\bibinfo  {journal} {Science Advances}\ }\textbf {\bibinfo {volume} {6}},\
  \bibinfo {pages} {16} (\bibinfo {year} {2020})}\BibitemShut {NoStop}%
\bibitem [{\citenamefont {Iten}\ \emph {et~al.}(2020)\citenamefont {Iten},
  \citenamefont {Metger}, \citenamefont {Wilming}, \citenamefont {del Rio},\
  and\ \citenamefont {Renner}}]{iten2020discovering}%
  \BibitemOpen
  \bibfield  {author} {\bibinfo {author} {\bibfnamefont {R.}~\bibnamefont
  {Iten}}, \bibinfo {author} {\bibfnamefont {T.}~\bibnamefont {Metger}},
  \bibinfo {author} {\bibfnamefont {H.}~\bibnamefont {Wilming}}, \bibinfo
  {author} {\bibfnamefont {L.}~\bibnamefont {del Rio}},\ and\ \bibinfo {author}
  {\bibfnamefont {R.}~\bibnamefont {Renner}},\ }\bibfield  {title} {\bibinfo
  {title} {Discovering physical concepts with neural networks},\ }\href
  {https://journals.aps.org/prl/abstract/10.1103/PhysRevLett.124.010508}
  {\bibfield  {journal} {\bibinfo  {journal} {Physical Review Letters}\
  }\textbf {\bibinfo {volume} {124}},\ \bibinfo {pages} {010508} (\bibinfo
  {year} {2020})}\BibitemShut {NoStop}%
\bibitem [{\citenamefont {Sarra}\ \emph {et~al.}(2021)\citenamefont {Sarra},
  \citenamefont {Aiello},\ and\ \citenamefont
  {Marquardt}}]{sarra2021renormalized}%
  \BibitemOpen
  \bibfield  {author} {\bibinfo {author} {\bibfnamefont {L.}~\bibnamefont
  {Sarra}}, \bibinfo {author} {\bibfnamefont {A.}~\bibnamefont {Aiello}},\ and\
  \bibinfo {author} {\bibfnamefont {F.}~\bibnamefont {Marquardt}},\ }\bibfield
  {title} {\bibinfo {title} {{Renormalized Mutual Information for Artificial
  Scientific Discovery}},\ }\href
  {https://journals.aps.org/prl/abstract/10.1103/PhysRevLett.126.200601}
  {\bibfield  {journal} {\bibinfo  {journal} {Physical Review Letters}\
  }\textbf {\bibinfo {volume} {126}},\ \bibinfo {pages} {200601} (\bibinfo
  {year} {2021})}\BibitemShut {NoStop}%
\bibitem [{\citenamefont {Lennon}\ \emph {et~al.}(2019)\citenamefont {Lennon},
  \citenamefont {Moon}, \citenamefont {Camenzind}, \citenamefont {Yu},
  \citenamefont {Zumb{\"u}hl}, \citenamefont {Briggs}, \citenamefont {Osborne},
  \citenamefont {Laird},\ and\ \citenamefont {Ares}}]{lennon2019efficiently}%
  \BibitemOpen
  \bibfield  {author} {\bibinfo {author} {\bibfnamefont {D.~T.}\ \bibnamefont
  {Lennon}}, \bibinfo {author} {\bibfnamefont {H.}~\bibnamefont {Moon}},
  \bibinfo {author} {\bibfnamefont {L.~C.}\ \bibnamefont {Camenzind}}, \bibinfo
  {author} {\bibfnamefont {L.}~\bibnamefont {Yu}}, \bibinfo {author}
  {\bibfnamefont {D.~M.}\ \bibnamefont {Zumb{\"u}hl}}, \bibinfo {author}
  {\bibfnamefont {G.~A.~D.}\ \bibnamefont {Briggs}}, \bibinfo {author}
  {\bibfnamefont {M.~A.}\ \bibnamefont {Osborne}}, \bibinfo {author}
  {\bibfnamefont {E.~A.}\ \bibnamefont {Laird}},\ and\ \bibinfo {author}
  {\bibfnamefont {N.}~\bibnamefont {Ares}},\ }\bibfield  {title} {\bibinfo
  {title} {Efficiently measuring a quantum device using machine learning},\
  }\href {https://doi.org/https://doi.org/10.1038/s41534-019-0193-4} {\bibfield
   {journal} {\bibinfo  {journal} {npj Quantum Information}\ }\textbf {\bibinfo
  {volume} {5}},\ \bibinfo {pages} {79} (\bibinfo {year} {2019})}\BibitemShut
  {NoStop}%
\bibitem [{\citenamefont {Moon}\ \emph {et~al.}(2020)\citenamefont {Moon},
  \citenamefont {Lennon}, \citenamefont {Kirkpatrick}, \citenamefont {van
  Esbroeck}, \citenamefont {Camenzind}, \citenamefont {Yu}, \citenamefont
  {Vigneau}, \citenamefont {Zumb{\"u}hl}, \citenamefont {Briggs}, \citenamefont
  {Osborne} \emph {et~al.}}]{moon2020machine}%
  \BibitemOpen
  \bibfield  {author} {\bibinfo {author} {\bibfnamefont {H.}~\bibnamefont
  {Moon}}, \bibinfo {author} {\bibfnamefont {D.~T.}\ \bibnamefont {Lennon}},
  \bibinfo {author} {\bibfnamefont {J.}~\bibnamefont {Kirkpatrick}}, \bibinfo
  {author} {\bibfnamefont {N.~M.}\ \bibnamefont {van Esbroeck}}, \bibinfo
  {author} {\bibfnamefont {L.~C.}\ \bibnamefont {Camenzind}}, \bibinfo {author}
  {\bibfnamefont {L.}~\bibnamefont {Yu}}, \bibinfo {author} {\bibfnamefont
  {F.}~\bibnamefont {Vigneau}}, \bibinfo {author} {\bibfnamefont {D.~M.}\
  \bibnamefont {Zumb{\"u}hl}}, \bibinfo {author} {\bibfnamefont {G.~A.~D.}\
  \bibnamefont {Briggs}}, \bibinfo {author} {\bibfnamefont {M.~A.}\
  \bibnamefont {Osborne}}, \emph {et~al.},\ }\bibfield  {title} {\bibinfo
  {title} {Machine learning enables completely automatic tuning of a quantum
  device faster than human experts},\ }\href
  {https://doi.org/https://doi.org/10.1038/s41467-020-17835-9} {\bibfield
  {journal} {\bibinfo  {journal} {Nature communications}\ }\textbf {\bibinfo
  {volume} {11}},\ \bibinfo {pages} {4161} (\bibinfo {year}
  {2020})}\BibitemShut {NoStop}%
\bibitem [{\citenamefont {Duris}\ \emph {et~al.}(2020)\citenamefont {Duris},
  \citenamefont {Kennedy}, \citenamefont {Hanuka}, \citenamefont {Shtalenkova},
  \citenamefont {Edelen}, \citenamefont {Baxevanis}, \citenamefont {Egger},
  \citenamefont {Cope}, \citenamefont {McIntire}, \citenamefont {Ermon} \emph
  {et~al.}}]{duris2020bayesian}%
  \BibitemOpen
  \bibfield  {author} {\bibinfo {author} {\bibfnamefont {J.}~\bibnamefont
  {Duris}}, \bibinfo {author} {\bibfnamefont {D.}~\bibnamefont {Kennedy}},
  \bibinfo {author} {\bibfnamefont {A.}~\bibnamefont {Hanuka}}, \bibinfo
  {author} {\bibfnamefont {J.}~\bibnamefont {Shtalenkova}}, \bibinfo {author}
  {\bibfnamefont {A.}~\bibnamefont {Edelen}}, \bibinfo {author} {\bibfnamefont
  {P.}~\bibnamefont {Baxevanis}}, \bibinfo {author} {\bibfnamefont
  {A.}~\bibnamefont {Egger}}, \bibinfo {author} {\bibfnamefont
  {T.}~\bibnamefont {Cope}}, \bibinfo {author} {\bibfnamefont {M.}~\bibnamefont
  {McIntire}}, \bibinfo {author} {\bibfnamefont {S.}~\bibnamefont {Ermon}},
  \emph {et~al.},\ }\bibfield  {title} {\bibinfo {title} {Bayesian optimization
  of a free-electron laser},\ }\href
  {https://doi.org/https://doi.org/10.1103/PhysRevLett.124.124801} {\bibfield
  {journal} {\bibinfo  {journal} {Physical Review Letters}\ }\textbf {\bibinfo
  {volume} {124}},\ \bibinfo {pages} {124801} (\bibinfo {year}
  {2020})}\BibitemShut {NoStop}%
\bibitem [{\citenamefont {Krenn}\ \emph {et~al.}(2016)\citenamefont {Krenn},
  \citenamefont {Malik}, \citenamefont {Fickler}, \citenamefont {Lapkiewicz},\
  and\ \citenamefont {Zeilinger}}]{krenn2016automated}%
  \BibitemOpen
  \bibfield  {author} {\bibinfo {author} {\bibfnamefont {M.}~\bibnamefont
  {Krenn}}, \bibinfo {author} {\bibfnamefont {M.}~\bibnamefont {Malik}},
  \bibinfo {author} {\bibfnamefont {R.}~\bibnamefont {Fickler}}, \bibinfo
  {author} {\bibfnamefont {R.}~\bibnamefont {Lapkiewicz}},\ and\ \bibinfo
  {author} {\bibfnamefont {A.}~\bibnamefont {Zeilinger}},\ }\bibfield  {title}
  {\bibinfo {title} {Automated search for new quantum experiments},\ }\href
  {https://journals.aps.org/prl/abstract/10.1103/PhysRevLett.116.090405}
  {\bibfield  {journal} {\bibinfo  {journal} {Physical Review Letters}\
  }\textbf {\bibinfo {volume} {116}},\ \bibinfo {pages} {090405} (\bibinfo
  {year} {2016})}\BibitemShut {NoStop}%
\bibitem [{\citenamefont {Krenn}\ \emph {et~al.}(2020)\citenamefont {Krenn},
  \citenamefont {Erhard},\ and\ \citenamefont {Zeilinger}}]{krenn2020computer}%
  \BibitemOpen
  \bibfield  {author} {\bibinfo {author} {\bibfnamefont {M.}~\bibnamefont
  {Krenn}}, \bibinfo {author} {\bibfnamefont {M.}~\bibnamefont {Erhard}},\ and\
  \bibinfo {author} {\bibfnamefont {A.}~\bibnamefont {Zeilinger}},\ }\bibfield
  {title} {\bibinfo {title} {Computer-inspired quantum experiments},\ }\href
  {https://www.nature.com/articles/s42254-020-0230-4} {\bibfield  {journal}
  {\bibinfo  {journal} {Nature Reviews Physics}\ }\textbf {\bibinfo {volume}
  {2}},\ \bibinfo {pages} {649} (\bibinfo {year} {2020})}\BibitemShut {NoStop}%
\bibitem [{\citenamefont {Menke}\ \emph {et~al.}(2021)\citenamefont {Menke},
  \citenamefont {Häse}, \citenamefont {Gustavsson}, \citenamefont {Kerman},
  \citenamefont {Oliver},\ and\ \citenamefont
  {Aspuru-Guzik}}]{menke_automated_2021}%
  \BibitemOpen
  \bibfield  {author} {\bibinfo {author} {\bibfnamefont {T.}~\bibnamefont
  {Menke}}, \bibinfo {author} {\bibfnamefont {F.}~\bibnamefont {Häse}},
  \bibinfo {author} {\bibfnamefont {S.}~\bibnamefont {Gustavsson}}, \bibinfo
  {author} {\bibfnamefont {A.~J.}\ \bibnamefont {Kerman}}, \bibinfo {author}
  {\bibfnamefont {W.~D.}\ \bibnamefont {Oliver}},\ and\ \bibinfo {author}
  {\bibfnamefont {A.}~\bibnamefont {Aspuru-Guzik}},\ }\bibfield  {title}
  {{\selectlanguage {en}\bibinfo {title} {Automated design of superconducting
  circuits and its application to 4-local couplers}},\ }\href
  {https://doi.org/10.1038/s41534-021-00382-6} {\bibfield  {journal} {\bibinfo
  {journal} {npj Quantum Information}\ }\textbf {\bibinfo {volume} {7}},\
  \bibinfo {pages} {49} (\bibinfo {year} {2021})}\BibitemShut {NoStop}%
\bibitem [{\citenamefont {Krenn}\ \emph {et~al.}(2023)\citenamefont {Krenn},
  \citenamefont {Drori},\ and\ \citenamefont {Adhikari}}]{krenn2023digital}%
  \BibitemOpen
  \bibfield  {author} {\bibinfo {author} {\bibfnamefont {M.}~\bibnamefont
  {Krenn}}, \bibinfo {author} {\bibfnamefont {Y.}~\bibnamefont {Drori}},\ and\
  \bibinfo {author} {\bibfnamefont {R.~X.}\ \bibnamefont {Adhikari}},\
  }\bibfield  {title} {\bibinfo {title} {Digital discovery of interferometric
  gravitational wave detectors},\ }\href
  {https://doi.org/10.48550/arXiv.2312.04258} {\bibfield  {journal} {\bibinfo
  {journal} {arXiv:2312.04258}\ } (\bibinfo {year} {2023})}\BibitemShut
  {NoStop}%
\bibitem [{\citenamefont {MacLellan}\ \emph {et~al.}(2024)\citenamefont
  {MacLellan}, \citenamefont {Roztocki}, \citenamefont {Belleville},
  \citenamefont {Romero~Cort{\'e}s}, \citenamefont {Ruscitti}, \citenamefont
  {Fischer}, \citenamefont {Aza{\~n}a},\ and\ \citenamefont
  {Morandotti}}]{maclellan2024inverse}%
  \BibitemOpen
  \bibfield  {author} {\bibinfo {author} {\bibfnamefont {B.}~\bibnamefont
  {MacLellan}}, \bibinfo {author} {\bibfnamefont {P.}~\bibnamefont {Roztocki}},
  \bibinfo {author} {\bibfnamefont {J.}~\bibnamefont {Belleville}}, \bibinfo
  {author} {\bibfnamefont {L.}~\bibnamefont {Romero~Cort{\'e}s}}, \bibinfo
  {author} {\bibfnamefont {K.}~\bibnamefont {Ruscitti}}, \bibinfo {author}
  {\bibfnamefont {B.}~\bibnamefont {Fischer}}, \bibinfo {author} {\bibfnamefont
  {J.}~\bibnamefont {Aza{\~n}a}},\ and\ \bibinfo {author} {\bibfnamefont
  {R.}~\bibnamefont {Morandotti}},\ }\bibfield  {title} {\bibinfo {title}
  {Inverse design of photonic systems},\ }\href
  {https://doi.org/https://doi.org/10.1002/lpor.202300500} {\bibfield
  {journal} {\bibinfo  {journal} {Laser \& Photonics Reviews}\ }\textbf
  {\bibinfo {volume} {2024}},\ \bibinfo {pages} {2300500} (\bibinfo {year}
  {2024})}\BibitemShut {NoStop}%
\bibitem [{\citenamefont {Yariv}(1973)}]{yariv1973coupled}%
  \BibitemOpen
  \bibfield  {author} {\bibinfo {author} {\bibfnamefont {A.}~\bibnamefont
  {Yariv}},\ }\bibfield  {title} {\bibinfo {title} {Coupled-mode theory for
  guided-wave optics},\ }\href
  {https://doi.org/https://doi.org/10.1109/JQE.1973.1077767} {\bibfield
  {journal} {\bibinfo  {journal} {IEEE Journal of Quantum Electronics}\
  }\textbf {\bibinfo {volume} {9}},\ \bibinfo {pages} {919} (\bibinfo {year}
  {1973})}\BibitemShut {NoStop}%
\bibitem [{\citenamefont {Haus}\ and\ \citenamefont
  {Huang}(1991)}]{haus1991coupled}%
  \BibitemOpen
  \bibfield  {author} {\bibinfo {author} {\bibfnamefont {H.~A.}\ \bibnamefont
  {Haus}}\ and\ \bibinfo {author} {\bibfnamefont {W.}~\bibnamefont {Huang}},\
  }\bibfield  {title} {\bibinfo {title} {Coupled-mode theory},\ }\href
  {https://doi.org/https://doi.org/10.1109/5.104225} {\bibfield  {journal}
  {\bibinfo  {journal} {Proceedings of the IEEE}\ }\textbf {\bibinfo {volume}
  {79}},\ \bibinfo {pages} {1505} (\bibinfo {year} {1991})}\BibitemShut
  {NoStop}%
\bibitem [{\citenamefont {Jensen}\ and\ \citenamefont
  {Sigmund}(2011)}]{jensen2011topology}%
  \BibitemOpen
  \bibfield  {author} {\bibinfo {author} {\bibfnamefont {J.~S.}\ \bibnamefont
  {Jensen}}\ and\ \bibinfo {author} {\bibfnamefont {O.}~\bibnamefont
  {Sigmund}},\ }\bibfield  {title} {\bibinfo {title} {Topology optimization for
  nano-photonics},\ }\href
  {https://doi.org/https://doi.org/10.1002/lpor.201000014} {\bibfield
  {journal} {\bibinfo  {journal} {Laser \& Photonics Reviews}\ }\textbf
  {\bibinfo {volume} {5}},\ \bibinfo {pages} {308} (\bibinfo {year}
  {2011})}\BibitemShut {NoStop}%
\bibitem [{\citenamefont {Piggott}\ \emph {et~al.}(2015)\citenamefont
  {Piggott}, \citenamefont {Lu}, \citenamefont {Lagoudakis}, \citenamefont
  {Petykiewicz}, \citenamefont {Babinec},\ and\ \citenamefont
  {Vuckovic}}]{piggott2015inverse}%
  \BibitemOpen
  \bibfield  {author} {\bibinfo {author} {\bibfnamefont {A.~Y.}\ \bibnamefont
  {Piggott}}, \bibinfo {author} {\bibfnamefont {J.}~\bibnamefont {Lu}},
  \bibinfo {author} {\bibfnamefont {K.~G.}\ \bibnamefont {Lagoudakis}},
  \bibinfo {author} {\bibfnamefont {J.}~\bibnamefont {Petykiewicz}}, \bibinfo
  {author} {\bibfnamefont {T.~M.}\ \bibnamefont {Babinec}},\ and\ \bibinfo
  {author} {\bibfnamefont {J.}~\bibnamefont {Vuckovic}},\ }\bibfield  {title}
  {\bibinfo {title} {Inverse design and demonstration of a compact and
  broadband on-chip wavelength demultiplexer},\ }\href
  {https://www.nature.com/articles/nphoton.2015.69} {\bibfield  {journal}
  {\bibinfo  {journal} {Nature Photonics}\ }\textbf {\bibinfo {volume} {9}},\
  \bibinfo {pages} {374} (\bibinfo {year} {2015})}\BibitemShut {NoStop}%
\bibitem [{\citenamefont {Molesky}\ \emph {et~al.}(2018)\citenamefont
  {Molesky}, \citenamefont {Lin}, \citenamefont {Piggott}, \citenamefont {Jin},
  \citenamefont {Vuckovic},\ and\ \citenamefont
  {Rodriguez}}]{molesky2018inverse}%
  \BibitemOpen
  \bibfield  {author} {\bibinfo {author} {\bibfnamefont {S.}~\bibnamefont
  {Molesky}}, \bibinfo {author} {\bibfnamefont {Z.}~\bibnamefont {Lin}},
  \bibinfo {author} {\bibfnamefont {A.~Y.}\ \bibnamefont {Piggott}}, \bibinfo
  {author} {\bibfnamefont {W.}~\bibnamefont {Jin}}, \bibinfo {author}
  {\bibfnamefont {J.}~\bibnamefont {Vuckovic}},\ and\ \bibinfo {author}
  {\bibfnamefont {A.~W.}\ \bibnamefont {Rodriguez}},\ }\bibfield  {title}
  {\bibinfo {title} {Inverse design in nanophotonics},\ }\href
  {https://www.nature.com/articles/s41566-018-0246-9} {\bibfield  {journal}
  {\bibinfo  {journal} {Nature Photonics}\ }\textbf {\bibinfo {volume} {12}},\
  \bibinfo {pages} {659} (\bibinfo {year} {2018})}\BibitemShut {NoStop}%
\bibitem [{\citenamefont {Ma}\ \emph {et~al.}(2021)\citenamefont {Ma},
  \citenamefont {Liu}, \citenamefont {Kudyshev}, \citenamefont {Boltasseva},
  \citenamefont {Cai},\ and\ \citenamefont {Liu}}]{ma2021deep}%
  \BibitemOpen
  \bibfield  {author} {\bibinfo {author} {\bibfnamefont {W.}~\bibnamefont
  {Ma}}, \bibinfo {author} {\bibfnamefont {Z.}~\bibnamefont {Liu}}, \bibinfo
  {author} {\bibfnamefont {Z.~A.}\ \bibnamefont {Kudyshev}}, \bibinfo {author}
  {\bibfnamefont {A.}~\bibnamefont {Boltasseva}}, \bibinfo {author}
  {\bibfnamefont {W.}~\bibnamefont {Cai}},\ and\ \bibinfo {author}
  {\bibfnamefont {Y.}~\bibnamefont {Liu}},\ }\bibfield  {title} {\bibinfo
  {title} {Deep learning for the design of photonic structures},\ }\href
  {https://doi.org/https://doi.org/10.1038/s41566-020-0685-y} {\bibfield
  {journal} {\bibinfo  {journal} {Nature Photonics}\ }\textbf {\bibinfo
  {volume} {15}},\ \bibinfo {pages} {77} (\bibinfo {year} {2021})}\BibitemShut
  {NoStop}%
\bibitem [{\citenamefont {Ranzani}\ and\ \citenamefont
  {Aumentado}(2015)}]{ranzani_graph-based_2015}%
  \BibitemOpen
  \bibfield  {author} {\bibinfo {author} {\bibfnamefont {L.}~\bibnamefont
  {Ranzani}}\ and\ \bibinfo {author} {\bibfnamefont {J.}~\bibnamefont
  {Aumentado}},\ }\bibfield  {title} {{\selectlanguage {en}\bibinfo {title}
  {Graph-based analysis of nonreciprocity in coupled-mode systems}},\ }\href
  {https://doi.org/10.1088/1367-2630/17/2/023024} {\bibfield  {journal}
  {\bibinfo  {journal} {New Journal of Physics}\ }\textbf {\bibinfo {volume}
  {17}},\ \bibinfo {pages} {023024} (\bibinfo {year} {2015})}\BibitemShut
  {NoStop}%
\bibitem [{\citenamefont {Naaman}\ and\ \citenamefont
  {Aumentado}(2022)}]{naaman_synthesis_2022}%
  \BibitemOpen
  \bibfield  {author} {\bibinfo {author} {\bibfnamefont {O.}~\bibnamefont
  {Naaman}}\ and\ \bibinfo {author} {\bibfnamefont {J.}~\bibnamefont
  {Aumentado}},\ }\bibfield  {title} {{\selectlanguage {en}\bibinfo {title}
  {Synthesis of {Parametrically} {Coupled} {Networks}}},\ }\href
  {https://doi.org/10.1103/PRXQuantum.3.020201} {\bibfield  {journal} {\bibinfo
   {journal} {PRX Quantum}\ }\textbf {\bibinfo {volume} {3}},\ \bibinfo {pages}
  {020201} (\bibinfo {year} {2022})}\BibitemShut {NoStop}%
\bibitem [{\citenamefont {Lecocq}\ \emph {et~al.}(2017)\citenamefont {Lecocq},
  \citenamefont {Ranzani}, \citenamefont {Peterson}, \citenamefont {Cicak},
  \citenamefont {Simmonds}, \citenamefont {Teufel},\ and\ \citenamefont
  {Aumentado}}]{lecocq_nonreciprocal_2017}%
  \BibitemOpen
  \bibfield  {author} {\bibinfo {author} {\bibfnamefont {F.}~\bibnamefont
  {Lecocq}}, \bibinfo {author} {\bibfnamefont {L.}~\bibnamefont {Ranzani}},
  \bibinfo {author} {\bibfnamefont {G.~A.}\ \bibnamefont {Peterson}}, \bibinfo
  {author} {\bibfnamefont {K.}~\bibnamefont {Cicak}}, \bibinfo {author}
  {\bibfnamefont {R.~W.}\ \bibnamefont {Simmonds}}, \bibinfo {author}
  {\bibfnamefont {J.~D.}\ \bibnamefont {Teufel}},\ and\ \bibinfo {author}
  {\bibfnamefont {J.}~\bibnamefont {Aumentado}},\ }\bibfield  {title}
  {{\selectlanguage {en}\bibinfo {title} {Nonreciprocal {Microwave} {Signal}
  {Processing} with a {Field}-{Programmable} {Josephson} {Amplifier}}},\ }\href
  {https://doi.org/10.1103/PhysRevApplied.7.024028} {\bibfield  {journal}
  {\bibinfo  {journal} {Physical Review Applied}\ }\textbf {\bibinfo {volume}
  {7}},\ \bibinfo {pages} {024028} (\bibinfo {year} {2017})}\BibitemShut
  {NoStop}%
\bibitem [{\citenamefont {Habraken}\ \emph {et~al.}(2012)\citenamefont
  {Habraken}, \citenamefont {Stannigel}, \citenamefont {Lukin}, \citenamefont
  {Zoller},\ and\ \citenamefont {Rabl}}]{habraken_continuous_2012}%
  \BibitemOpen
  \bibfield  {author} {\bibinfo {author} {\bibfnamefont {S.~J.~M.}\
  \bibnamefont {Habraken}}, \bibinfo {author} {\bibfnamefont {K.}~\bibnamefont
  {Stannigel}}, \bibinfo {author} {\bibfnamefont {M.~D.}\ \bibnamefont
  {Lukin}}, \bibinfo {author} {\bibfnamefont {P.}~\bibnamefont {Zoller}},\ and\
  \bibinfo {author} {\bibfnamefont {P.}~\bibnamefont {Rabl}},\ }\bibfield
  {title} {{\selectlanguage {en}\bibinfo {title} {Continuous mode cooling and
  phonon routers for phononic quantum networks}},\ }\href
  {https://dx.doi.org/10.1088/1367-2630/14/11/115004} {\bibfield  {journal}
  {\bibinfo  {journal} {New Journal of Physics}\ }\textbf {\bibinfo {volume}
  {14}},\ \bibinfo {pages} {115004} (\bibinfo {year} {2012})}\BibitemShut
  {NoStop}%
\bibitem [{\citenamefont {Koch}\ \emph {et~al.}(2010)\citenamefont {Koch},
  \citenamefont {Houck}, \citenamefont {Hur},\ and\ \citenamefont
  {Girvin}}]{koch_time-reversal-symmetry_2010}%
  \BibitemOpen
  \bibfield  {author} {\bibinfo {author} {\bibfnamefont {J.}~\bibnamefont
  {Koch}}, \bibinfo {author} {\bibfnamefont {A.~A.}\ \bibnamefont {Houck}},
  \bibinfo {author} {\bibfnamefont {K.~L.}\ \bibnamefont {Hur}},\ and\ \bibinfo
  {author} {\bibfnamefont {S.~M.}\ \bibnamefont {Girvin}},\ }\bibfield  {title}
  {\bibinfo {title} {Time-reversal-symmetry breaking in circuit-{QED}-based
  photon lattices},\ }\href {https://doi.org/10.1103/PhysRevA.82.043811}
  {\bibfield  {journal} {\bibinfo  {journal} {Physical Review A}\ }\textbf
  {\bibinfo {volume} {82}},\ \bibinfo {pages} {043811} (\bibinfo {year}
  {2010})}\BibitemShut {NoStop}%
\bibitem [{\citenamefont {Metelmann}\ and\ \citenamefont
  {Clerk}(2015)}]{metelmann_nonreciprocal_2015}%
  \BibitemOpen
  \bibfield  {author} {\bibinfo {author} {\bibfnamefont {A.}~\bibnamefont
  {Metelmann}}\ and\ \bibinfo {author} {\bibfnamefont {A.~A.}\ \bibnamefont
  {Clerk}},\ }\bibfield  {title} {\bibinfo {title} {Nonreciprocal {Photon}
  {Transmission} and {Amplification} via {Reservoir} {Engineering}},\ }\href
  {https://doi.org/10.1103/PhysRevX.5.021025} {\bibfield  {journal} {\bibinfo
  {journal} {Physical Review X}\ }\textbf {\bibinfo {volume} {5}},\ \bibinfo
  {pages} {021025} (\bibinfo {year} {2015})}\BibitemShut {NoStop}%
\bibitem [{\citenamefont {Malz}\ \emph {et~al.}(2018)\citenamefont {Malz},
  \citenamefont {Tóth}, \citenamefont {Bernier}, \citenamefont {Feofanov},
  \citenamefont {Kippenberg},\ and\ \citenamefont
  {Nunnenkamp}}]{malz_quantum-limited_2018}%
  \BibitemOpen
  \bibfield  {author} {\bibinfo {author} {\bibfnamefont {D.}~\bibnamefont
  {Malz}}, \bibinfo {author} {\bibfnamefont {L.~D.}\ \bibnamefont {Tóth}},
  \bibinfo {author} {\bibfnamefont {N.~R.}\ \bibnamefont {Bernier}}, \bibinfo
  {author} {\bibfnamefont {A.~K.}\ \bibnamefont {Feofanov}}, \bibinfo {author}
  {\bibfnamefont {T.~J.}\ \bibnamefont {Kippenberg}},\ and\ \bibinfo {author}
  {\bibfnamefont {A.}~\bibnamefont {Nunnenkamp}},\ }\bibfield  {title}
  {\bibinfo {title} {Quantum-{Limited} {Directional} {Amplifiers} with
  {Optomechanics}},\ }\href {https://doi.org/10.1103/PhysRevLett.120.023601}
  {\bibfield  {journal} {\bibinfo  {journal} {Physical Review Letters}\
  }\textbf {\bibinfo {volume} {120}},\ \bibinfo {pages} {023601} (\bibinfo
  {year} {2018})}\BibitemShut {NoStop}%
\bibitem [{\citenamefont {Abdo}\ \emph {et~al.}(2013)\citenamefont {Abdo},
  \citenamefont {Sliwa}, \citenamefont {Frunzio},\ and\ \citenamefont
  {Devoret}}]{abdo2013directional}%
  \BibitemOpen
  \bibfield  {author} {\bibinfo {author} {\bibfnamefont {B.}~\bibnamefont
  {Abdo}}, \bibinfo {author} {\bibfnamefont {K.}~\bibnamefont {Sliwa}},
  \bibinfo {author} {\bibfnamefont {L.}~\bibnamefont {Frunzio}},\ and\ \bibinfo
  {author} {\bibfnamefont {M.}~\bibnamefont {Devoret}},\ }\bibfield  {title}
  {\bibinfo {title} {{Directional amplification with a Josephson circuit}},\
  }\href {https://journals.aps.org/prx/abstract/10.1103/PhysRevX.3.031001}
  {\bibfield  {journal} {\bibinfo  {journal} {Physical Review X}\ }\textbf
  {\bibinfo {volume} {3}},\ \bibinfo {pages} {031001} (\bibinfo {year}
  {2013})}\BibitemShut {NoStop}%
\bibitem [{\citenamefont {Liu}\ \emph {et~al.}(2024)\citenamefont {Liu},
  \citenamefont {Lingenfelter}, \citenamefont {Joshi}, \citenamefont
  {Frattini}, \citenamefont {Sivak}, \citenamefont {Shankar},\ and\
  \citenamefont {Devoret}}]{liu_fully_2023}%
  \BibitemOpen
  \bibfield  {author} {\bibinfo {author} {\bibfnamefont {G.}~\bibnamefont
  {Liu}}, \bibinfo {author} {\bibfnamefont {A.}~\bibnamefont {Lingenfelter}},
  \bibinfo {author} {\bibfnamefont {V.~R.}\ \bibnamefont {Joshi}}, \bibinfo
  {author} {\bibfnamefont {N.~E.}\ \bibnamefont {Frattini}}, \bibinfo {author}
  {\bibfnamefont {V.~V.}\ \bibnamefont {Sivak}}, \bibinfo {author}
  {\bibfnamefont {S.}~\bibnamefont {Shankar}},\ and\ \bibinfo {author}
  {\bibfnamefont {M.~H.}\ \bibnamefont {Devoret}},\ }\bibfield  {title}
  {\bibinfo {title} {Fully directional quantum-limited phase-preserving
  amplifier},\ }\href {https://doi.org/10.1103/PhysRevApplied.21.014021}
  {\bibfield  {journal} {\bibinfo  {journal} {Phys. Rev. Appl.}\ }\textbf
  {\bibinfo {volume} {21}},\ \bibinfo {pages} {014021} (\bibinfo {year}
  {2024})}\BibitemShut {NoStop}%
\bibitem [{\citenamefont {Clerk}\ \emph {et~al.}(2010)\citenamefont {Clerk},
  \citenamefont {Devoret}, \citenamefont {Girvin}, \citenamefont {Marquardt},\
  and\ \citenamefont {Schoelkopf}}]{clerk2010introduction}%
  \BibitemOpen
  \bibfield  {author} {\bibinfo {author} {\bibfnamefont {A.~A.}\ \bibnamefont
  {Clerk}}, \bibinfo {author} {\bibfnamefont {M.~H.}\ \bibnamefont {Devoret}},
  \bibinfo {author} {\bibfnamefont {S.~M.}\ \bibnamefont {Girvin}}, \bibinfo
  {author} {\bibfnamefont {F.}~\bibnamefont {Marquardt}},\ and\ \bibinfo
  {author} {\bibfnamefont {R.~J.}\ \bibnamefont {Schoelkopf}},\ }\bibfield
  {title} {\bibinfo {title} {Introduction to quantum noise, measurement, and
  amplification},\ }\href
  {https://journals.aps.org/rmp/abstract/10.1103/RevModPhys.82.1155} {\bibfield
   {journal} {\bibinfo  {journal} {Reviews of Modern Physics}\ }\textbf
  {\bibinfo {volume} {82}},\ \bibinfo {pages} {1155} (\bibinfo {year}
  {2010})}\BibitemShut {NoStop}%
\bibitem [{\citenamefont {Bernier}\ \emph {et~al.}(2017)\citenamefont
  {Bernier}, \citenamefont {T\'{o}th}, \citenamefont {Koottandavida},
  \citenamefont {Ioannou}, \citenamefont {Malz}, \citenamefont {Nunnenkamp},
  \citenamefont {Feofanov},\ and\ \citenamefont
  {Kippenberg}}]{bernier2017nonreciprocal}%
  \BibitemOpen
  \bibfield  {author} {\bibinfo {author} {\bibfnamefont {N.~R.}\ \bibnamefont
  {Bernier}}, \bibinfo {author} {\bibfnamefont {L.~D.}\ \bibnamefont
  {T\'{o}th}}, \bibinfo {author} {\bibfnamefont {A.}~\bibnamefont
  {Koottandavida}}, \bibinfo {author} {\bibfnamefont {M.~A.}\ \bibnamefont
  {Ioannou}}, \bibinfo {author} {\bibfnamefont {D.}~\bibnamefont {Malz}},
  \bibinfo {author} {\bibfnamefont {A.}~\bibnamefont {Nunnenkamp}}, \bibinfo
  {author} {\bibfnamefont {A.~K.}\ \bibnamefont {Feofanov}},\ and\ \bibinfo
  {author} {\bibfnamefont {T.~J.}\ \bibnamefont {Kippenberg}},\ }\bibfield
  {title} {\bibinfo {title} {Nonreciprocal reconfigurable microwave
  optomechanical circuit},\ }\href
  {https://www.nature.com/articles/s41467-017-00447-1} {\bibfield  {journal}
  {\bibinfo  {journal} {Nature Communications}\ }\textbf {\bibinfo {volume}
  {8}},\ \bibinfo {pages} {604} (\bibinfo {year} {2017})}\BibitemShut {NoStop}%
\bibitem [{\citenamefont {Kwende}\ \emph {et~al.}(2023)\citenamefont {Kwende},
  \citenamefont {White},\ and\ \citenamefont {Naaman}}]{kwende2023josephson}%
  \BibitemOpen
  \bibfield  {author} {\bibinfo {author} {\bibfnamefont {R.}~\bibnamefont
  {Kwende}}, \bibinfo {author} {\bibfnamefont {T.}~\bibnamefont {White}},\ and\
  \bibinfo {author} {\bibfnamefont {O.}~\bibnamefont {Naaman}},\ }\bibfield
  {title} {\bibinfo {title} {Josephson parametric circulator with
  same-frequency signal ports, 200 mhz bandwidth, and high dynamic range},\
  }\href {https://doi.org/10.1063/5.0150427} {\bibfield  {journal} {\bibinfo
  {journal} {Applied Physics Letters}\ }\textbf {\bibinfo {volume} {122}}
  (\bibinfo {year} {2023})}\BibitemShut {NoStop}%
\bibitem [{\citenamefont {Aspelmeyer}\ \emph {et~al.}(2014)\citenamefont
  {Aspelmeyer}, \citenamefont {Kippenberg},\ and\ \citenamefont
  {Marquardt}}]{aspelmeyer2014cavity}%
  \BibitemOpen
  \bibfield  {author} {\bibinfo {author} {\bibfnamefont {M.}~\bibnamefont
  {Aspelmeyer}}, \bibinfo {author} {\bibfnamefont {T.~J.}\ \bibnamefont
  {Kippenberg}},\ and\ \bibinfo {author} {\bibfnamefont {F.}~\bibnamefont
  {Marquardt}},\ }\bibfield  {title} {\bibinfo {title} {Cavity optomechanics},\
  }\href {https://journals.aps.org/rmp/abstract/10.1103/RevModPhys.86.1391}
  {\bibfield  {journal} {\bibinfo  {journal} {Reviews of Modern Physics}\
  }\textbf {\bibinfo {volume} {86}},\ \bibinfo {pages} {1391} (\bibinfo {year}
  {2014})}\BibitemShut {NoStop}%
\bibitem [{\citenamefont {Hafezi}\ and\ \citenamefont
  {Rabl}(2012)}]{Hafezi_Optomechanically_2012}%
  \BibitemOpen
  \bibfield  {author} {\bibinfo {author} {\bibfnamefont {M.}~\bibnamefont
  {Hafezi}}\ and\ \bibinfo {author} {\bibfnamefont {P.}~\bibnamefont {Rabl}},\
  }\bibfield  {title} {\bibinfo {title} {Optomechanically induced
  non-reciprocity in microring resonators},\ }\href
  {https://doi.org/10.1364/OE.20.007672} {\bibfield  {journal} {\bibinfo
  {journal} {Optics Express}\ }\textbf {\bibinfo {volume} {20}},\ \bibinfo
  {pages} {7672} (\bibinfo {year} {2012})}\BibitemShut {NoStop}%
\bibitem [{\citenamefont {Krenn}\ \emph {et~al.}(2021)\citenamefont {Krenn},
  \citenamefont {Kottmann}, \citenamefont {Tischler},\ and\ \citenamefont
  {Aspuru-Guzik}}]{krenn_conceptual_2021}%
  \BibitemOpen
  \bibfield  {author} {\bibinfo {author} {\bibfnamefont {M.}~\bibnamefont
  {Krenn}}, \bibinfo {author} {\bibfnamefont {J.~S.}\ \bibnamefont {Kottmann}},
  \bibinfo {author} {\bibfnamefont {N.}~\bibnamefont {Tischler}},\ and\
  \bibinfo {author} {\bibfnamefont {A.}~\bibnamefont {Aspuru-Guzik}},\
  }\bibfield  {title} {\bibinfo {title} {Conceptual {Understanding} through
  {Efficient} {Automated} {Design} of {Quantum} {Optical} {Experiments}},\
  }\href {https://doi.org/10.1103/PhysRevX.11.031044} {\bibfield  {journal}
  {\bibinfo  {journal} {Physical Review X}\ }\textbf {\bibinfo {volume} {11}},\
  \bibinfo {pages} {031044} (\bibinfo {year} {2021})}\BibitemShut {NoStop}%
\bibitem [{\citenamefont {Ruiz-Gonzalez}\ \emph {et~al.}(2022)\citenamefont
  {Ruiz-Gonzalez}, \citenamefont {Arlt}, \citenamefont {Petermann},
  \citenamefont {Sayyad}, \citenamefont {Jaouni}, \citenamefont {Karimi},
  \citenamefont {Tischler}, \citenamefont {Gu},\ and\ \citenamefont
  {Krenn}}]{ruiz-gonzalez_digital_2022}%
  \BibitemOpen
  \bibfield  {author} {\bibinfo {author} {\bibfnamefont {C.}~\bibnamefont
  {Ruiz-Gonzalez}}, \bibinfo {author} {\bibfnamefont {S.}~\bibnamefont {Arlt}},
  \bibinfo {author} {\bibfnamefont {J.}~\bibnamefont {Petermann}}, \bibinfo
  {author} {\bibfnamefont {S.}~\bibnamefont {Sayyad}}, \bibinfo {author}
  {\bibfnamefont {T.}~\bibnamefont {Jaouni}}, \bibinfo {author} {\bibfnamefont
  {E.}~\bibnamefont {Karimi}}, \bibinfo {author} {\bibfnamefont
  {N.}~\bibnamefont {Tischler}}, \bibinfo {author} {\bibfnamefont
  {X.}~\bibnamefont {Gu}},\ and\ \bibinfo {author} {\bibfnamefont
  {M.}~\bibnamefont {Krenn}},\ }\bibfield  {title} {\bibinfo {title} {Digital
  {Discovery} of 100 diverse {Quantum} {Experiments} with {PyTheus}},\ }\href
  {http://arxiv.org/abs/2210.09980} {\bibfield  {journal} {\bibinfo  {journal}
  {arXiv:2210.09980}\ } (\bibinfo {year} {2022})}\BibitemShut {NoStop}%
\bibitem [{\citenamefont {Arlt}\ \emph {et~al.}(2022)\citenamefont {Arlt},
  \citenamefont {Ruiz-Gonzalez},\ and\ \citenamefont
  {Krenn}}]{arlt2022digital}%
  \BibitemOpen
  \bibfield  {author} {\bibinfo {author} {\bibfnamefont {S.}~\bibnamefont
  {Arlt}}, \bibinfo {author} {\bibfnamefont {C.}~\bibnamefont
  {Ruiz-Gonzalez}},\ and\ \bibinfo {author} {\bibfnamefont {M.}~\bibnamefont
  {Krenn}},\ }\bibfield  {title} {\bibinfo {title} {Digital discovery of a
  scientific concept at the core of experimental quantum optics},\ }\href
  {https://arxiv.org/abs/2210.09981} {\bibfield  {journal} {\bibinfo  {journal}
  {arXiv:2210.09981}\ } (\bibinfo {year} {2022})}\BibitemShut {NoStop}%
\bibitem [{\citenamefont {Cerezo}\ \emph {et~al.}(2021)\citenamefont {Cerezo},
  \citenamefont {Arrasmith}, \citenamefont {Babbush}, \citenamefont {Benjamin},
  \citenamefont {Endo}, \citenamefont {Fujii}, \citenamefont {McClean},
  \citenamefont {Mitarai}, \citenamefont {Yuan}, \citenamefont {Cincio} \emph
  {et~al.}}]{cerezo2021variational}%
  \BibitemOpen
  \bibfield  {author} {\bibinfo {author} {\bibfnamefont {M.}~\bibnamefont
  {Cerezo}}, \bibinfo {author} {\bibfnamefont {A.}~\bibnamefont {Arrasmith}},
  \bibinfo {author} {\bibfnamefont {R.}~\bibnamefont {Babbush}}, \bibinfo
  {author} {\bibfnamefont {S.~C.}\ \bibnamefont {Benjamin}}, \bibinfo {author}
  {\bibfnamefont {S.}~\bibnamefont {Endo}}, \bibinfo {author} {\bibfnamefont
  {K.}~\bibnamefont {Fujii}}, \bibinfo {author} {\bibfnamefont {J.~R.}\
  \bibnamefont {McClean}}, \bibinfo {author} {\bibfnamefont {K.}~\bibnamefont
  {Mitarai}}, \bibinfo {author} {\bibfnamefont {X.}~\bibnamefont {Yuan}},
  \bibinfo {author} {\bibfnamefont {L.}~\bibnamefont {Cincio}}, \emph
  {et~al.},\ }\bibfield  {title} {\bibinfo {title} {Variational quantum
  algorithms},\ }\href
  {https://doi.org/https://doi.org/10.1038/s42254-021-00348-9} {\bibfield
  {journal} {\bibinfo  {journal} {Nature Reviews Physics}\ }\textbf {\bibinfo
  {volume} {3}},\ \bibinfo {pages} {625} (\bibinfo {year} {2021})}\BibitemShut
  {NoStop}%
\bibitem [{\citenamefont {Clarke}\ and\ \citenamefont
  {Braginski}(2006)}]{SQUID_handbook}%
  \BibitemOpen
  \bibfield  {author} {\bibinfo {author} {\bibfnamefont {J.}~\bibnamefont
  {Clarke}}\ and\ \bibinfo {author} {\bibfnamefont {A.~I.}\ \bibnamefont
  {Braginski}},\ }\href@noop {} {\emph {\bibinfo {title} {The SQUID handbook:
  Applications of SQUIDs and SQUID systems}}}\ (\bibinfo  {publisher} {John
  Wiley \& Sons},\ \bibinfo {year} {2006})\BibitemShut {NoStop}%
\bibitem [{\citenamefont {Hafezi}\ \emph {et~al.}(2011)\citenamefont {Hafezi},
  \citenamefont {Demler}, \citenamefont {Lukin},\ and\ \citenamefont
  {Taylor}}]{hafezi_robust_2011}%
  \BibitemOpen
  \bibfield  {author} {\bibinfo {author} {\bibfnamefont {M.}~\bibnamefont
  {Hafezi}}, \bibinfo {author} {\bibfnamefont {E.~A.}\ \bibnamefont {Demler}},
  \bibinfo {author} {\bibfnamefont {M.~D.}\ \bibnamefont {Lukin}},\ and\
  \bibinfo {author} {\bibfnamefont {J.~M.}\ \bibnamefont {Taylor}},\ }\bibfield
   {title} {{\selectlanguage {en}\bibinfo {title} {Robust optical delay lines
  with topological protection}},\ }\href {https://doi.org/10.1038/nphys2063}
  {\bibfield  {journal} {\bibinfo  {journal} {Nature Physics}\ }\textbf
  {\bibinfo {volume} {7}},\ \bibinfo {pages} {907} (\bibinfo {year}
  {2011})}\BibitemShut {NoStop}%
\bibitem [{\citenamefont {Ruesink}\ \emph {et~al.}(2016)\citenamefont
  {Ruesink}, \citenamefont {Miri}, \citenamefont {Al\`{u}},\ and\ \citenamefont
  {Verhagen}}]{ruesink2016nonreciprocity}%
  \BibitemOpen
  \bibfield  {author} {\bibinfo {author} {\bibfnamefont {F.}~\bibnamefont
  {Ruesink}}, \bibinfo {author} {\bibfnamefont {M.-A.}\ \bibnamefont {Miri}},
  \bibinfo {author} {\bibfnamefont {A.}~\bibnamefont {Al\`{u}}},\ and\ \bibinfo
  {author} {\bibfnamefont {E.}~\bibnamefont {Verhagen}},\ }\bibfield  {title}
  {\bibinfo {title} {Nonreciprocity and magnetic-free isolation based on
  optomechanical interactions},\ }\href
  {https://doi.org/https://doi.org/10.1038/ncomms13662} {\bibfield  {journal}
  {\bibinfo  {journal} {Nature Communications}\ }\textbf {\bibinfo {volume}
  {7}},\ \bibinfo {pages} {13662} (\bibinfo {year} {2016})}\BibitemShut
  {NoStop}%
\bibitem [{\citenamefont {Chan}\ \emph {et~al.}(2012)\citenamefont {Chan},
  \citenamefont {Safavi-Naeini}, \citenamefont {Hill}, \citenamefont
  {Meenehan},\ and\ \citenamefont {Painter}}]{chan2012optimized}%
  \BibitemOpen
  \bibfield  {author} {\bibinfo {author} {\bibfnamefont {J.}~\bibnamefont
  {Chan}}, \bibinfo {author} {\bibfnamefont {A.~H.}\ \bibnamefont
  {Safavi-Naeini}}, \bibinfo {author} {\bibfnamefont {J.~T.}\ \bibnamefont
  {Hill}}, \bibinfo {author} {\bibfnamefont {S.}~\bibnamefont {Meenehan}},\
  and\ \bibinfo {author} {\bibfnamefont {O.}~\bibnamefont {Painter}},\
  }\bibfield  {title} {\bibinfo {title} {Optimized optomechanical crystal
  cavity with acoustic radiation shield},\ }\href
  {https://doi.org/10.1063/1.4747726} {\bibfield  {journal} {\bibinfo
  {journal} {Applied Physics Letters}\ }\textbf {\bibinfo {volume} {101}}
  (\bibinfo {year} {2012})}\BibitemShut {NoStop}%
\bibitem [{\citenamefont {Asano}\ and\ \citenamefont
  {Noda}(2018)}]{asano2018optimization}%
  \BibitemOpen
  \bibfield  {author} {\bibinfo {author} {\bibfnamefont {T.}~\bibnamefont
  {Asano}}\ and\ \bibinfo {author} {\bibfnamefont {S.}~\bibnamefont {Noda}},\
  }\bibfield  {title} {\bibinfo {title} {Optimization of photonic crystal
  nanocavities based on deep learning},\ }\href
  {https://doi.org/10.1364/OE.26.032704} {\bibfield  {journal} {\bibinfo
  {journal} {Optics express}\ }\textbf {\bibinfo {volume} {26}},\ \bibinfo
  {pages} {32704} (\bibinfo {year} {2018})}\BibitemShut {NoStop}%
\bibitem [{\citenamefont {Asano}\ and\ \citenamefont
  {Noda}(2019)}]{asano2019iterative}%
  \BibitemOpen
  \bibfield  {author} {\bibinfo {author} {\bibfnamefont {T.}~\bibnamefont
  {Asano}}\ and\ \bibinfo {author} {\bibfnamefont {S.}~\bibnamefont {Noda}},\
  }\bibfield  {title} {\bibinfo {title} {Iterative optimization of photonic
  crystal nanocavity designs by using deep neural networks},\ }\href
  {https://doi.org/10.1515/nanoph-2019-0308} {\bibfield  {journal} {\bibinfo
  {journal} {Nanophotonics}\ }\textbf {\bibinfo {volume} {8}},\ \bibinfo
  {pages} {2243} (\bibinfo {year} {2019})}\BibitemShut {NoStop}%
\bibitem [{\citenamefont {Nocedal}\ and\ \citenamefont
  {Wright}(2006)}]{BFGS_reference}%
  \BibitemOpen
  \bibfield  {author} {\bibinfo {author} {\bibfnamefont {J.}~\bibnamefont
  {Nocedal}}\ and\ \bibinfo {author} {\bibfnamefont {S.~J.}\ \bibnamefont
  {Wright}},\ }\href@noop {} {\emph {\bibinfo {title} {Numerical
  Optimization}}},\ \bibinfo {edition} {2nd}\ ed.\ (\bibinfo  {publisher}
  {Springer},\ \bibinfo {address} {New York, NY, USA},\ \bibinfo {year}
  {2006})\BibitemShut {NoStop}%
\end{thebibliography}%


\begin{thebibliography}{11}%
\makeatletter
\providecommand \@ifxundefined [1]{%
 \@ifx{#1\undefined}
}%
\providecommand \@ifnum [1]{%
 \ifnum #1\expandafter \@firstoftwo
 \else \expandafter \@secondoftwo
 \fi
}%
\providecommand \@ifx [1]{%
 \ifx #1\expandafter \@firstoftwo
 \else \expandafter \@secondoftwo
 \fi
}%
\providecommand \natexlab [1]{#1}%
\providecommand \enquote  [1]{``#1''}%
\providecommand \bibnamefont  [1]{#1}%
\providecommand \bibfnamefont [1]{#1}%
\providecommand \citenamefont [1]{#1}%
\providecommand \href@noop [0]{\@secondoftwo}%
\providecommand \href [0]{\begingroup \@sanitize@url \@href}%
\providecommand \@href[1]{\@@startlink{#1}\@@href}%
\providecommand \@@href[1]{\endgroup#1\@@endlink}%
\providecommand \@sanitize@url [0]{\catcode `\\12\catcode `\$12\catcode
  `\&12\catcode `\#12\catcode `\^12\catcode `\_12\catcode `\%12\relax}%
\providecommand \@@startlink[1]{}%
\providecommand \@@endlink[0]{}%
\providecommand \url  [0]{\begingroup\@sanitize@url \@url }%
\providecommand \@url [1]{\endgroup\@href {#1}{\urlprefix }}%
\providecommand \urlprefix  [0]{URL }%
\providecommand \Eprint [0]{\href }%
\providecommand \doibase [0]{https://doi.org/}%
\providecommand \selectlanguage [0]{\@gobble}%
\providecommand \bibinfo  [0]{\@secondoftwo}%
\providecommand \bibfield  [0]{\@secondoftwo}%
\providecommand \translation [1]{[#1]}%
\providecommand \BibitemOpen [0]{}%
\providecommand \bibitemStop [0]{}%
\providecommand \bibitemNoStop [0]{.\EOS\space}%
\providecommand \EOS [0]{\spacefactor3000\relax}%
\providecommand \BibitemShut  [1]{\csname bibitem#1\endcsname}%
\let\auto@bib@innerbib\@empty
\bibitem [{\citenamefont {Hafezi}\ and\ \citenamefont
  {Rabl}(2012)}]{Hafezi_Optomechanically_2012}%
  \BibitemOpen
  \bibfield  {author} {\bibinfo {author} {\bibfnamefont {M.}~\bibnamefont
  {Hafezi}}\ and\ \bibinfo {author} {\bibfnamefont {P.}~\bibnamefont {Rabl}},\
  }\href {https://doi.org/10.1364/OE.20.007672} {\bibfield  {journal} {\bibinfo
   {journal} {Optics Express}\ }\textbf {\bibinfo {volume} {20}},\ \bibinfo
  {pages} {7672} (\bibinfo {year} {2012})}\BibitemShut {NoStop}%
\bibitem [{\citenamefont {Kim}\ \emph {et~al.}(2015)\citenamefont {Kim},
  \citenamefont {Kuzyk}, \citenamefont {Han}, \citenamefont {Wang},\ and\
  \citenamefont {Bahl}}]{kim_non-reciprocal_2015}%
  \BibitemOpen
  \bibfield  {author} {\bibinfo {author} {\bibfnamefont {J.}~\bibnamefont
  {Kim}}, \bibinfo {author} {\bibfnamefont {M.~C.}\ \bibnamefont {Kuzyk}},
  \bibinfo {author} {\bibfnamefont {K.}~\bibnamefont {Han}}, \bibinfo {author}
  {\bibfnamefont {H.}~\bibnamefont {Wang}},\ and\ \bibinfo {author}
  {\bibfnamefont {G.}~\bibnamefont {Bahl}},\ }\href
  {https://doi.org/10.1038/nphys3236} {\bibfield  {journal} {\bibinfo
  {journal} {Nature Physics}\ }\textbf {\bibinfo {volume} {11}},\ \bibinfo
  {pages} {275} (\bibinfo {year} {2015})}\BibitemShut {NoStop}%
\bibitem [{\citenamefont {Ruesink}\ \emph {et~al.}(2016)\citenamefont
  {Ruesink}, \citenamefont {Miri}, \citenamefont {Al\`{u}},\ and\ \citenamefont
  {Verhagen}}]{ruesink2016nonreciprocity}%
  \BibitemOpen
  \bibfield  {author} {\bibinfo {author} {\bibfnamefont {F.}~\bibnamefont
  {Ruesink}}, \bibinfo {author} {\bibfnamefont {M.-A.}\ \bibnamefont {Miri}},
  \bibinfo {author} {\bibfnamefont {A.}~\bibnamefont {Al\`{u}}},\ and\ \bibinfo
  {author} {\bibfnamefont {E.}~\bibnamefont {Verhagen}},\ }\href
  {https://doi.org/https://doi.org/10.1038/ncomms13662} {\bibfield  {journal}
  {\bibinfo  {journal} {Nature Communications}\ }\textbf {\bibinfo {volume}
  {7}},\ \bibinfo {pages} {13662} (\bibinfo {year} {2016})}\BibitemShut
  {NoStop}%
\bibitem [{\citenamefont {Shen}\ \emph {et~al.}(2016)\citenamefont {Shen},
  \citenamefont {Zhang}, \citenamefont {Chen}, \citenamefont {Zou},
  \citenamefont {Xiao}, \citenamefont {Zou}, \citenamefont {Sun}, \citenamefont
  {Guo},\ and\ \citenamefont {Dong}}]{shen_experimental_2016}%
  \BibitemOpen
  \bibfield  {author} {\bibinfo {author} {\bibfnamefont {Z.}~\bibnamefont
  {Shen}}, \bibinfo {author} {\bibfnamefont {Y.-L.}\ \bibnamefont {Zhang}},
  \bibinfo {author} {\bibfnamefont {Y.}~\bibnamefont {Chen}}, \bibinfo {author}
  {\bibfnamefont {C.-L.}\ \bibnamefont {Zou}}, \bibinfo {author} {\bibfnamefont
  {Y.-F.}\ \bibnamefont {Xiao}}, \bibinfo {author} {\bibfnamefont {X.-B.}\
  \bibnamefont {Zou}}, \bibinfo {author} {\bibfnamefont {F.-W.}\ \bibnamefont
  {Sun}}, \bibinfo {author} {\bibfnamefont {G.-C.}\ \bibnamefont {Guo}},\ and\
  \bibinfo {author} {\bibfnamefont {C.-H.}\ \bibnamefont {Dong}},\ }\href
  {https://doi.org/10.1038/nphoton.2016.161} {\bibfield  {journal} {\bibinfo
  {journal} {Nature Photonics}\ }\textbf {\bibinfo {volume} {10}},\ \bibinfo
  {pages} {657} (\bibinfo {year} {2016})}\BibitemShut {NoStop}%
\bibitem [{\citenamefont {Hafezi}\ \emph {et~al.}(2011)\citenamefont {Hafezi},
  \citenamefont {Demler}, \citenamefont {Lukin},\ and\ \citenamefont
  {Taylor}}]{hafezi_robust_2011}%
  \BibitemOpen
  \bibfield  {author} {\bibinfo {author} {\bibfnamefont {M.}~\bibnamefont
  {Hafezi}}, \bibinfo {author} {\bibfnamefont {E.~A.}\ \bibnamefont {Demler}},
  \bibinfo {author} {\bibfnamefont {M.~D.}\ \bibnamefont {Lukin}},\ and\
  \bibinfo {author} {\bibfnamefont {J.~M.}\ \bibnamefont {Taylor}},\ }\href
  {https://doi.org/10.1038/nphys2063} {\bibfield  {journal} {\bibinfo
  {journal} {Nature Physics}\ }\textbf {\bibinfo {volume} {7}},\ \bibinfo
  {pages} {907} (\bibinfo {year} {2011})}\BibitemShut {NoStop}%
\bibitem [{\citenamefont {Metelmann}\ and\ \citenamefont
  {Clerk}(2015)}]{metelmann_nonreciprocal_2015}%
  \BibitemOpen
  \bibfield  {author} {\bibinfo {author} {\bibfnamefont {A.}~\bibnamefont
  {Metelmann}}\ and\ \bibinfo {author} {\bibfnamefont {A.~A.}\ \bibnamefont
  {Clerk}},\ }\href {https://doi.org/10.1103/PhysRevX.5.021025} {\bibfield
  {journal} {\bibinfo  {journal} {Physical Review X}\ }\textbf {\bibinfo
  {volume} {5}},\ \bibinfo {pages} {021025} (\bibinfo {year}
  {2015})}\BibitemShut {NoStop}%
\bibitem [{\citenamefont {Lecocq}\ \emph {et~al.}(2017)\citenamefont {Lecocq},
  \citenamefont {Ranzani}, \citenamefont {Peterson}, \citenamefont {Cicak},
  \citenamefont {Simmonds}, \citenamefont {Teufel},\ and\ \citenamefont
  {Aumentado}}]{lecocq_nonreciprocal_2017}%
  \BibitemOpen
  \bibfield  {author} {\bibinfo {author} {\bibfnamefont {F.}~\bibnamefont
  {Lecocq}}, \bibinfo {author} {\bibfnamefont {L.}~\bibnamefont {Ranzani}},
  \bibinfo {author} {\bibfnamefont {G.~A.}\ \bibnamefont {Peterson}}, \bibinfo
  {author} {\bibfnamefont {K.}~\bibnamefont {Cicak}}, \bibinfo {author}
  {\bibfnamefont {R.~W.}\ \bibnamefont {Simmonds}}, \bibinfo {author}
  {\bibfnamefont {J.~D.}\ \bibnamefont {Teufel}},\ and\ \bibinfo {author}
  {\bibfnamefont {J.}~\bibnamefont {Aumentado}},\ }\href
  {https://doi.org/10.1103/PhysRevApplied.7.024028} {\bibfield  {journal}
  {\bibinfo  {journal} {Physical Review Applied}\ }\textbf {\bibinfo {volume}
  {7}},\ \bibinfo {pages} {024028} (\bibinfo {year} {2017})}\BibitemShut
  {NoStop}%
\bibitem [{\citenamefont {Sliwa}\ \emph {et~al.}(2015)\citenamefont {Sliwa},
  \citenamefont {Hatridge}, \citenamefont {Narla}, \citenamefont {Shankar},
  \citenamefont {Frunzio}, \citenamefont {Schoelkopf},\ and\ \citenamefont
  {Devoret}}]{sliwa2015reconfigurable}%
  \BibitemOpen
  \bibfield  {author} {\bibinfo {author} {\bibfnamefont {K.~M.}\ \bibnamefont
  {Sliwa}}, \bibinfo {author} {\bibfnamefont {M.}~\bibnamefont {Hatridge}},
  \bibinfo {author} {\bibfnamefont {A.}~\bibnamefont {Narla}}, \bibinfo
  {author} {\bibfnamefont {S.}~\bibnamefont {Shankar}}, \bibinfo {author}
  {\bibfnamefont {L.}~\bibnamefont {Frunzio}}, \bibinfo {author} {\bibfnamefont
  {R.~J.}\ \bibnamefont {Schoelkopf}},\ and\ \bibinfo {author} {\bibfnamefont
  {M.~H.}\ \bibnamefont {Devoret}},\ }\href
  {https://journals.aps.org/prx/abstract/10.1103/PhysRevX.5.041020} {\bibfield
  {journal} {\bibinfo  {journal} {Physical Review X}\ }\textbf {\bibinfo
  {volume} {5}},\ \bibinfo {pages} {041020} (\bibinfo {year}
  {2015})}\BibitemShut {NoStop}%
\bibitem [{\citenamefont {Ranzani}\ and\ \citenamefont
  {Aumentado}(2015)}]{ranzani_graph-based_2015}%
  \BibitemOpen
  \bibfield  {author} {\bibinfo {author} {\bibfnamefont {L.}~\bibnamefont
  {Ranzani}}\ and\ \bibinfo {author} {\bibfnamefont {J.}~\bibnamefont
  {Aumentado}},\ }\href {https://doi.org/10.1088/1367-2630/17/2/023024}
  {\bibfield  {journal} {\bibinfo  {journal} {New Journal of Physics}\ }\textbf
  {\bibinfo {volume} {17}},\ \bibinfo {pages} {023024} (\bibinfo {year}
  {2015})}\BibitemShut {NoStop}%
\bibitem [{\citenamefont {Pozar}(2012)}]{Pozar_microwave_engineering}%
  \BibitemOpen
  \bibfield  {author} {\bibinfo {author} {\bibfnamefont {D.~M.}\ \bibnamefont
  {Pozar}},\ }\href@noop {} {\emph {\bibinfo {title} {{Microwave engineering;
  4th ed.}}}}\ (\bibinfo  {publisher} {Wiley},\ \bibinfo {address} {Hoboken,
  NJ},\ \bibinfo {year} {2012})\BibitemShut {NoStop}%
\bibitem [{\citenamefont {Shekel}(1953)}]{shekel1953gyrator}%
  \BibitemOpen
  \bibfield  {author} {\bibinfo {author} {\bibfnamefont {J.}~\bibnamefont
  {Shekel}},\ }\href
  {https://doi.org/https://ieeexplore.ieee.org/abstract/document/4051433}
  {\bibfield  {journal} {\bibinfo  {journal} {Proceedings of the IRE}\ }\textbf
  {\bibinfo {volume} {41}},\ \bibinfo {pages} {1014} (\bibinfo {year}
  {1953})}\BibitemShut {NoStop}%
\end{thebibliography}%

\end{document}


\title{Supplemental Material for ``Automated Discovery of Coupled Mode Setups''}

\author{Jonas~Landgraf}\email{Jonas.Landgraf@mpl.mpg.de}
\affiliation{Max Planck Institute for the Science of Light, Staudtstr.~2, 91058 Erlangen, Germany}
\affiliation{Physics Department, University of Erlangen-Nuremberg, Staudstr.~5, 91058 Erlangen, Germany}
\author{Vittorio~Peano}
\affiliation{Max Planck Institute for the Science of Light, Staudtstr.~2, 91058 Erlangen, Germany}
\author{Florian~Marquardt}
\affiliation{Max Planck Institute for the Science of Light, Staudtstr.~2, 91058 Erlangen, Germany}
\affiliation{Physics Department, University of Erlangen-Nuremberg, Staudstr.~5, 91058 Erlangen, Germany}

\date{\today}

\maketitle

\begin{appendix}
\onecolumngrid
\renewcommand\thefigure{S\arabic{figure}}
\renewcommand\thetable{S\arabic{table}}

\section{Discussion of coupling in transmission}
\begin{figure}
	\centering \includegraphics[width=0.5\textwidth]{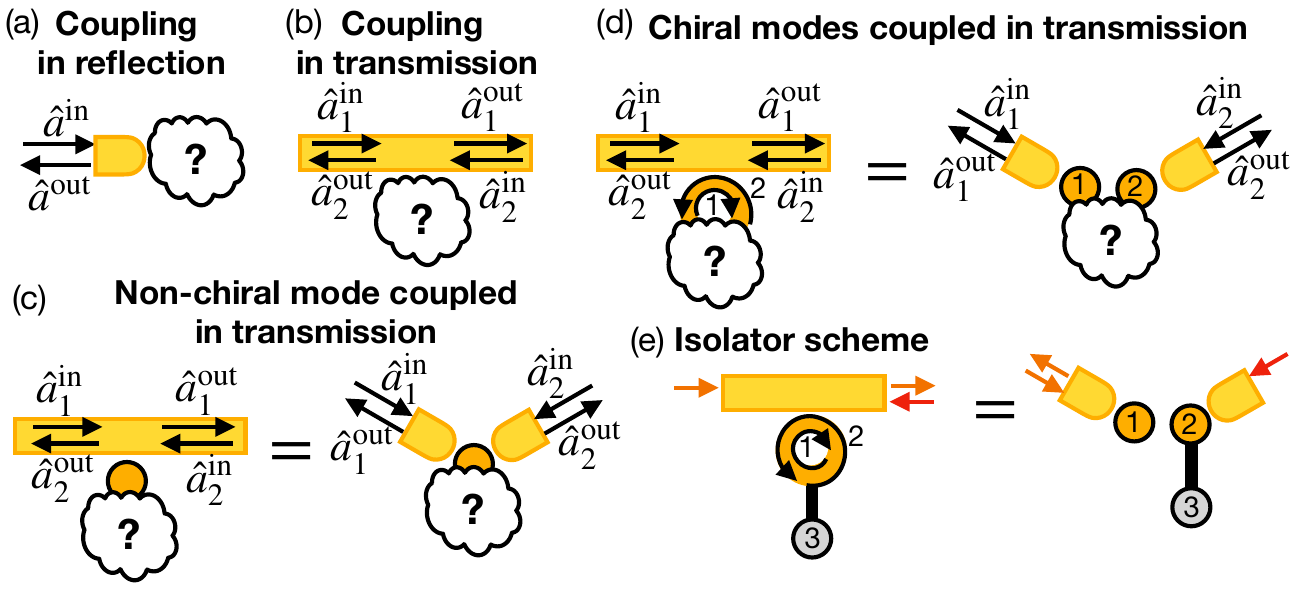}
	\caption{(a) and (b) Sketch of two possible waveguide coupling configurations. (c) Coupling in transmission to a non-chiral mode and equivalent model (described by the same Langevin equations) with two waveguides coupled in reflection to the same mode.   (d) and (e)  (left panels) Waveguide  coupled in transmission to a racetrack resonator  supporting two degenerate modes with opposite chirality (clockwise and anti-clockwise arrows)  and (right panels) equivalent model  with two  modes  coupled in reflection. 
 In (a-d) the arrows indicate the directions of the input and output fields. 
(e)  Shows a solution realizing the target scattering matrix Eq.~(\ref{eq:target_scatt_isolator}) for an isolator with coupling in transmission. The arrows of the same color (yellow or red) visualize the signal entering from one side (left or right) and the corresponding emitted signal. In both cases, only the signal entering from the left is re-emitted. However, the direction of the emitted signal differs in the two coupling configurations,  reflecting the different direction of the output field $a^{\rm out}_1$, see panel (d). Hence,  the device functions as an isolator only for the coupling in transmission.  }
	\label{fig:two_ports}
\end{figure}

In the main text, we have assumed that each port mode is coupled in ``reflection" to the waveguide relaying the input and output signals, see Fig.~\ref{fig:two_ports}(a). In this scenario, the port mode is coupled to a single decay channel. In this section, we discuss how to adapt our method to account for  coupling in ``transmission", see Fig.~\ref{fig:two_ports}(b).

We  distinguish between two different scenarios. In the first scenario, the port mode of interest  is  non-chiral, e.g.~a localized defect mode in a photonic crystal, see Fig.~\ref{fig:two_ports}(c). In the second scenario, the waveguide is coupled to a pair of degenerate modes that are mapped one into the other via time-reversal and, thus, have opposite chirality, e.g.~a pair of whispering gallery modes in a microresonator or a race-track resonator, see Fig.~\ref{fig:two_ports}(d).

In the non-chiral mode scenario, the interaction of the mode with the waveguide can be modeled by introducing  two decay channels, one for  each propagation direction, see Fig.~\ref{fig:two_ports}(c). In the presence of time-reversal symmetry, both decay channels have the same coupling. It is straightforward to extend our method to include port modes coupled to two input-output decay channels. 

In the chiral modes scenario, the interaction region typically  spans several wavelengths, and mode matching ensures that  each chiral mode is coupled only with the radiation propagating in the same direction inside the waveguide. As a consequence, the interaction between each mode and the waveguide can be modeled by introducing only one decay channel, as for coupling in reflection, see Fig.~\ref{fig:two_ports}(d). Thus, our method  directly applies to this scenario. Special considerations are required when setting up the target scattering matrix and interpreting  the couplings in the dimensionless Hamiltonian, e.g.~distinguishing between passive and active coupling, see below.
\section{Application to chiral modes}

Chiral modes are usually coupled in transmission to exploit their directionality.  Analogously to any mode coupled in reflection, their interaction with the waveguide
can be modeled using a single decay channel, see discussion in the previous section. Thus, a pair of chiral modes coupled in transmission to a waveguide (and coupled to other modes) are described by the same Langevin equations as two port modes coupled in reflection to two different waveguides (and with the same coupling to other modes). However,  transmission and reflection are exchanged compared to the equivalent system with two port modes coupled in reflection. For a fixed target functionality, e.g. an isolator, the target scattering matrix should reflect this change, $S^{\rm target}_{ji}\leftrightarrow S^{\rm target}_{ii}$ and $S^{\rm target}_{ij}\leftrightarrow S^{\rm target}_{jj}$ with  $i$, and $j$ labeling the pair of chiral port modes.


As an example  consider an isolator with the input fields to be injected on opposite ends of the same waveguide coupled in transmission to a pair of chiral modes. The target scattering matrix for an isolator with the  two port modes coupled in reflection to the respective waveguides would be,
\begin{equation}
	 S^{\rm target}=
 \begin{pmatrix}
 0&0\\
1&0\\
 \end{pmatrix}.
\end{equation}
Instead, for the setup coupled in transmission we exchange transmission and reflection as described above and use the target scattering matrix
\begin{equation}\label{eq:target_scatt_isolator}
	 S^{\rm target}=
 \begin{pmatrix}
 1&0\\
0&0\\
 \end{pmatrix}.
\end{equation}
By using our standard method, we find that the simplest solution is a disconnected graph with a single auxiliary mode, see Fig.~\ref{fig:two_ports}(d) right-hand side. Here,  the auxiliary mode interacts  only with the anti-clockwise port mode $2$ with dimensionless coupling $\tilde{g}_{23}=1/2$ or, equivalently, cooperativity $C_{23}=1$. This scheme  was originally proposed in \cite{Hafezi_Optomechanically_2012} to implement an optical isolator using a mechanical mode as an auxiliary mode, as  experimentally realized in \cite{kim_non-reciprocal_2015,ruesink2016nonreciprocity,shen_experimental_2016}. 

We note that the  method, presented in Section VIII of the main text, to identify couplings  that can be implemented passively does not apply to the chiral modes discussed here. As we explained in the main text, our method is tailored  to experimental settings in which no phase (synthetic gauge) is acquired on a loop formed by passive couplings.   On the other hand, for a set of chiral modes with the same chirality a  synthetic gauge flux can be  implemented purely passively,  by coupling them via asymmetric couplers \cite{hafezi_robust_2011}. Also in this experimental setting, graphs that do not include a flux might require an active coupling, e.g. when an auxiliary mode is coupled to only one of the chiral modes and not its time-reversed partner as in the isolator scheme used in \cite{kim_non-reciprocal_2015,ruesink2016nonreciprocity,shen_experimental_2016}.  

\section{Effective dissipative model  for the limiting case of overdamped auxiliary modes}
\label{sec:effective_lindblad}

Here, we derive the  effective Markovian dissipative  model obtained after integrating out a fast-decaying auxiliary mode $\hat{a}_f$. Such a model can be formulated in terms of a set of Langevin equations for the remaining modes or, equivalently, a Lindblad master equation.

We assume that the dimensionless BdG Hamiltonian $H$ remains fixed and derive the resulting Langevin equations and the Lindbladian  in terms of  $H$ and the decay rates $\kappa_i$.  
For this purpose, it is convenient to introduce the rescaled couplings $\tilde{g}$ and $\tilde{\nu}$,
\begin{equation}
\tilde{g}=\frac{1}{\sqrt{\kappa}}g\frac{1}{\sqrt{\kappa}}, \quad \tilde{\nu}=\frac{1}{\sqrt{\kappa}}\nu\frac{1}{\sqrt{\kappa}}.
\end{equation}
These couplings  are blocks of the rescaled Hamiltonian 
\begin{equation}
	 H=
 \begin{pmatrix}
 \tilde{g}&\tilde{\nu}\\
 \tilde{\nu}^{*}&\tilde{g}^{*}\\
 \end{pmatrix}
\end{equation}
and their numerical values are directly obtained using our method. The decay rates $\kappa_i$ can be viewed as free parameters. For simplicity, we consider the special case of a fast-decaying mode $\hat{a}_f$ not subject to a single-mode squeezing interaction. This scenario is quite general as it applies to all phase-preserving devices as well as many phase-sensitive devices.

We start by solving the Langevin equation for the fast-decaying auxiliary mode:
\begin{align}
    \dot{\hat{a}}_f(t) &=\left(i\Delta_f- \frac{\kappa_f}{2}\right) \hat{a}_f(t)  -\mathrm{i} \sum_{j\neq f} g_{fj} \hat{a}_j(t) - \mathrm{i} \sum_{j\neq f} \nu_{fj} \hat{a}_j^\dag (t)  - \sqrt{\kappa}_f \hat{a}_f^\mathrm{in} (t) 
\end{align}
to find
\begin{equation}
    \hat{a}_f(t)=-\int_{-\infty}^tdt' e^{(i\Delta_f-\kappa_f/2)(t-t')}\left(\mathrm{i} \sum_{j\neq f} g_{fj} \hat{a}_j(t') + \mathrm{i} \sum_{j\neq f} \nu_{fj} \hat{a}_j^\dag (t')  +\sqrt{\kappa}_f \hat{a}_f^\mathrm{in} (t') \right).
\end{equation}
In the limit  $\kappa_f/\kappa_i\to\infty$ we can set $\hat{a}_j(t')=\hat{a}_j(t)$, $\hat{a}^\dagger_j(t')=\hat{a}^\dagger_j(t)$, and $\hat{a}_f^\mathrm{in} (t')=\hat{a}_f^\mathrm{in} (t)$  into the integral. With this Markovian approximation, we can calculate the integral explicitly
\begin{subequations}
\label{eq:a_l_fast_decay}
    \begin{align}
        \hat{a}_f(t)=&\frac{1}{i\Delta_f-\kappa_f/2}\left(\mathrm{i} \sum_{j\neq f} g_{fj} \hat{a}_j(t) + \mathrm{i} \sum_{j\neq f} \nu_{fj} \hat{a}_j^\dag (t)  +\sqrt{\kappa}_f \hat{a}_f^\mathrm{in} (t) \right)\\
        =&-\frac{2}{\kappa_f}\frac{1+2i\Delta_f/\kappa_f}{1+4(\Delta_f/\kappa_f)^2}\left(\mathrm{i} \sum_{j\neq f} g_{fj} \hat{a}_j(t) + \mathrm{i} \sum_{j\neq f} \nu_{fj} \hat{a}_j^\dag (t)  +\sqrt{\kappa}_f \hat{a}_f^\mathrm{in} (t) \right).
    \end{align}
\end{subequations}

By plugging \cref{eq:a_l_fast_decay} into the input-output relation
\begin{equation}
\hat{a}_f^{\rm out}=\hat{\tilde{a}}_f^{\rm in}+\sqrt{\kappa_f}\hat{a}_f.
\end{equation}
and expressing in terms of $\tilde{g}_{fk}$ and $\tilde{\nu}_{fk}$ we find
\begin{equation}
\hat{a}_f^{\rm out}=-e^{2i\varphi_f}\hat{a}_f^{\rm in}-2i\frac{\exp(i\varphi_f)}{\sqrt{1+4\tilde{g}_{ff}^2}}\sum_{k\neq f}\sqrt{\kappa_k}\left(\tilde{g}_{fk}\hat{a}_k+\tilde{\nu}_{fk}\hat{a}^\dagger_k\right),
\end{equation}
with $\varphi_f=\arg(1-2i\tilde{g}_{ff})$. 
It is convenient to apply a gauge transformation to the input field, $\hat{\tilde{a}}_f^{\rm in}=-e^{2i\varphi_f}\hat{a}_f^{\rm in}$ and define the Bogoliubov  operator
\begin{equation}\label{eq:z_of_g_tilde}
\hat{z}=\sum_{k\neq f}c^-_k\hat{a}_k+c^+_k\hat{a}^\dagger_k,\quad [\hat{z},\hat{z}^\dagger]={\rm sgn} ({\cal N}_z)
\end{equation}
with
\begin{equation}\label{eq:z_of_g_tilde_2}
c^-_k=-ie^{i\varphi_f}\sqrt{\frac{\kappa_k}{|{\cal N}_z|}}\tilde{g}_{fk},\quad c^+_k=-ie^{i\varphi_f}\sqrt{\frac{\kappa_k}{|{\cal N}_z|}}\tilde{\nu}_{fk},\quad {\cal N}_z=\sum_{j\neq f}\kappa_j\left(|\tilde{g}_{fj}|^2-|\tilde{\nu}_{fj}|^2\right).
\end{equation}
We note that   $\hat{z}$  is an annihilation (creation) operator for ${\cal N}_z>0$ (${\cal N}_z<0$).
The above definitions allow one to rewrite the input-output relations as
\begin{equation}
\hat{a}_f^{\rm out}=\hat{\tilde{a}}_f^{\rm in}+\sqrt{\Gamma^{\rm eff}}\hat{z},
\end{equation}
with
\begin{equation}\label{eq:gamma_of_g_tilde}
\Gamma^{\rm eff}=\frac{4|{\cal N}_z|} {1+4\tilde{g}^2_{ff}}.  
\end{equation}
We anticipate that  $\hat{z}$ can be interpreted as a jump operator and   $\Gamma^{\rm eff}$ as the corresponding  dissipative rate. More precisely, the dissipative dynamics is described by the Lindblad master equation
\begin{equation}\label{eq:lidblad_master}
    \dot{\hat{\rho}}=-i[\hat{H}^{\rm eff},\hat{\rho}]+\sum_{i\neq f}\kappa_i {\cal L}[\hat{a}_i](\hat{\rho})
       +\Gamma^{\rm eff}{\cal L}[\hat{z}](\hat{\rho}).
\end{equation}
where $\hat{\rho}$ is the density matrix,  $\hat{H}^{\rm eff}$  an effective second-quantized Hamiltonian in the form,
\begin{equation}
    \hat{H}^{\rm eff} = \frac{1}{2}\sum_{j,k\neq f} \left(g^{\rm eff}_{jk} \hat{a}_j^\dag \hat{a}_k + \nu^{\rm eff}_{jk} \hat{a}_j^\dag \hat{a}_k^\dag \right)+ \mathrm{H.c.},
\end{equation}
 and ${\cal L}[\hat{A}]$ is the Lindblad superoperator, 
\begin{equation}
    {\cal L}[\hat{A}](\hat{\rho})=\hat{A}\hat{\rho}\hat{A}^\dagger-\frac{1}{2}\hat{\rho}\hat{A}^\dagger \hat{A}-\frac{1}{2}\hat{A}^\dagger \hat{A}\hat{\rho}.
\end{equation}
We note that for ${\cal N}_z>0$ (${\cal N}_z<0$)   the collective Bogoliubov excitations decay (are absorbed)  into (from) the effective Markovian bath. We note that Eq.~(\ref{eq:lidblad_master}) does not depend on ${\cal N}_z$, cf  Eqs.~(\ref{eq:z_of_g_tilde_2}) and Eq.~(\ref{eq:gamma_of_g_tilde}). Thus,  it is still well defined  in the special limiting case ${\cal N}_z=0$ even though Eq.~(\ref{eq:z_of_g_tilde_2}) is not. In this case, the jump operator $\hat{z}$ can not be chosen to be a Bogoliubov operator. An alternative choice is to choose it  to be a multimode quadrature
\begin{equation}
\hat{z}=\sum_{k\neq f}c^-_k\hat{a}_k+c^+_k\hat{a}^\dagger_k,\quad \hat{p}=i\sum_{k\neq f}c^{-*}_k\hat{a}^\dagger_k-c^{+*}_k\hat{a}^\dagger_k,\quad[\hat{z},\hat{p}]=i,
\end{equation}
with
\begin{equation}
c^-_k=-ie^{i\varphi_f}\sqrt{\frac{\kappa_k}{|{\cal N}_z|}}\tilde{g}_{fk},\quad c^+_k=-ie^{i\varphi_f}\sqrt{\frac{\kappa_k}{|{\cal N}_z|}}\tilde{\nu}_{fk},\quad {\cal N}_z=\sum_{j\neq f}\kappa_j\left(|\tilde{g}_{fj}|^2+|\tilde{\nu}_{fj}|^2\right).
\end{equation}

To further substantiate our interpretations of $\hat{z}$ and $\Gamma^{\rm eff}$ we  plug \cref{eq:a_l_fast_decay}  into the Langevin equations for the remaining modes  
\begin{align}
    \dot{\hat{a}}_j(t) &= -\mathrm{i}  \sum_k  g_{jk} \hat{a}_k(t) - \mathrm{i} \sum_k \nu_{jk} \hat{a}_k^\dag (t) - \frac{\kappa_j+\Gamma_j}{2} \hat{a}_j(t) - \sqrt{\kappa}_j \hat{a}_j^\mathrm{in} (t) - \sqrt{\Gamma}_j \hat{a}_j^\mathrm{noise} (t) 
\end{align}
to find the effective Langevin equation
\begin{align}
    \label{equ:equation_of_motion_dissipative}
    \dot{\hat{a}}_j(t) =& - \sum_{k\neq f} [(c^{-*}_jc^-_k-c^{+}_jc^{+*}_k)\Gamma^{\rm eff}/2+\mathrm{i}  g^{\rm eff}_{jk}] \hat{a}_k(t) -  \sum_{k\neq f} [(c^{-*}_jc^+_k-c^{+}_jc^{-*}_k)\Gamma^{\rm eff}/2+\mathrm{i}\nu^{\rm eff}_{jk}] \hat{a}_k^\dag (t) \nonumber\\
    &- \frac{\kappa_j+\Gamma_j}{2} \hat{a}_j(t) - \sqrt{\kappa}_j \hat{a}_j^\mathrm{in} (t) - \sqrt{\Gamma}_j \hat{a}_j^\mathrm{noise} (t)- \sqrt{\Gamma^{\rm eff}}c_j^{-*} \hat{\tilde{a}}_f^\mathrm{in} (t)  - \sqrt{\Gamma^{\rm eff}}c^+_j \hat{\tilde{a}}_f^{\mathrm{in}\dagger} (t), 
\end{align}
with the effective coherent couplings
\begin{eqnarray}
g^{\rm eff}_{jk}&=&g_{jk}-\sqrt{\kappa_j\kappa_k}\frac{4\tilde{g}_{ff}}{1+4\tilde{g}_{ff}^2} \left(\tilde{g}_{jf}\tilde{g}_{fk}+\tilde{\nu}_{jf}\tilde{\nu}^*_{fk}\right)\\ 
  \nu^{\rm eff}_{jk}&=&g_{jk}-\sqrt{\kappa_j\kappa_k}\frac{4\tilde{g}_{ff}}{1+4\tilde{g}_{ff}^2} \left(\tilde{g}_{jf}\tilde{\nu}_{fk}+\tilde{\nu}_{jf}\tilde{g}^*_{fk}\right) .
\end{eqnarray}
The terms proportional to the decay rate $\Gamma^{\rm eff}$ describe  dissipative interactions. By comparing the corresponding equations  for the mean fields $\langle \hat{a}_i\rangle$ and those straightforwardly calculated from the Lindblad master Eq.~(\ref{eq:lidblad_master}), one can  verify that the two sets of equations are identical  and, thus, confirm our interpretation of $\hat{z}$ and $\Gamma^{\rm eff}$.

To illustrate our general method we apply it to some specific  examples of interest. The simplest possible scenario features a resonant  fast-decaying auxiliary mode ($\tilde{g}_{ff}=0$) coupled to a single port mode $j$ with a beamsplitter interaction, see e.g. the setups  (IV) in Fig.~\ref{fig:amplifier_all_solutions}. Applying Eqs.~(\ref{eq:z_of_g_tilde}-\ref{eq:gamma_of_g_tilde})  we find $\hat{z}=-i\tilde{g}_{fj}/\abs{\tilde{g}_{fj}}\hat{a}_j$ and $\Gamma^{\rm eff}=4\kappa_j|\tilde{g}_{fj}|^2=4|g_{fj}|^2/\kappa_f$. Thus, the auxiliary mode can be replaced with a simple loss channel recovering the solution corresponding to graph (V)  in Fig.~\ref{fig:amplifier_all_solutions}.  If the fast-decaying mode and the port mode are coupled via a pair-creation interaction, as in the setups  (I) in Fig.~\ref{fig:amplifier_all_solutions}, the auxiliary mode can be replaced by a gain channel corresponding to $\hat{z}=-i\tilde{\nu}_{fj}/\abs{\tilde{\nu}_{fj}}\hat{a}^\dagger_j$ and $\Gamma^{\rm eff}=4\kappa_j|\tilde{\nu}_{fj}|^2=4|\nu_{fj}|^2/\kappa_f$.

Next, we consider the setups discussed in \cite{metelmann_nonreciprocal_2015} with a resonant  fast-decaying auxiliary mode coupled to two port modes. If the modes are coupled pairwise via beamsplitter couplings to realize an isolator [see Fig.~4(a) of the main text] the matrix elements of the dimensionless Hamiltonian obtained using our method read $\tilde{g}_{12}=i/2$, $\tilde{g}_{23}=1/2$, and $\tilde{g}_{31}=1/2$. Applying Eqs.~(\ref{eq:z_of_g_tilde}-\ref{eq:gamma_of_g_tilde})  we find $$\hat{z}=-\frac{i}{\sqrt{\kappa_1+\kappa_2}}(\sqrt{\kappa_1}\hat{a}_1+\sqrt{\kappa_2}\hat{a}_2),\quad \Gamma^{\rm eff}=\kappa_1+\kappa_2.$$ This is the same solution derived in \cite{metelmann_nonreciprocal_2015} for the special case $\kappa_1=\kappa_2$.

One can also operate the same three modes setup  as a quantum-limited amplifier. Using our method, we recover the same solution also used in \cite{metelmann_nonreciprocal_2015,lecocq_nonreciprocal_2017,sliwa2015reconfigurable}. The required  dimensionless matrix elements read  $\tilde{\nu}_{12}=\tilde{\nu}_{21}=iF(G)/2$, $\tilde{\nu}_{23}=\tilde{\nu}_{32}=F(G)/2$, and $\tilde{g}_{31}=\tilde{g}_{13}=1/2$ with $F(G)=\sqrt{G-1}/(1+\sqrt{G})$ (all other matrix elements are zero). As in \cite{metelmann_nonreciprocal_2015} we consider $\kappa_1=\kappa_2$ which corresponds to a stable system. Applying Eqs.~(\ref{eq:z_of_g_tilde}-\ref{eq:gamma_of_g_tilde})  we find $$\hat{z}=-\frac{i}{\sqrt{1-F^2(G)}}(\hat{a}_1+F(G)\hat{a}^\dagger_2),\quad \Gamma^{\rm eff}=\kappa_1(1-F^2(G)),$$
recovering the same dissipative scheme as in \cite{metelmann_nonreciprocal_2015}.

\section{Fully-directional phase-preserving quantum-limited amplifier}
\label{app:directional_amplifier}
\begin{figure}
	\centering \includegraphics[width=0.4\textwidth]{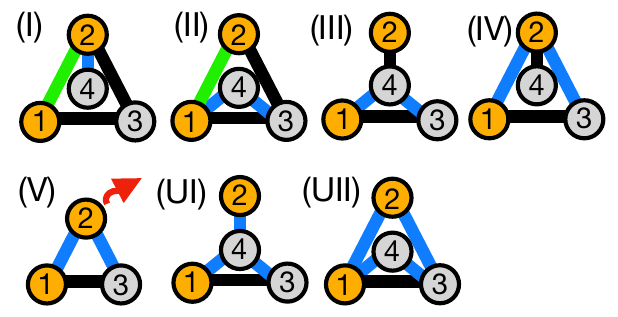}
	\caption{All irreducible graphs identified for the fully-directional phase preserving quantum-limited amplifier. Mode 1 is the input mode, mode 2 the output. Graph (I-V) have a stable parameter regime. For graph (UI) and (UII) we did not identify a stable regime.}
	\label{fig:amplifier_all_solutions}
\end{figure}
Here, we give more details on all irreducible-graph solutions   for the fully directional phase-preserving quantum-limited amplifier, see  Fig.~4(c). In \cref{fig:amplifier_all_solutions}, we display  all discovered graphs, here, including also those for which we did not find a stable parameter regime.

The analytical expressions for the scattering matrix $S(G)$ 
given below are inferred by analytical regression starting from a dataset of solutions created by applying our method for a set of $G$ values. The corresponding analytical expressions for the dimensionless Bogoliubov Hamiltonian $H(G)$ are then obtained by solving Eq.~(F3a) in Appendix F in the main text.

\subsection{Graph I}
The scattering matrix for Graph (I) reads
\begin{equation}
	S=\begin{pmatrix}0 & 0 & - \mathrm{i} & 0\\\sqrt{G} & 0 & 0 & \mathrm{i} \sqrt{G - 1} e^{\mathrm{i} \varphi_{24}}\\\mathrm{i} \left(G - 1\right) & - \mathrm{i} \sqrt{G} & 0 & - \sqrt{G} \sqrt{G - 1} e^{\mathrm{i} \varphi_{24}}\\\mathrm{i} \sqrt{G} \sqrt{G - 1} e^{- \mathrm{i} \varphi_{24}} & - \mathrm{i} \sqrt{G - 1} e^{- \mathrm{i} \varphi_{24}} & 0 & - G\end{pmatrix}.
\end{equation}
The underlying matrix elements of the dimensionless Hamiltonian read  $g_{12}/\sqrt{\kappa_1\kappa_2}=\mathrm{i}\sqrt{G}/2$, $g_{13}/\sqrt{\kappa_1\kappa_3}=-1/2$, $g_{23}/\sqrt{\kappa_2\kappa_3}=-\sqrt{G}/2$, and $\nu_{24}/\sqrt{\kappa_2\kappa_4}=\sqrt{G-1}\mathrm{e}^{\mathrm{i}\varphi_{24}}/2$.
We note that the phase $\varphi_{24}\equiv\arg{\nu_{24}}$ can be fixed to any arbitrary value. In this solution, the phase acquired  from the input  mode $1$ to the output mode $2$ assumes the value  $\pi/2$ while the remaining coupling amplitudes are real, corresponding to an overall flux $\Phi_{123}=\pi/2$. We note that the free phases $\gamma^{\rm out/in}_j$ in the cost function allow one to arbitrarily change the coupling phases  and still find a valid solution as long as the flux $\Phi_{123}$ remains the same. In this way, one can change the number of green edges or arrive at a different valid irreducible graph with a single green edge connecting sites 1 and 3 or  2 and 3. We view all these solutions as equivalent.
We have checked that for $\kappa_1=\kappa_2=\kappa_3=\kappa_4$ the system is stable for all values of $G$.

\subsection{Graph II}
The scattering matrix for Graph (II) reads
\begin{equation}
	S = \begin{pmatrix}0 & 0 & - \mathrm{i} & 0\\\sqrt{G} & 0 & 0 & - \mathrm{i} \sqrt{G - 1} e^{\mathrm{i} \varphi_{14}}\\\mathrm{i} \left(1 - G\right) & - \mathrm{i} \sqrt{G} & 0 & - \sqrt{G} \sqrt{G - 1} e^{\mathrm{i} \varphi_{14}}\\- \mathrm{i} \sqrt{G} \sqrt{G - 1} e^{- \mathrm{i} \varphi_{14}} & - \mathrm{i} \sqrt{G - 1} e^{- \mathrm{i} \varphi_{14}} & 0 & - G\end{pmatrix}.
\end{equation}
The underlying matrix elements of the dimensionless Hamiltonian read $g_{12}/\sqrt{\kappa_1\kappa_2}=\mathrm{i}/(2\sqrt{G})$, $g_{13}/\sqrt{\kappa_1\kappa_3}=-1/2$, $\nu_{14}/\sqrt{\kappa_1\kappa_4}=\sqrt{G-1}\mathrm{e}^{\mathrm{i}\varphi_{14}}/(2\sqrt{G})$, $g_{23}/\sqrt{\kappa_2\kappa_3}=-1/(2\sqrt{G})$, $\nu_{34}/\sqrt{\kappa_3\kappa_4}=\mathrm{i}\sqrt{G-1}\mathrm{e}^{\mathrm{i}\varphi_{14}}/(2\sqrt{G})$. 
Here, the phase $\varphi_{14}\equiv\arg\nu_{14}$ is a free parameter. Equivalent irreducible graphs have a single green edge connecting modes 1 and 3 or 2 and 3. We have checked that for $\kappa_1=\kappa_2=\kappa_3=\kappa_4$, the system is stable for all values of $G$.

\subsection{Graph III}
The scattering matrix for Graph (III) reads
\begin{equation}
	S=\begin{pmatrix}0 & 0 & - \mathrm{i} & 0\\- \sqrt{G} e^{- \mathrm{i} \varphi_{14}} & 0 & 0 & \mathrm{i}\sqrt{G+1}\\\mathrm{i} \left(- G - 1\right) & \mathrm{i} \sqrt{G} e^{\mathrm{i} \varphi_{14}} & 0 & - \sqrt{G} \sqrt{G + 1} e^{\mathrm{i} \varphi_{14}}\\- \mathrm{i} \sqrt{G} \sqrt{G + 1} e^{- \mathrm{i} \varphi_{14}} & \mathrm{i}\sqrt{G+1} & 0 & - G\end{pmatrix}.
\end{equation}
The underlying matrix elements of the dimensionless Hamiltonian read $g_{13}/\sqrt{\kappa_1\kappa_3}=-1/2$, $\nu_{14}/\sqrt{\kappa_1\kappa_4}=\sqrt{G}\mathrm{e}^{\mathrm{i}\varphi{14}}/(2\sqrt{G+1})$, $g_{24}/\sqrt{\kappa_2\kappa_4}=-1/(2\sqrt{G+1})$, $\nu_{34}/\sqrt{\kappa_3\kappa_4}=\mathrm{i}\sqrt{G}\mathrm{e}^{\mathrm{i}\varphi{14}}/(2\sqrt{G+1})$. The phase $\varphi_{14}\equiv\arg\nu_{14}$ is a free parameter. We have checked that for $\kappa_1=\kappa_2=\kappa_3=\kappa_4$ the system is stable for all values of $G$.

\subsection{Graph IV}
The scattering matrix for Graph (IV) reads
\begin{equation}
    \label{equ:scattering_matrix_graph_IV}
	S  = \begin{pmatrix}0 & 0 & - \mathrm{i} & 0\\- \mathrm{i} \sqrt{G} e^{- \mathrm{i} \varphi_{12}} & 0 & 0 & \mathrm{i}\sqrt{ G + 1}\\\mathrm{i} \left(- G - 1\right) & - \sqrt{G} e^{\mathrm{i} \varphi_{12}} & 0 & \mathrm{i} \sqrt{G} \sqrt{G + 1} e^{\mathrm{i} \varphi_{12}}\\- \sqrt{G} \sqrt{G + 1} e^{- \mathrm{i} \varphi_{12}} & \mathrm{i}\sqrt{G + 1} & 0 & G\end{pmatrix}.
\end{equation}
The underlying matrix elements of the dimensionless Hamiltonian read  $\nu_{12}/\sqrt{\kappa_1\kappa_2}=\sqrt{G}\mathrm{e}^{\mathrm{i}\varphi_{12}}/2$, $g_{13}/\sqrt{\kappa_1\kappa_3}=-1/2$, $\nu_{23}/\sqrt{\kappa_2\kappa_3}=\mathrm{i}\sqrt{G}\mathrm{e}^{\mathrm{i}\varphi_{12}}/2$, $g_{24}/\sqrt{\kappa_2\kappa_4}=-\sqrt{G+1}/2$ with the free parameter $\varphi_{12}$. We have checked that for $\kappa_1=\kappa_2=\kappa_3$ and $\kappa_4/\kappa_1=G$ the system is stable for all values of $G$.

\subsection{Graph V}
The scattering matrix for Graph (V) reads
\begin{equation}
	S=\begin{pmatrix}0 & 0 & - \mathrm{i}\\- \mathrm{i} \sqrt{G} e^{- \mathrm{i} \varphi_{12}} & 0 & 0\\\mathrm{i} \left(- G - 1\right) & - \sqrt{G} e^{\mathrm{i} \varphi_{12}} & 0\end{pmatrix}.
\end{equation}
The underlying matrix elements of the dimensionless Hamiltonian read $\nu_{12}/\sqrt{\kappa_1\kappa_2}=\sqrt{G}\mathrm{e}^{\mathrm{i}\varphi_{12}}/2$, $g_{13}/\sqrt{\kappa_1\kappa_3}=-1/2$ and $\nu_{23}/\sqrt{\kappa_2\kappa_3}=\mathrm{i}\sqrt{G}\mathrm{e}^{\mathrm{i}\varphi_{12}}/2$. In addition, the intrinsic loss rate of the output mode is  $\Gamma_2/\kappa_2=G+1$. The phase $\varphi_{12}$ is a free parameter. We have checked that for $\kappa_1=\kappa_2=\kappa_3$ the system is stable for all values of $G$.

As mentioned above solution (V) can be  recovered  starting from solution (IV) by applying our general method outlined in the Supplementary Section~\ref{sec:effective_lindblad} to  integrate out auxiliary mode $4$.
\subsection{Graph UI}
The scattering matrix for Graph (UI) reads
\begin{equation}
	S=\begin{pmatrix}0 & 0 & \mathrm{i} & 0\\\sqrt{G} e^{- \mathrm{i} \left(\varphi_{14} - \varphi_{24}\right)} & 0 & 0 & - \mathrm{i} \sqrt{G - 1} e^{\mathrm{i} \varphi_{24}}\\\mathrm{i} \left(1 - G\right) & - \mathrm{i} \sqrt{G} e^{\mathrm{i} \left(\varphi_{14} - \varphi_{24}\right)} & 0 & - \sqrt{G} \sqrt{G - 1} e^{\mathrm{i} \varphi_{14}}\\\mathrm{i} \sqrt{G} \sqrt{G - 1} e^{- \mathrm{i} \varphi_{14}} & \mathrm{i} \sqrt{G - 1} e^{- \mathrm{i} \varphi_{24}} & 0 & G\end{pmatrix}.
\end{equation}
The underlying matrix elements of the dimensionless Hamiltonian read $g_{13}/\sqrt{\kappa_1\kappa_3}=1/2$, $\nu_{14}/\sqrt{\kappa_1\kappa_4}=\sqrt{G}\mathrm{e}^{\mathrm{i}\varphi_{14}}/(2\sqrt{G-1})$, $\nu_{24}/\sqrt{\kappa_2\kappa_4}=\mathrm{e}^{\mathrm{i}\varphi_{24}}/(2\sqrt{G-1})$, $\nu_{34}/\sqrt{\kappa_3\kappa_4}=-\mathrm{i}\sqrt{G}\mathrm{e}^{\mathrm{i}\varphi_{14}}/(2\sqrt{G-1})$. $\varphi_{14}$ and $\varphi_{24}$ are coupling phases and can be chosen arbitrarily. We did not identify any stable parameter regime for this graph.

\subsection{Graph UII}
The scattering matrix for Graph (UII) reads
\begin{equation}
    S=\begin{pmatrix}0 & 0 & - \mathrm{i} & 0\\- \mathrm{i} \sqrt{G} & 0 & 0 & \sqrt{G + 1}\\\mathrm{i} \left(G + 1\right) & - \sqrt{G} & 0 & - \sqrt{G^{2} + G}\\- \mathrm{i} \sqrt{G^{2} + G} & \sqrt{G + 1} & 0 & G\end{pmatrix}.
\end{equation}
The underlying matrix elements of the dimensionless Hamiltonian read $\nu_{12}/\sqrt{\kappa_1\kappa_2}=-1/(2\sqrt{G})$, $g_{13}/\sqrt{\kappa_1\kappa_3}=-1/2$, $\nu_{14}/\sqrt{\kappa_1\kappa_4}=-\sqrt{G+1}/(2\sqrt{G})$, $\nu_{23}/\sqrt{\kappa_2\kappa_3}=-\mathrm{i}/(2\sqrt{G})$, and $\nu_{34}/\sqrt{\kappa_3\kappa_4}=-\mathrm{i}\sqrt{G+1}/(2\sqrt{G})$. We chose the coupling phases of $\varphi_{12}$ and $\varphi_{14}$ to be $\pi$. As long as the sum of phases of each loop is conserved, the phases of all the parametric couplings can be set to any value. We were not able to identify any stable parameter regime for this graph.

\subsection{Noise analysis}

For completeness, we summarise here the noise behavior of all graphs discussed in \cref{fig:amplifier_all_solutions}, also including the graphs without a stable solution. For Graph (I,II,UI) the number of added photons on the output channel turns out to equal $N_\mathrm{output}^\mathrm{add}=(G-1)/2$, for the graphs (III--V,UII) $N_\mathrm{output}^\mathrm{add}=(G+1)/2$. All graphs fulfill $N_\mathrm{input}^\mathrm{add}=1/2$.

\subsection{Comparison between graph (V) and three-mode directional amplifiers from the literature}
\begin{figure}
    \centering
    \includegraphics[width=0.8\textwidth]{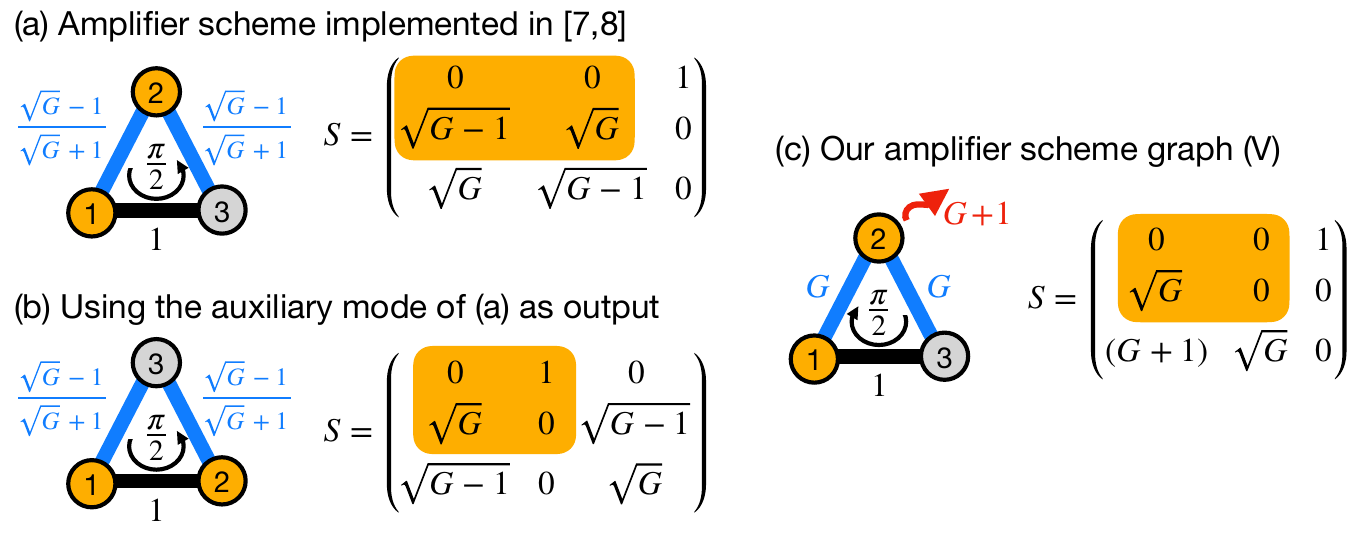}
    \caption{ 
    Comparison between our amplifier scheme graph (V) and other three-mode amplifier setups. The edge labels correspond to the respective cooperativities. The part of the scattering matrix describing the scattering between input port (mode 1) and output port (mode 2) is highlighted in orange. To simplify the comparison, we ignore any phases in the scattering matrices. (a) Amplifier scheme implemented in \cite{sliwa2015reconfigurable,lecocq_nonreciprocal_2017}. (b) Same setup as in (a) with interchanged roles of the output mode and the auxiliary mode. (c) Our proposed amplifier scheme graph (V). The red arrow in (b) corresponds to the intrinsic loss channel of the output mode, and its label is the rescaled intrinsic loss rate, see Eq.~(F4) in Appendix F.}
    \label{fig:compare_graph_V}
\end{figure}


In this section, we compare our graph (V) with previously proposed three-mode amplifier schemes. In \cref{fig:compare_graph_V}(a,b), we show the amplifier scheme implemented in \cite{sliwa2015reconfigurable,lecocq_nonreciprocal_2017}. The schemes in (a) and (b) use the same setup, however, the roles of the output mode and auxiliary mode are interchanged. 

The structure of our graph (V), see \cref{fig:compare_graph_V}(c), is similar to the schemes in (a) and (b). However, there are important differences:
(i) Only in our graph the output is coupled to an intrinsic loss channel whose decay is by a factor of $G+1$ stronger than the coupling to the waveguide;
(ii) The cooperativites  have different functional dependencies on the gain rate $G$;
(iii) In comparison to (b), different modes (auxiliary mode or output mode) are coupled to the input mode via the beamsplitter interaction.

The scheme in (a) reflects noise at the output port, and the scheme in (b) scatters this noise to the input port. Therefore, when considering a hot output port, the schemes in (a) and (b) suffer from non-quantum-limited noise being scattered towards either the output or input port. In contrast, the added noise in our graph (V) is fully quantum limited as the noise at the output port is scattered exclusively to the intrinsic loss channel of the output port and the loss channel of the auxiliary mode. This can be seen best by comparing graph (V) to graph (IV), see \cref{fig:amplifier_all_solutions}.
In graph (IV), effectively, the same intrinsic loss  as in graph (V) is achieved by coupling the output mode to an additional auxiliary mode (mode 4 in graph (IV)). Indeed, graph (V) can be derived from graph (IV) by integrating out mode 4, see Supplementary Section \ref{sec:effective_lindblad}. 
The interpretability offered by our automated design approach allows us to use symbolic regression to derive closed analytical expressions for the general behavior of the discovered amplifiers. In \cref{equ:scattering_matrix_graph_IV}, we provide the full scattering matrix for graph (IV), which also describes the scattering from and to the auxiliary modes and their intrinsic loss channels. From this expression, we see that the noise from the hot output port (mode 2) is scattered exclusively to those intrinsinc loss channels (mode 3 and 4).

For completeness, we note that the amplifier scheme proposed in \cite{metelmann_nonreciprocal_2015} represents the special limiting case of \cref{fig:compare_graph_V}(a) in which the auxiliary mode is fastly decaying ($\kappa_3 \to \infty$) and acts as Markovian baths, see also \cref{sec:effective_lindblad}. The scheme presented in \cite{ranzani_graph-based_2015} is a generalisation of \cref{fig:compare_graph_V}(a). Also here, the noise at the output port is back-reflected.

\section{Bus modes} 
In this section, we give more details on how we have defined asymptotic building blocks and how we have incorporated them into {\sc AutoScatter} to discover asymptotic solutions.

\subsection{Far-detuned bus modes}

To discover schemes in which one or more bus modes mediate purely coherent interactions between the port modes we introduce as an additional building block of our discrete optimization a far-detuned bus mode.  Formally, we set the decay rates of this type of mode to zero and at the same time assume that their detuning is not zero. This allows us to eliminate all far-detuned modes and incorporate their effect into an effective Hamiltonian for the remaining modes, as shown below.

For $N$  standard modes $\hat{a}_j$  and $M$ far-detuned modes $\hat{b}_j$ the most general quadratic Hamiltonian reads:
\begin{equation}
    \hat{H} =\frac{1}{2} \sum_{jk=1}^{NN} \left(g^{(aa)}_{jk} \hat{a}_j^\dag \hat{a}_k +  \nu^{(aa)}_{jk} \hat{a}_j^\dag \hat{a}_k^\dag\right) + \frac{1}{2}\sum_{jk=1}^{MM} \left(g^{(bb)}_{jk} \hat{b}_j^\dag \hat{b}_k +  \nu^{(bb)}_{jk} \hat{b}_j^\dag \hat{b}_k^\dag \right)+\sum_{jk=1}^{NM} \left(g^{(ab)}_{jk} \hat{a}_j^\dag \hat{b}_k +  \nu^{(ab)}_{jk} \hat{a}_j^\dag \hat{b}_k^\dag\right) +\mathrm{H.c.},
\end{equation}
where $g^{(aa)}$ and $g^{(bb)}$ are hermitian matrices, $\nu^{(aa)}$ and $\nu^{(bb)}$ are symmetric matrices, and $g^{(ab)}$ and $\nu^{(ab)}$ are arbitrary matrices.
To write the effective Hamiltonian compactly, it is convenient to define the matrices
\begin{equation}
	H^{(ss')}_{\rm BdG}=
 \begin{pmatrix}
 g^{(ss')}&\nu^{(ss')}\\
 \nu^{(ss')*}&g^{(ss')*}\\
 \end{pmatrix}
\end{equation}
with  $s,s'=a$ or $b$ and $g^{(ba)}=g^{(ab)*}$ and $\nu^{(ba)}=\nu^{(ab)}$. These matrices can be viewed as blocks of the underlying Bogoliubov de Gennes Hamiltonian. The effective Bogoliubov de Gennes Hamiltonian can be written as
\begin{equation}    H^\mathrm{({\rm eff})}_{{\rm BdG}} = H^\mathrm{(aa)}_{{\rm BdG}} - H^\mathrm{(ab)}_{{\rm BdG}}  (H^\mathrm{(bb)}_{{\rm BdG}} )^{-1} H^\mathrm{(ba)}_{{\rm BdG}}. 
\end{equation}
The corresponding  second-quantized Hamiltonian reads 
\begin{equation}
    \hat{H} =\frac{1}{2} \sum_{jk=1}^{NN} \left(g^{(\rm eff)}_{jk} \hat{a}_j^\dag \hat{a}_k +  \nu^{(\rm eff)}_{jk} \hat{a}_j^\dag \hat{a}_k^\dag\right)  + \mathrm{H.c.}
\end{equation}
with
\begin{equation}
	H^{({\rm eff})}_{\rm BdG}=
 \begin{pmatrix}
 g^{({\rm eff})}&\nu^{({\rm eff})}\\
 \nu^{({\rm eff})*}&g^{({\rm eff})*}\\
 \end{pmatrix}
\end{equation}
Next, we assume that the  Bogoliubov de Gennes Hamiltonian $H^{(bb)}_{\rm BdG}$ projected on the far-detuned modes  had been diagonalized to eliminate any coupling between the bus modes. In other words, $\nu^{(bb)}$ is zero and  $g^{(bb)}$ is diagonal with  $g^{(bb)}_{ii}=-\Delta^{(b)}_i$. In this case, we find
\begin{eqnarray}
g_{ij}^{(\rm eff)}&=&g_{ij}^{(aa)}+\sum_{k}\frac{g^{(ab)}_{ik}g^{(ab)*}_{kj}}{\Delta^{(b)}_k}+\frac{\nu^{(ab)}_{ik}\nu^{(ab)*}_{kj}}{\Delta^{(b)}_k}\\
\nu_{ij}^{(\rm eff)}&=&\nu_{ij}^{(aa)}+\sum_{k}\frac{g^{(ab)}_{ik}\nu^{(ab)}_{kj}}{\Delta^{(b)}_k}+\frac{\nu^{(ab)}_{ik}g^{(ab)}_{kj}}{\Delta^{(b)}_k}.
\end{eqnarray}
Thus, the rescaled Bogoliubov Hamiltonian in Eq.~(11a) is modified as following
\begin{eqnarray}
H_{ij}&=&\frac{g_{ij}^{(aa)}}{\sqrt{\kappa_i\kappa_j}}+\sum_{k}{\rm sgn}(\Delta_k)\left(\frac{g^{(ab)}_{ik}}{\sqrt{\kappa_i\abs{\Delta^{(b)}_k}}}\frac{g^{(ab)*}_{kj}}{\sqrt{\kappa_j\abs{\Delta^{(b)}_k}}}+\frac{\nu^{(ab)}_{ik}}{\sqrt{\kappa_i\abs{\Delta^{(b)}_k}}}\frac{\nu^{(ab)*}_{kj}}{\sqrt{\kappa_j\abs{\Delta^{(b)}_k}}}\right).
\end{eqnarray}
or
\begin{eqnarray}
H_{ij}&=&\frac{\nu_{ij}^{(aa)}}{\sqrt{\kappa_i\kappa_j}}+\sum_{k}{\rm sgn}(\Delta_k)\left(\frac{g^{(ab)}_{ik}}{\sqrt{\kappa_i\abs{\Delta^{(b)}_k}}}\frac{\nu^{(ab)}_{kj}}{\sqrt{\kappa_j\abs{\Delta^{(b)}_k}}}+\frac{\nu^{(ab)}_{ik}}{\sqrt{\kappa_i\abs{\Delta^{(b)}_k}}}\frac{g^{(ab)}_{kj}}{\sqrt{\kappa_j\abs{\Delta^{(b)}_k}}}\right).
\end{eqnarray}
We use the dimensionless parameters  and $g^{(ab)}_{ik}/\sqrt{\kappa_i\Delta^{(b)}_k}$ with $k=1,...,M$ and $i=1,...,N$ as additional learning parameters of our continuous optimization. The signs ${\rm sgn}(\Delta_k)$ remain fixed during the continuous optimization and are, instead, chosen within the discrete optimization.  Note the different scaling of the learning parameters encoding the interaction with a bus mode with the detuning $\Delta^{(b)}_k$ playing the role of the decay rate $\kappa^{(b)}_k$. This scaling reflects that the coherent interactions mediated by the bus modes become independent of the bus mode decay $\kappa^{(b)}_k$ in the asymptotic limit of large cooperativities,  $4|g^{(ab)}_{ik}|^2/|\kappa_i\kappa^{(b)}_k|\to\infty$. Similar to  the decay rates $\kappa_i$ of the standard modes, we can view the detunings $\Delta_i^{(b)}$ of the far-detuned bus mode as  free parameters (possibly subject to some stability constraints).

\section{Details on the the analysis of the search space reduction}
\begin{figure}
    \centering
    \includegraphics[width=0.95\textwidth]{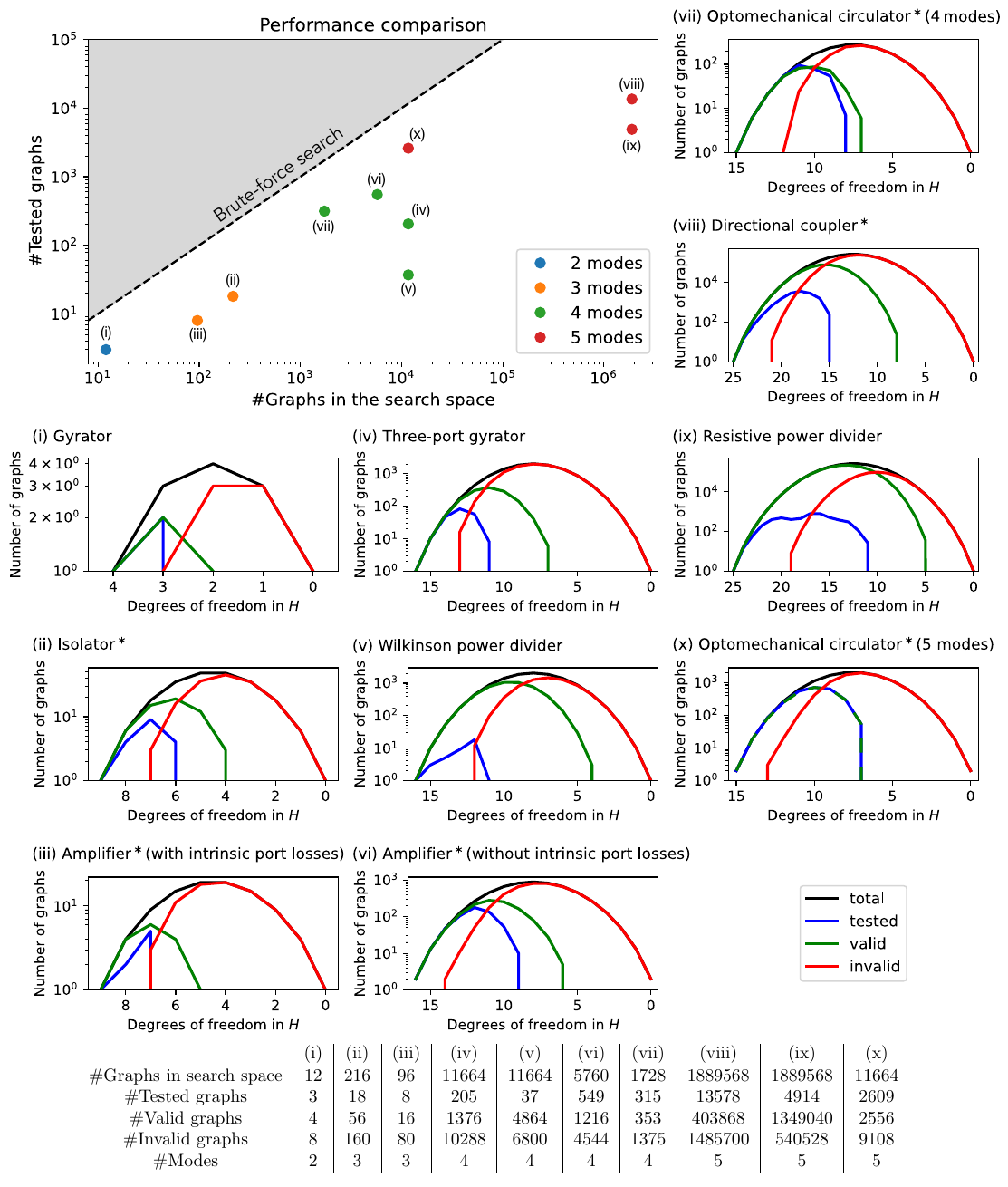}
    \caption{Performance of our optimization approach during the graph pruning. We show the number of graphs tested by our optimization in comparison to the total number of graphs within the search space. A brute-force search would have to test every graph (dashed black line). Each data point corresponds to a different target functionality/device, see (i--x) and \cref{sup:summary_examples} for the description of each device. (i--x) Structure of the search space for each target device. 
    Shown is the total number of graphs (black), the number valid and invalid graphs (green and red), and the number of graphs tested by our optimization scheme (blue) as a function of the number of graph weights or, equivalently, the free parameters in the dimensionless Bogoliubov de Gennes Hamiltonian $H$.  All devices marked with an asterisk are also discussed in the main text. 
    The dashing in (x) is meant to make both the curve of valid and tested graphs visible as both lie almost on top of each other. The table in the bottom summarizes the overall number of graphs of each type for different devices.
    }
    \label{fig:compare_to_brute_force}
\end{figure}

In Fig.~7 in the main text, we analyze how much our optimization scheme reduces the size of the search space for a variety of target functionalities.
In this section, we provide more details on this comparison and discuss the structure of the search spaces of the compared target functionalities.


In \cref{fig:compare_to_brute_force}, we show again the number of graphs tested by our optimization scheme in comparison to the total number of graphs within the search space (the top left plot is identical with Fig.~7 in the main text). Each data point corresponds to a different 
target device, as identified by the target scattering matrix. 
As described in the main text, we have defined the performance gain as the ratio of the number of graphs in the search space over the number of tested graphs. This ratio varied in a broad range from $\sim 5$ (see, e.g., (vii)) to $\sim 400$ in our experiments (see (ix)). Given the large variations of the data for an equal number of modes and search space dimensions, a general quantitative analysis seems not feasible.

To give more insight into these variations 
we have plotted the number of graphs (black), the number valid and invalid graphs (green and red), and the number of graphs 
tested by our optimization scheme (blue) as a function of the number of free parameters in the dimensionless Hamiltonian $H$, which collects the graph weights, see \cref{fig:compare_to_brute_force}(i--x). The number of free parameters in $H$ is maximum at the beginning of the pruning algorithm and is progressively reduced following a breadth-first algorithm. Our optimization skips every graph, which is a subgraph of an invalid graph. Therefore, if there are many invalid graphs in the top part of the search space, these invalid graphs will have a lot of invalid subgraphs later on, which our optimization can skip. For example, the search spaces of (iv) and (vi) have many invalid graphs at their top, which leads to a large performance gain in our optimization scheme. Another mechanism that can lead to large performance gain occurs whenever our continuous optimization scheme finds a solution in which one or more free parameters turn out to be zero. This can occur because of constraints in the ideal solutions, e.g., dictated by some underlying symmetry. In this scenario, our algorithm removes the corresponding graph edges from the tested graph and adds this smaller graph to the library of valid graphs, see Appendix A. Simultaneously, the remaining space of graphs to be tested is reduced, as all extensions of the smaller graph are automatically added to the list of valid graphs, and, thus, will be skipped. Since we use a breadth-first algorithm, this is the only scenario leading to the number of tested graphs being smaller than the number of valid graphs. Hence, in cases in which this mechanism is not relevant (just one case in our experiments), the number of tested graphs is strictly larger than the number of valid graphs, see \cref{fig:compare_to_brute_force}(x). On the other hand, for all but one of our experiments, the number of tested graphs is smaller indicating that this mechanism is important. This is particularly true in the later stage of the discrete optimization after many edges have already been removed, see \cref{fig:compare_to_brute_force}(i--ix).  

\section{Brief description of the devices shown in Fig.~7}
\label{sup:summary_examples}
Here, we provide a brief description of the illustrative examples we have considered  in Fig.~7 and \cref{fig:compare_to_brute_force}. We put a special emphasis on those examples that are not yet described in the main text.

\textbf{(i) Gyrator:} A gyrator transmits signals between its two ports. It is nonreciprocal and applies a phase shift of $\pi$ depending on the direction of the transmission. Its scattering matrix reads \cite{Pozar_microwave_engineering}
\begin{equation}
    S_\mathrm{target}=
    \begin{pmatrix}
        0 & 1 \\
        -1 & 0
    \end{pmatrix}
\end{equation}
No auxiliary modes are required to realize this device.

\textbf{(ii) Isolator:} An isolator is a two-port device with perfect transmission from one port to the other and zero transmission in the reverse direction, see Section III in the main text for more details.
One auxiliary mode is required to realize this device.

\textbf{(iii, vi) Directional amplifier:} This two-port device amplifies signals from the input to the output port, with zero transmission in the reverse direction. Furthermore, the noise scattered towards the input and output ports has to be quantum limited. See Section V for an extended discussion of our results. When allowing for an additional intrinsic loss channel on the output port, one auxiliary mode is sufficient to realize the target behavior, otherwise, two auxiliary modes are required. The performance of the graph pruning process is shown for both situations in \cref{fig:compare_to_brute_force}(iii) and (vi), respectively.

\textbf{(iv) Three-port gyrator:} The authors of \cite{shekel1953gyrator} have proposed a three-port generalization of the gyrator, example  (i). Their device has the scattering matrix
\begin{equation}
    S_\mathrm{target}=
    \begin{pmatrix}
        0 & t & -t \\
        -t & 0 & t\\
        t & -t & 0
    \end{pmatrix}
\end{equation}
Here, the transmission rate $t$ is a learnable parameter during the continuous optimization. One auxiliary mode is required to realize this target behavior.

\textbf{(v) Wilkinson power divider:} This three-port device splits a  signal from the first port equally. The two halves are fed into the second and third ports, respectively. In the reverse direction, it can also be used as a coupler, since the incoming signals at the second and third ports are transmitted equally to the first port. The scattering matrix of a  Wilkinson power divider reads \cite{Pozar_microwave_engineering}
\begin{equation}
    S_\mathrm{target}=-\frac{\mathrm{i}}{\sqrt{2}}
    \begin{pmatrix}
        0 & 1 & 1 \\
        1 & 0 & 0\\
        1 & 0 & 0
    \end{pmatrix}    
\end{equation}
We require one auxiliary mode to realise this scattering behavior.

\textbf{(vii, x) Optomechanical circulator:} The goal is to realize a circulator between three optical modes whose couplings have different constraints. In (vii), we allow only for passive couplings between the optical modes, meaning that the beamsplitter interactions have to be real-valued. To realize the circulator, at least one far-detuned mode is required to mediate the interactions. In (x), no direct couplings are allowed between the optical modes. With this more stringent constraint, at least two far-detuned modes are needed. The cooperativities describing the coupling to the far-detuned modes have to diverge asymptotically, see Section VI in the main text for more details.

\textbf{(viii) Directional coupler:} This three-port device couples incoming signals from two ports and sends the summed signal to an output port. Unlike the Wilkinson power divider, signals arriving at the output port are not scattered to the input ports, see Section IV in the main text for more details.

\textbf{(ix) Resistive power divider:} Similar to the Wilkinson power divider, this device divides an incoming signal from an input port and scatters the split signals to two output ports. Unlike the Wilkinson power divider, this device is not lossless, and a portion of the signal must be lost to intrinsic loss channels. Its scattering matrix has the form \cite{Pozar_microwave_engineering}:
\begin{equation}
    S_\mathrm{target}=
    \begin{pmatrix}
        0 & t & t \\
        t & 0 & t\\
        t & t & 0
    \end{pmatrix}      
\end{equation}
with $t$ being a transmission coefficient. Ideally, this transmission coefficient assumes the value $1/2$, meaning that $S$ is not unitary and loss is required to realize this scattering matrix. There a two ways to add the required loss channels: One can add two auxiliary modes, whose loss channels will dissipate the energy. We run our optimisation approach for this scenario in \cref{fig:compare_to_brute_force}(ix). Alternatively, one can allow for intrinsic losses on the port modes. In this case, no auxiliary modes are needed.

\end{appendix}
\bibliography{artificial_discovery.bib}